\pdfoutput=1

\documentclass[acmsmall]{acmart}
\AtBeginDocument{%
  }

\setcopyright{none} %
\renewcommand\footnotetextcopyrightpermission[1]{} %
\acmConference[Preprint]{}{}{}
\acmBooktitle{}
\acmPrice{}
\acmISBN{}
\acmDOI{}
\acmYear{}
\acmMonth{}
\acmVolume{}
\acmNumber{}
\acmArticle{}

\UseRawInputEncoding

\usepackage{latexsym}
\usepackage{multirow}
\usepackage{colortbl}
\usepackage{array}
\usepackage{tabularx} 
\usepackage{makecell}
\usepackage{fancyvrb}
\usepackage{ragged2e}
\usepackage{tikz}
\usetikzlibrary{positioning, arrows.meta, shapes.geometric, calc, fit, shadows}
\usepackage{graphicx}
\usepackage{tcolorbox}
\usepackage{pgfplots}
\usepackage{listings}
\usepackage{xspace}
\usepackage{setspace}
\usetikzlibrary{patterns}
\pgfplotsset{compat=1.18} 
\usepackage{fancyhdr} %

\newtcolorbox{prompt}[1]{colback=gray!20,colframe=gray!50!black,fonttitle=\bfseries,title=#1}

\newcommand{\dataset}{CRUMB\xspace}
\newcommand{\stackexchange}{StackExchange\xspace}
\newcommand{\paperretrieval}{Paper Retrieval\xspace}

\newcolumntype{L}[1]{>{\RaggedRight\arraybackslash}p{#1}}
\newcolumntype{R}{>{\RaggedRight\arraybackslash}X} %
\newcommand{\rotcol}[1]{\rotatebox{90}{\parbox{3cm}{\raggedright #1}}}
\newcolumntype{C}{>{\centering\arraybackslash}X}

\definecolor{colorMain}{RGB}{31, 119, 180} 
\definecolor{colorQTA}{RGB}{255, 127, 14}
\definecolor{colorQTD}{RGB}{44, 160, 44}
\definecolor{colorQAR}{RGB}{214, 39, 40}

\definecolor{PaperRetrievalTop}{RGB}{70 143 175} 
\definecolor{PaperRetrievalBottom}{RGB}{224 237 243}

\definecolor{TipOfTheTongueTop}{RGB}{70 143 175} 
\definecolor{TipOfTheTongueBottom}{RGB}{224 237 243}

\definecolor{CodeRetrievalTop}{RGB}{97 165 194} 
\definecolor{CodeRetrievalBottom}{RGB}{224 238 243}

\definecolor{ClinicalTrialTop}{RGB}{97 165 194} 
\definecolor{ClinicalTrialBottom}{RGB}{224 238 243}

\definecolor{TheoremRetrievalTop}{RGB}{137 194 217} 
\definecolor{TheoremRetrievalBottom}{RGB}{233 244 248}

\definecolor{StackExchangeTop}{RGB}{137 194 217} 
\definecolor{StackExchangeBottom}{RGB}{233 244 248}

\definecolor{SetOpsTop}{RGB}{169 214 229} 
\definecolor{SetOpsBottom}{RGB}{237 247 250}

\definecolor{LegalQATop}{RGB}{169 214 229} 
\definecolor{LegalQABottom}{RGB}{237 247 250}

\pgfplotsset{
    mybarstyle/.style 2 args={
        ybar,
        bar width=5pt,
        draw=#1,      
        fill=#1!60,
        pattern color=#1,   
        postaction={pattern=#2},
    }
}

\tikzset{
  info card/.pic={
    \tikzset{
      /info card/.cd,
      #1 %
    }
    \def\cardwidth{\pgfkeysvalueof{/info card/width}}
    \def\cardheight{\pgfkeysvalueof{/info card/height}}
    \def\topcolor{\pgfkeysvalueof{/info card/top color}}
    \def\bottomcolor{\pgfkeysvalueof{/info card/bottom color}}
    \def\theemoji{\pgfkeysvalueof{/info card/emoji}}
    \def\theicon{\pgfkeysvalueof{/info card/icon}}
    \def\thetitle{\pgfkeysvalueof{/info card/title}}
    \def\thesubtitle{\pgfkeysvalueof{/info card/subtitle}}
    \def\thebody{\pgfkeysvalueof{/info card/body}}
    \node (-base) [
      fill=\bottomcolor, 
      rounded corners=10pt, 
      text width=\cardwidth, 
      minimum height=\cardheight, 
      font=\sffamily,
      anchor=north,
    ] {};
    \node (-top) [
      fill=\topcolor, 
      rounded corners=10pt, 
      anchor=north, 
      minimum height=1.0cm, 
      text width=\cardwidth, %
      font=\sffamily
    ] at (-base.north) {};
    \ifx\theicon\empty
      \node at (-top.west) [
        anchor=west, 
        xshift=5mm, 
        font=\Huge
      ] {\setmainfont{Noto Color Emoji}\theemoji}; %
    \else
      \node at (-top.west) [
        anchor=west, 
        xshift=1mm,
        inner sep=0pt %
      ] {\includegraphics[height=0.8cm]{\theicon}}; %
    \fi
    \node at (-top.west) [
      text width=\cardwidth, 
      align=left, 
      anchor=west, 
      xshift=0.8cm
    ] {
      \textbf{\textcolor{black}{\thetitle}} \\ \small \textcolor{black}{\thesubtitle}
    };
    \node at (-top.south west) [
      text width=\cardwidth - 2.1mm, 
      align=left, 
      anchor=north west, 
      xshift=2mm, 
      font={\fontsize{8pt}{9pt}\selectfont\sffamily}
    ] {
      \thebody
    };
  },
  /info card/.cd,
  width/.initial={7cm},
  height/.initial={2.4cm},
  top color/.initial={red!80},
  bottom color/.initial={red!40},
  emoji/.initial={⭐}, %
  icon/.initial={},      %
  title/.initial={Default Title},
  subtitle/.initial={Default Subtitle},
  body/.initial={Default body text.},
}

\begin{document}

\title{Benchmarking Information Retrieval Models on Complex Retrieval Tasks}

\author{Julian Killingback}
\email{jkillingback@cs.umass.edu}
\author{Hamed Zamani}
\email{zamani@cs.umass.edu}
\affiliation{%
  \institution{University of Massachusetts Amherst}
  \city{Amherst}
  \state{Massachusetts}
  \country{USA}
}

\begin{abstract}
Large language models (LLMs) are incredible and versatile tools for text-based tasks that have enabled countless, previously unimaginable, applications. Retrieval models, in contrast, have not yet seen such capable general-purpose models emerge. To achieve this goal, retrieval models must be able to perform \emph{complex retrieval tasks}, where queries contain multiple parts, constraints, or requirements in natural language. These tasks represent a natural progression from the simple, single-aspect queries that are used in the vast majority of existing, commonly used evaluation sets. Complex queries naturally arise as people expect search systems to handle more specific and often ambitious information requests, as is demonstrated by how people use LLM-based information systems. Despite the growing desire for retrieval models to expand their capabilities in complex retrieval tasks, there exist limited resources to assess the ability of retrieval models on a comprehensive set of diverse complex tasks. The few resources that do exist feature a limited scope and often lack realistic settings making it hard to know the true capabilities of retrieval models on complex real-world retrieval tasks. To address this shortcoming and spur innovation in next-generation retrieval models, we construct a diverse and realistic set of complex retrieval tasks and benchmark a representative set of state-of-the-art retrieval models. Additionally, we explore the impact of LLM-based query expansion and rewriting on retrieval quality. Our results show that even the best models struggle to produce high-quality retrieval results with the highest average nDCG@10 of only 0.346 and R@100 of only 0.587 across all tasks. Although LLM augmentation can help weaker models, the strongest model has decreased performance across all metrics with all rewriting techniques. We explore the potential causes for poor performance, as well as what features the best performing models have and why they might help. Our findings suggest that there is still work to be done to improve the quality of retrieval models on complex search tasks. We hope that by sharing our unified complex evaluation set we can set a new standard for complex retrieval evaluation and spur innovation in retrieval models. The data and artifacts are available at https://github.com/jfkback/crumb.
\end{abstract}

\maketitle
\pagestyle{fancy}     %
\fancyhf{}            %
\renewcommand{\headrulewidth}{0pt} %

\fancyhead[LE,RO]{\footnotesize  \thepage} %
\fancyhead[LO,RE]{\footnotesize \shortauthors}   %

\thispagestyle{plain}

\section{Introduction}
Information retrieval (IR) systems are fundamental to accessing and organizing the vast repositories of digital information available today. Traditional IR approaches have predominantly focused on matching queries with documents through keyword overlap or basic semantic similarity. These systems, while effective for simple search scenarios, frequently fall short when faced with complex information needs that contain multiple aspects \cite{quest, bright, tip_of_the_tongue_original}. In large part because these aspects can often require nuanced understanding to properly untangle and weight while determining relevance. This problem has become especially important recently, as the broad adoption of large language models (LLMs) is changing how users expect information systems to work \cite{comparing_traditional_and_llm_based_search, what_do_users_really_ask_large_language_models}. As user expectations for information access and retrieval systems increase, it is crucial to quantify and understand how well current state-of-the-art retrieval paradigms and models, such as LLM-based dense and sparse retrieval models, perform on a wide range of complex retrieval tasks. This paper provides insights which highlight the limitations of the current state-of-the-art in information retrieval systems and helps inform the development of the next generation of IR technology.

The first step towards this goal is to prepare a suitable comprehensive benchmark that focuses on \emph{diverse complex retrieval tasks}. We believe that existing resources are not sufficient. For example, standard retrieval benchmarks provided by TREC ad hoc retrieval tracks \cite{trec_robust_2004_overview, trec_common_core_2017_overview}, TREC Web Tracks \cite{trec_web_track_2013_overview, trec_web_track_2014_overview}, MSMARCO \cite{msmarco}, and TREC Deep Learning Track \cite{craswell2020overview, craswell2021overviewdl} predominantly feature keyword-based queries or simple natural language questions that generally have a singular focus. Although these datasets have driven substantial improvements in retrieval models, the frequent reliance on existing search systems to source queries results in collections that primarily capture the subset of queries where users expect existing systems to succeed \cite{students_mental_model_of_information_retrieval_systems, where_do_queries_come_from, relationship_between_query_characterisitics_and_retrieval_bias, what_do_users_really_ask_large_language_models, web_searchers_attributions_of_success_and_failure}. This bias results in the failure to capture the true breadth and complexity that is inherent to real-world information-seeking behavior \cite{what_do_users_really_ask_large_language_models, comparing_traditional_and_llm_based_search}.
The BEIR \cite{beir} benchmark attempted to aggregate diverse datasets to evaluate zero-shot capabilities of retrieval models. However, BEIR is still primarily composed of common retrieval tasks with limited query complexity. Additionally, some of BEIR's datasets have problems such as (1) using answers to questions as target documents despite answers for a specific question generally having higher lexical overlap, (2) including tasks that are significantly different from standard retrieval tasks such as citation prediction and fact verification, and (3) standardizing the labels from original datasets in a way that results in incorrect or unsupported labels \cite{rank1}. 
More recently, \citet{bright} introduced the BRIGHT benchmark to evaluate retrieval on reasoning-intensive tasks. While BRIGHT addresses some limitations by focusing on reasoning-intensive retrieval tasks that often contain multiple requirements or sub-questions, the queries tend to have limited breadth and exclude many types of complex retrieval tasks. 
For example, complex queries where multiple requirements are combined using logical operations such as \textit{and}, \textit{or}, and \textit{not} do not require reasoning and are thus missing from BRIGHT, but are an important group of complex queries that are relevant to many real-world search scenarios. Additionally, BRIGHT has some practical limitations, one major issue is that for the tasks derived from Stack Exchange each subject only has between 500-2000 documents making full-document evaluation unrealistic when compared to most real-world document collections. Additionally, the naive scraping and chunking approach results in many chunks solely containing webpage boilerplate or chunks without sufficient context.

\begin{figure}
\centering
\begin{tikzpicture}
  \pic at (0,0) {info card={
      top color=PaperRetrievalTop,
      bottom color=PaperRetrievalBottom,
      width=(\textwidth - 8mm) / 2,
      icon={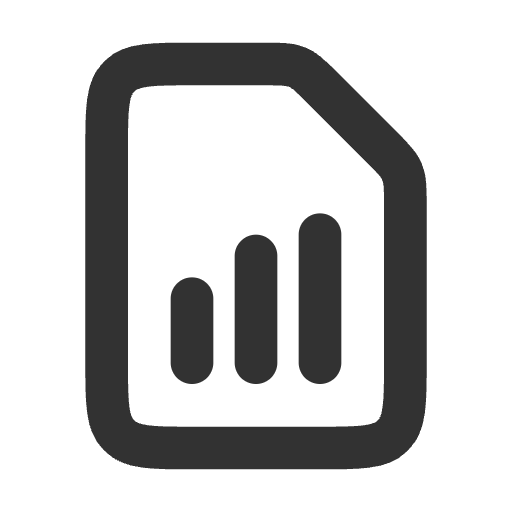},
      title={Paper Retrieval},
      subtitle={Multi-aspect Paper Criteria Queries},
      body={My goal is to develop a learning model that can handle multiple tasks simultaneously. This proposed learning model will function as a sub-model selector, meaning that when presented...}
    }
  };

  \pic at (\textwidth / 2, 0) {info card={
      top color=TipOfTheTongueTop,
      bottom color=TipOfTheTongueBottom,
      width=(\textwidth - 8mm) / 2,
      icon={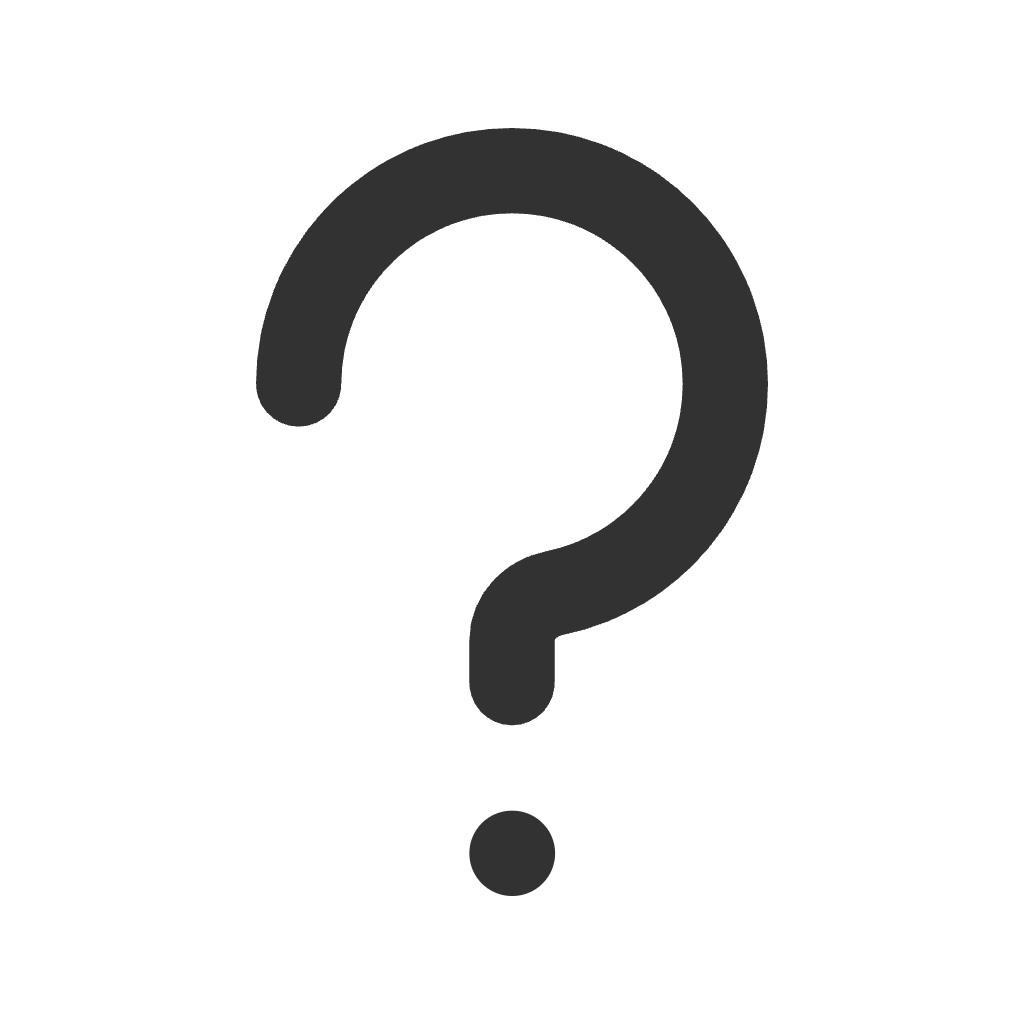},
      title={Tip-of-the-Tongue},
      subtitle={Vague, Multi-detail Queries},
      body={Cop's son needs blood transfusion from an inmate. A cop's son is very sick and needs a certain type of blood transfusion. He finds out an inmate has the same type and once agreed the inmate...}
    }
  };

  \pic at (0,-2.6) {info card={
      top color=CodeRetrievalTop,
      bottom color=CodeRetrievalBottom,
      width=(\textwidth - 8mm) / 2,
      icon={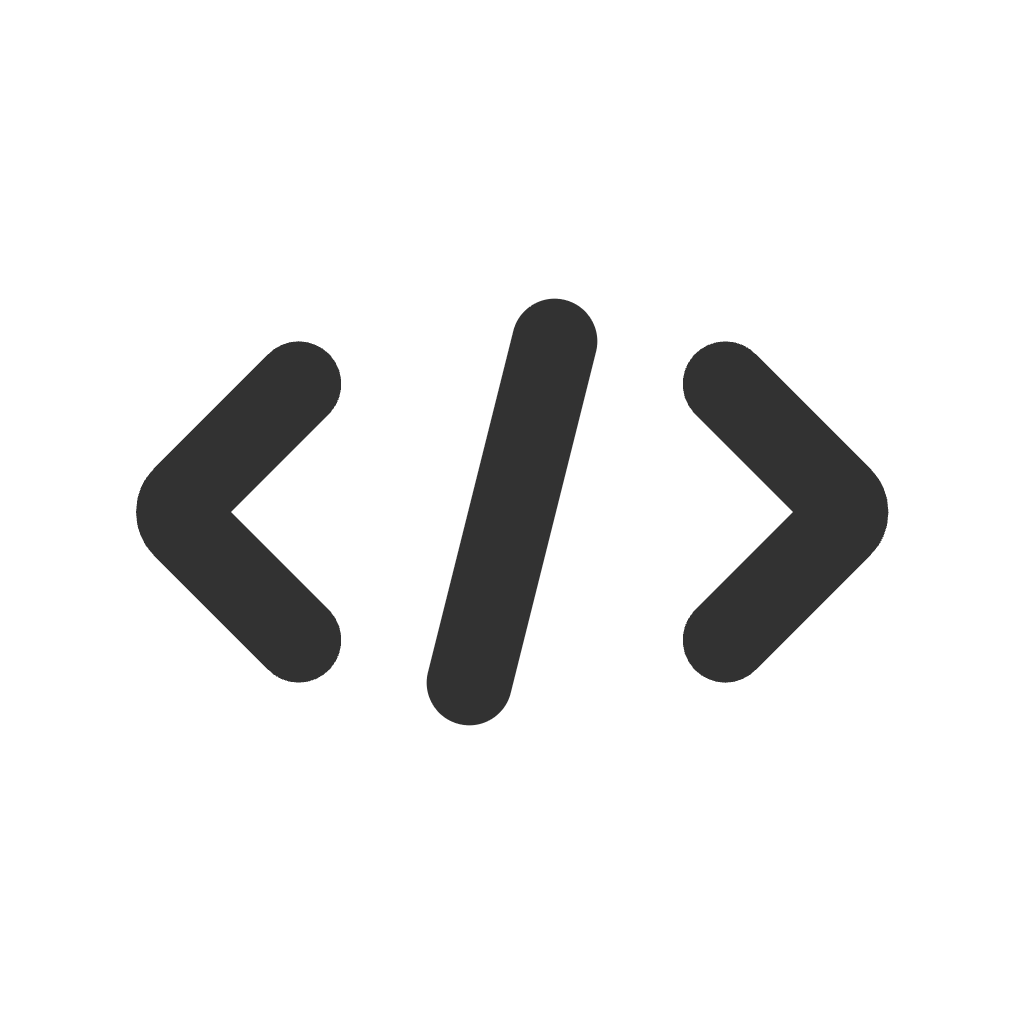},
      title={Code Retrieval},
      subtitle={Coding Problem Queries},
      body={On the way to school, Karen became fixated on the puzzle game on her phone! The game is played as follows. In each level, you have a grid with n rows and m columns. Each cell originally contains...}
    }
  };

  \pic at (\textwidth / 2, -2.6) {info card={
      top color=ClinicalTrialTop,
      bottom color=ClinicalTrialBottom,
      width=(\textwidth - 8mm) / 2,
      icon={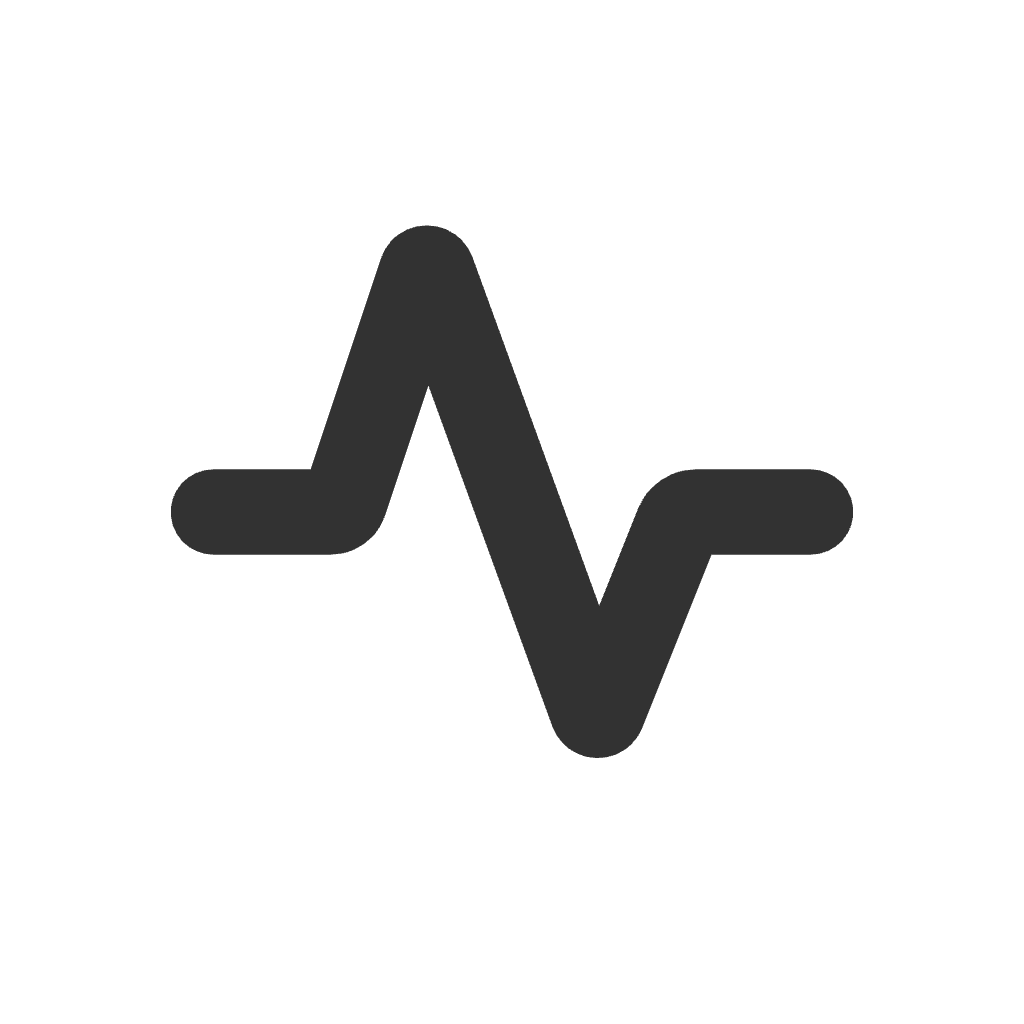},
      title={Clinical Trial Retrieval},
      subtitle={Patient History Queries},
      body={Patient is a 45-year-old man with a history of anaplastic astrocytoma of the spine complicated by severe lower extremity weakness and urinary retention s/p Foley catheter, high-dose steroids...}
    }
  };

  \pic at (0,-5.2) {info card={
      top color=TheoremRetrievalTop,
      bottom color=TheoremRetrievalBottom,
      width=(\textwidth - 8mm) / 2,
      icon={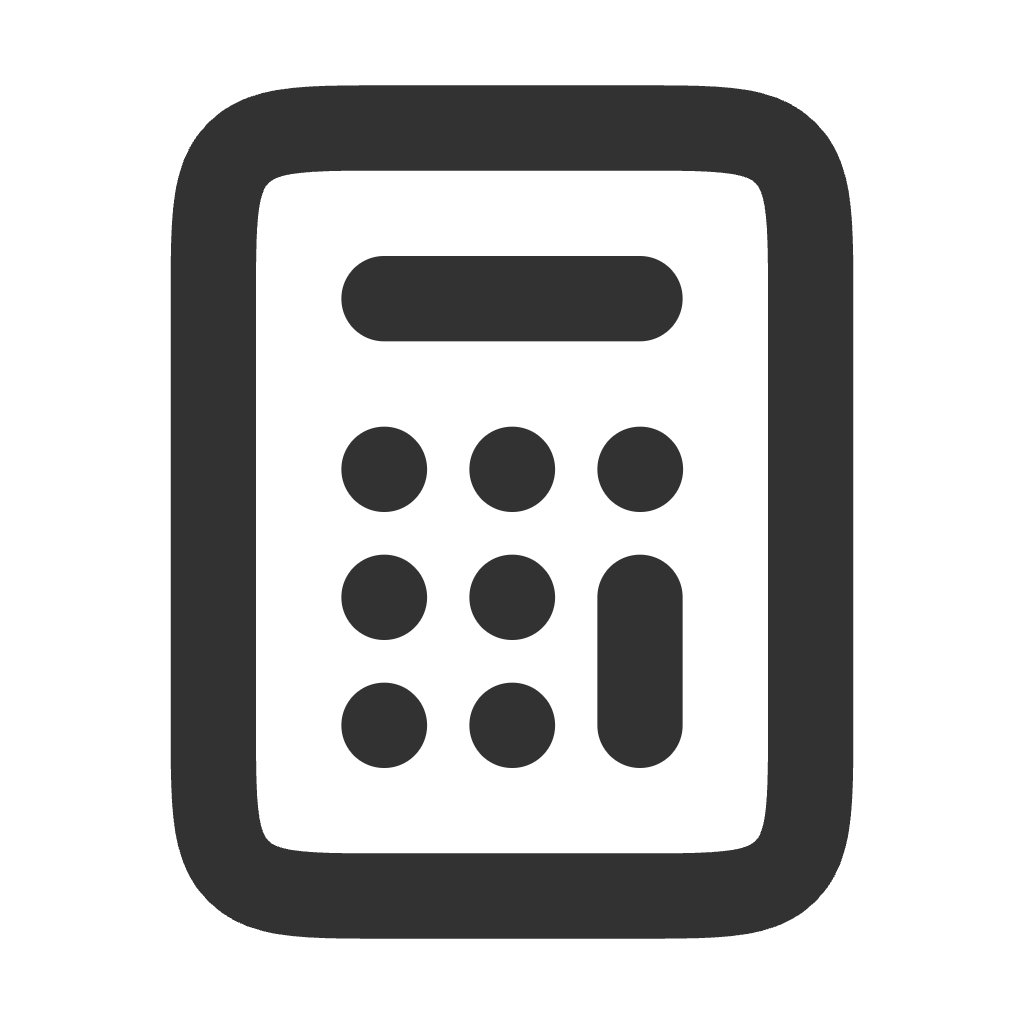},
      title={Theorem Retrieval},
      subtitle={Mathematical Problem Queries},
      body={Imagine you have a digital scale that can measure the weight of an infinite number of infinitely small digital dots. Each dot can either be on or off, and their weights are determined by a specific pattern...}
    }
  };

  \pic at (\textwidth / 2,-5.2) {info card={
      top color=StackExchangeTop,
      bottom color=StackExchangeBottom,
      width=(\textwidth - 8mm) / 2,
      icon={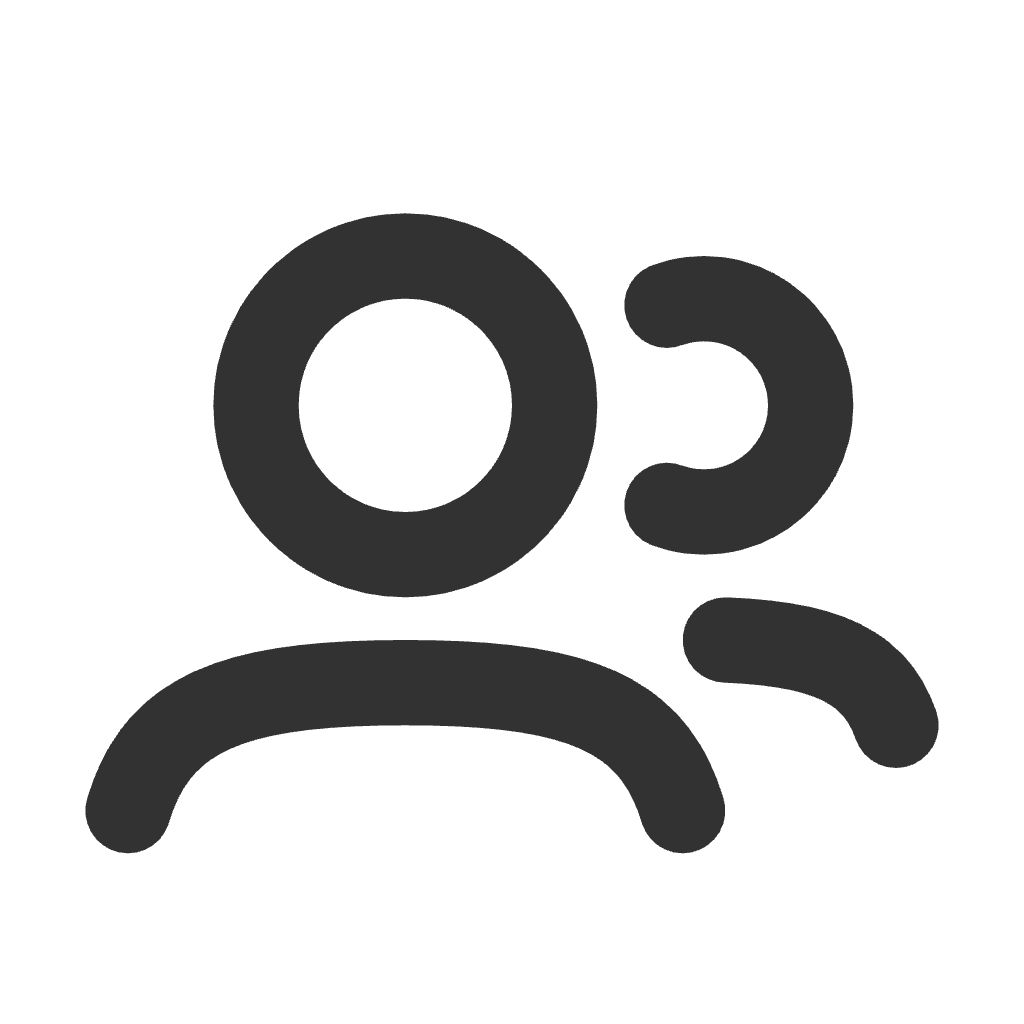},
      title={StackExchange QA},
      subtitle={Community Questions Requiring Reasoning},
      body={Does the genetic sequence of SARS-CoV-2 end with 33 A's? Looking at the DNA (or RNA?) sequence of the Covid-19 virus here: https://www.ncbi.nlm.nih.gov/nuccore/MN908947.3...}
    }
  };

  \pic at (0,-7.8) {info card={
      top color=LegalQATop,
      bottom color=LegalQABottom,
      width=(\textwidth - 8mm) / 2,
      height=1.8cm,
      icon={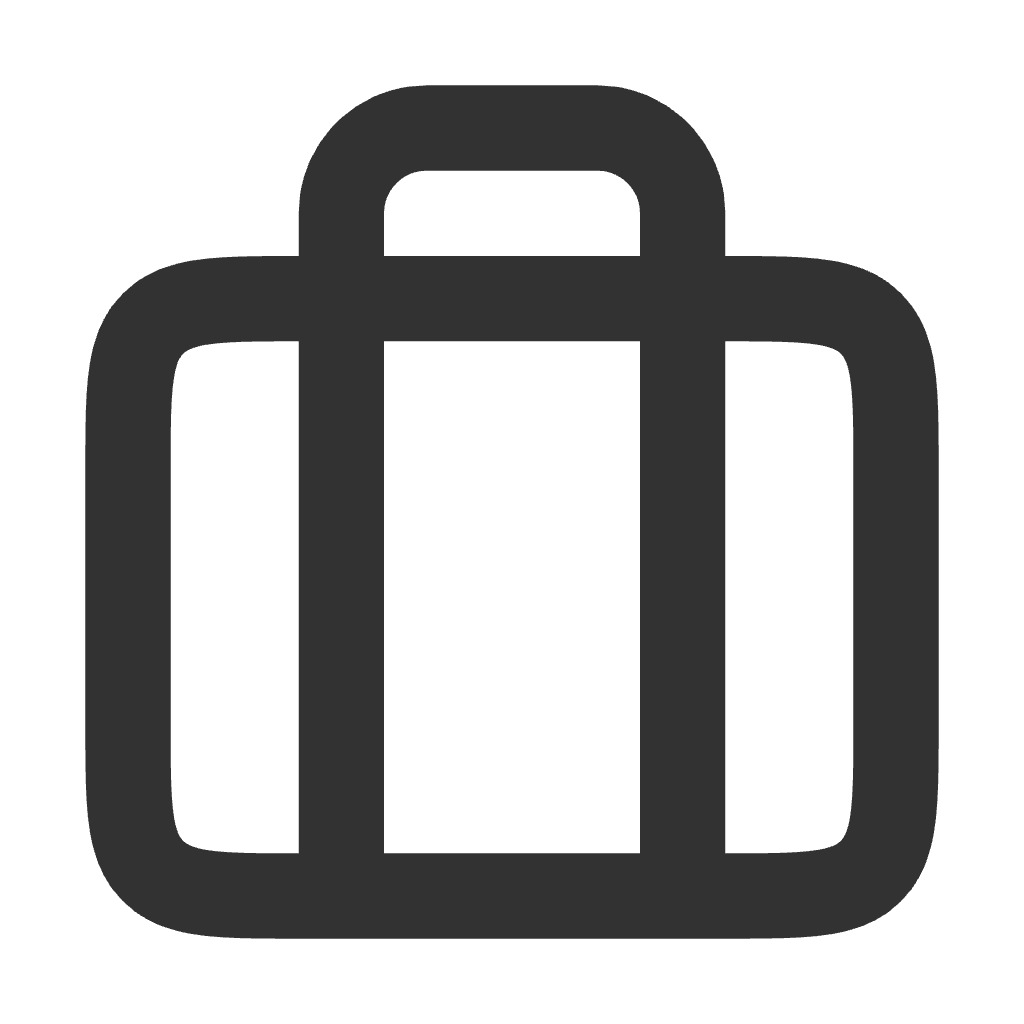},
      title={Legal QA},
      subtitle={Legal Queries with Geographic Constraints},
      body={Are eviction cases first heard in high court? In the state of Tennessee}
    }
  };

  \pic at (\textwidth / 2, -7.8) {info card={
      top color=SetOpsTop,
      bottom color=SetOpsBottom,
      width=(\textwidth - 8mm) / 2,
      height=1.8cm,
      icon={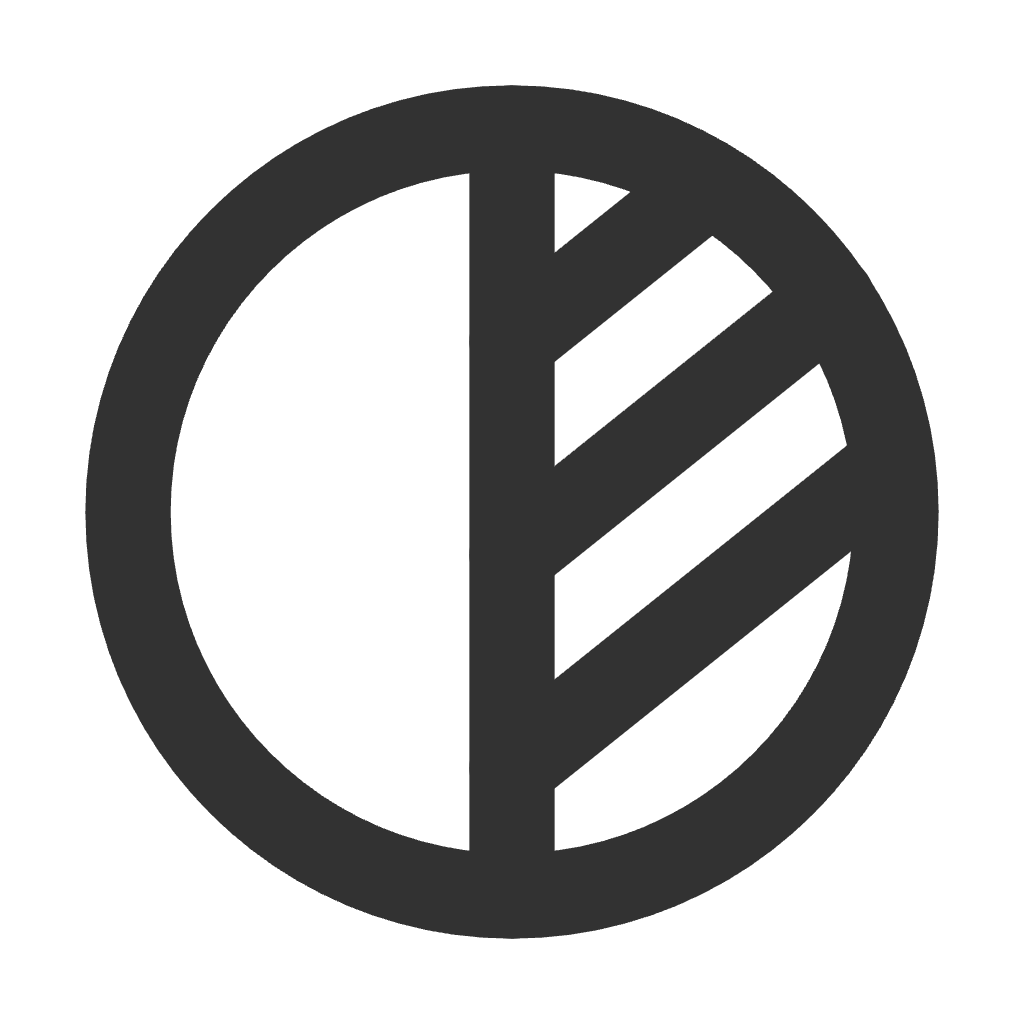},
      title={SetOps},
      subtitle={Entity Queries with Set-based Operations},
      body={German spy comedy films, or 2000s comedy-drama mystery films.}
    }
  };
\end{tikzpicture}
\caption{An overview of our proposed benchmark: \dataset. Each card is dedicated to one of the complex tasks in \dataset. The card header includes the name of the task and a short overview of the query type. The body of the card gives a truncated example query.}
\label{fig:crumb_overview}
\vspace{-20pt}
\end{figure}

In this paper, we introduce the \textbf{C}omplex \textbf{R}etrieval \textbf{U}nified \textbf{M}ulti-task \textbf{B}enchmark (\textbf{\dataset}) an evaluation suite of eight retrieval tasks that are meticulously curated from existing datasets. An overview of the eight tasks can be seen in Figure \ref{fig:crumb_overview}. Each task has multiple aspects per query expressed in unique ways as well as additional dataset-specific features that make the collection diverse and well rounded with a focus on realistic retrieval settings. These include: tip-of-the-tongue queries for movie retrieval, multi-aspect queries for scientific paper retrieval, set-based logical queries for entity retrieval, state-specific legal questions for statute retrieval, multi-constraint math problems as queries for theorem retrieval, varied Stack Exchange questions with related web pages, clinical trial search where patient histories are used as queries, and code retrieval where a multi-constraint code problem is the query and the code snippets are the documents. Some unique features of these tasks include different vocabularies between queries and documents, highly technical terms, and numerical comparisons. Note that although we feature many types of retrieval tasks which have complex queries there will always be some that are not accounted for, our aim was to cover several different variations and include other features that might impact retrieval quality to get a holistic view of how retrieval models perform on complex retrieval tasks. To promote the best retrieval performance, we use a unified markdown format for documents and include headings in the chunked versions to provide context. Our formatting allows for the future study of retrieval models that use document structure and provides important context for retrieval models to perform well. We believe that there is significant value in unifying these datasets so that it is simple to test retrieval systems on a broad range of complex tasks. Our selection of both datasets and the subset of data used in our final collection reflect a substantial investment in choosing realistic retrieval tasks that have high-quality relevance judgments. Additionally, several of the original datasets lack a standardized version that is well suited for modern retrieval models (e.g. the documents come in XML format with unnecessary fields) which our version solves.

Using \dataset, we evaluated a diverse and representative set of state-of-the-art neural retrieval models to assess how they perform on complex tasks and derive several insights about where current models struggle most and what characteristics the best models share. We find that even the best models struggle on these complex tasks with the best average nDCG@10 of 0.346 and R@100 of 0.587. We find that models tend to struggle with precision metrics for the top-ranked documents when semantic and keyword overlap between queries and documents is low or when such overlaps are weak signals for relevance. For example, in the tip-of-the-tongue retrieval \cite{tip_of_the_tongue_original} task where queries often have minimal term overlap has one of the lowest nDCG values. The task SetOps which has queries featuring set-based operations, such as \textit{and} and \textit{not}, with entity pages as documents has the lowest maximum nDCG across all baseline models. Due to the set-based operations in the query, often documents with relevant terms or semantics are either irrelevant or only partially relevant. These observations are true for the best performing models, while in general the weaker models struggle on datasets with significant differences from common retrieval training datasets. For instance, Theorem Retrieval and Code Retrieval both have very different document vocabularies than common retrieval models and see poor performance from the less capable models. We find that model performance is most impacted by four factors: (1) the ability of the model to follow instructions, (2) the size of the model, (3) the diversity and difficulty of training data on which the model was trained and (4) what base model was used. Experiments with LLM-based query rewriting techniques show that rewriting tends to harm better performing models while bringing notable improvements to weaker models. This finding suggests that there is a limit to the usefulness of query rewriting to improve performance on complex tasks, at least with current state-of-the-art rewriting techniques.

In summary, the main contributions of this work include:
\begin{itemize}
    \item The construction of a complex retrieval benchmark composed of eight diverse complex retrieval tasks.
    \item Benchmarking a wide range of top-performing retrieval models to uncover how they perform on complex tasks.
    \item An analysis of what qualities the best performing models have which allows them to do well on complex tasks.
    \item An analysis of what data features impact model performance the most.
\end{itemize}

\begin{table*}[]
\centering
\caption{Overview of dataset characteristics for the chunked-document (passage) version of \dataset. $\mathcal{Q}$ is the number of evaluation queries, $\mathcal{D}$ is the number of documents in the corpus. Avg $\mathcal{D}^+_{NB}$ and Avg $\mathcal{D}^+_{B}$ are the average number of relevant documents per query for non-binary and binary relevance judgments, respectively. Avg $|\mathcal{Q}|_{char}$ and Avg $|\mathcal{D}|_{char}$ are the average character lengths of queries and documents.}
\label{tab:my_dataset_summary}
\begin{tabular}{l !{\color{darkgray}\vrule} c !{\color{lightgray}\vrule} c !{\color{darkgray}\vrule} c !{\color{lightgray}\vrule} c !{\color{darkgray}\vrule} c !{\color{lightgray}\vrule} c}
\toprule
Dataset Name & $\mathcal{Q}$ & $\mathcal{D}$ & Avg $\mathcal{D}^+_{NB}$ & Avg $\mathcal{D}^+_{B}$ & Avg $|\mathcal{Q}|_{char}$ & Avg $|\mathcal{D}|_{char}$ \\
\midrule
Tip-of-the-tongue & 135 & 1,083,337 & 6.92 & - & 726 & 2,535 \\
\stackexchange & 107 & 40,956 & 2.20 & - & 876 & 2,661 \\
\paperretrieval & 72 & 363,133 & 90.72 & 9.67 & 951 & 1,258 \\
Set-Ops & 423 & 651,704 & 29.91 & - & 47 & 1,411 \\
Clinical Trial & 113 & 914,628 & 391.57 & - & 740 & 1,743 \\
Legal QA & 6,753 & 1,182,626 & 2.04 & - & 148 & 2,704 \\
Theorem Retrieval & 69 & 23,839 & 1.99 & - & 415 & 874 \\
Code Retrieval & 3,665 & 232,444 & 29.72 & - & 837 & 511 \\
\bottomrule
\end{tabular}
\end{table*}

\begin{table*}[]
\centering
\caption{Overview of dataset characteristics for the full-document version of \dataset. $\mathcal{Q}$ is the number of evaluation queries, $\mathcal{D}$ is the number of documents in the corpus. Avg $\mathcal{D}^+_{NB}$ and Avg $\mathcal{D}^+_{B}$ are the average number of relevant documents per query for non-binary and binary relevance judgments, respectively. Avg $|\mathcal{Q}|_{char}$ and Avg $|\mathcal{D}|_{char}$ are the average character lengths of queries and documents.}
\label{tab:dataset_characteristics_long}
\begin{tabular}{l !{\color{darkgray}\vrule} r !{\color{lightgray}\vrule} r !{\color{darkgray}\vrule} r !{\color{lightgray}\vrule} r !{\color{darkgray}\vrule} r !{\color{lightgray}\vrule} r}
\toprule
Dataset Name & $\mathcal{Q}$ & $\mathcal{D}$ & Avg $\mathcal{D}^+_{NB}$ & Avg $\mathcal{D}^+_{B}$ & Avg $|\mathcal{Q}|_{char}$ & Avg $|\mathcal{D}|_{char}$ \\
\midrule
Clinical Trial & 113 & 375,580 & 150.63 & - & 740 & 4,032 \\
Legal QA & 6,753 & 296,274 & 1.50 & - & 148 & 10,312 \\
Set-Ops & 423 & 325,505 & 8.22 & - & 47 & 2,812 \\
\stackexchange & 107 & 5,035 & 1.07 & - & 876 & 25,529 \\
Tip-of-the-tongue & 135 & 231,852 & 1.00 & - & 726 & 10,168 \\
\bottomrule
\end{tabular}
\end{table*}

\section{Related Works}
\subsection{General Evaluation Collections for Retrieval}

Previous work has used a collection of retrieval tasks to evaluate how retrieval models perform on a range of tasks. These collections are generally preferred over a single task because they provide a more complete picture of how retrieval models generalize. One of the most used evaluation collections is BEIR \cite{beir} which was constructed to evaluate the zero-shot performance of retrieval models on various domains. BEIR is constructed by aggregating existing datasets, including datasets that were existing retrieval datasets and those that originally were constructed for other tasks, but which were converted to retrieval datasets. Another popular collection with a broader scope is the Massive Text Embedding Benchmark (MTEB) \cite{mteb}. MTEB is not just a retrieval dataset as it also considers other text embedding uses such as clustering and sentence similarity as well as more classic IR tasks such as retrieval and reranking. Similar to BEIR, MTEB does not create new datasets; instead, it unifies and curates existing datasets. MTEB has spurred follow-up work which includes a second version which adds new tasks like long-document embedding and code retrieval \cite{multilingual_mteb}. As well as a multilingual variant \cite{multilingual_mteb}. There have also been several language-specific embedding benchmarks, such as for Vietnamese \cite{vietnamese_mteb} and Polish \cite{polish_mteb}.

Although BEIR has been a useful benchmark for generalization it has shortcomings which limit the validity of results when drawing general conclusions. A major limitation is that several of the tasks are not strictly retrieval tasks such as citation prediction and fact verification, evaluating retrieval models on these tasks likely provides little insight as they are out of domain for the models. Additionally, several of BEIR's datasets include answer retrieval where the relevant document is an exact answer to a question which acts as a query. Although this might be a valid retrieval task, it is likely far from realistic given that answers written for questions will likely include more similar terms which likely results in an overestimate of retrieval performance. Furthermore, the way BEIR converted some of the dataset labels to a standard range resulted in incorrect labels as mentioned by \citet{rank1}. As MTEB uses many retrieval tasks sourced from BEIR it inherits many of these problems.

ATEB \cite{atab_evaluating_embeddings_advanced_nlp_tasks} is another embedding benchmark, like MTEB, it focuses on evaluating embeddings on a range of tasks including ranking, classification, retrieval, and bi-text mining. Unlike MTEB, ATEB focuses on advanced NLP tasks that include reasoning-based retrieval and safety classification. ATEBs focus on advanced NLP tasks differs from our focus on complex retrieval tasks. Additionally, the ATEB reranking and retrieval tasks largely use existing NLP datasets in ways that produce unrealistic tasks. In contrast, our benchmark consists of realistic search tasks.

Another recent collection MAIR \cite{mair}, has a large number of retrieval tasks sourced from existing datasets. Unlike MTEB and BEIR, MAIR has a specific focus on instruction following with each task including handwritten instructions. MAIR shares several tasks with our proposed dataset, but it has some drawbacks. It includes many tasks from BEIR and thus includes many of the aforementioned problems. Furthermore, many of the tasks are also question and answer tasks that have the same lexical biases mentioned in the limitations of BEIR. As MAIR contains so many tasks each task has minimal attention without any standardization of documents. For instance, in the TREC Clinical Trial 2022 task their documents are the raw XML documents while for the HuggingfaceAPI task the documents are raw JSON files. Their documents are also not chunked, despite the large body of evidence that chunked documents result in better performance \cite{advanced_chunking_strategies}. Additionally, their dataset focuses on instructions, so many of the datasets are not complex unlike our dataset. We also ensure that the datasets have human written queries and that our documents are in a unified markdown format and include contextualized chunks to evaluate the best-case performance and provide a standardized version for future users.

\subsection{Retrieval and Reranking Models for Hard Retrieval Tasks}
Although complex retrieval has not become a common genre of retrieval collections, several modeling approaches have emerged to address complex queries in various forms. Some of these techniques are limited to specific types of complex queries while other approaches are broader approaches that would likely be successful over a range of complex query types.

As our focus is on unified models which can generalize to varied complex tasks, we mostly focus on approaches that are not tightly-coupled to a specific query type. We start with a popular approach to doing retrieval on multi-part or complex queries, using an explicit decomposition step either before retrieval or between iterative retrieval steps. 
\citet{generating_followup_questions_for_multihop_qa} was one of the earliest works to explicitly generate intermediate questions to address multi-hop queries. While \citet{whats_that_book_decomposing_complex_queries_for_tot_retrieval} used an LLM to decompose tip-of-the-tongue queries into various sections corresponding to fields in the target document before retrieval. \citet{handling_complex_queries_using_query_trees} uses a part-of-speech parsing to construct a "query tree" which can then be deconstructed into simpler queries to handle multi-hop queries, though it is unclear if their approach would work on complex queries that do not follow the specific multi-hop construction. 

Instruction following retrieval models has emerged as a way to adapt retrieval models without further fine-tuning. As a consequence, instruction-tuned models must understand subtle instructions and multiple relevance criteria in a way that most retrieval and reranking models cannot. Early work in instruction-tuned model had task-level instructions to adapt models for new tasks, but did not consider query-level instructions \cite{task_aware_retrieval_with_instructions, one_embedder_instruction_finetuned_embeddings}. \citet{followir} proposed an instruction-tuned ranker that works on query-level instructions instead of task-level instructions. The recent dense-retrieval model Promptriever \cite{promptriever} uses a large-scale instruction dataset and LLM backbone model to enable strong instructions following. Promptriever is the first retrieval model to show prompting behavior similar to generative LLMs.

An interest in ranking models that can handle reasoning-heavy tasks has recently gained popularity. These reasoning-heavy tasks often require more than lexical or semantic matching, instead requiring reasoning steps to connect a query to a relevant document. Though not always explicitly complex, reasoning intensive retrieval tasks often contain several questions and constraints which makes them complex. A recent work ReasonIR \cite{reason_ir} focuses on creating synthetic reasoning data to enable a dense-retrieval model to better handle reasoning-heavy retrieval tasks.

The prior approaches largely focus on using training data to modify model behavior to enable better performance on complex tasks, but there have also been a number of approaches to modify the traditional retrieval training techniques and model architecture to enable fundamentally more capable retrieval and reranking models. Inspired by generative reasoning models such as OpenAI's o3 model \cite{open_ai_o3_and_o4_mini} and DeepSeek's R1 model \cite{deepseek_r1}, which make use of extended reasoning traces with verifiable rewards, several ranking models have incorporated reasoning. The reranking model Rank1 \cite{rank1} uses reasoning traces from R1 to distill reranking reasoning to smaller LLMs such as Qwen 2.5 7B. Other approaches have used reinforcement learning such as Search R1 \cite{search_r1} and \cite{r1_searcher} for retrieval and for reranking \cite{rank_r1}. Beyond reasoning, Hypencoder \cite{hypencoder} which uses a hypernetwork to enable query-specific neural networks has also been proposed as a method to enable retrieval on more complex tasks.

\subsection{Complex Retrieval Evaluation Sets}
Evaluation sets and collections that feature complex queries that include multiple aspects have existed for some time but have seen an increase in recent years, as simpler single-aspect query evaluation sets have become increasingly saturated.
\subsubsection{Evaluation Sets}
One of the earliest examples of complex natural language queries is from the TREC Filtering Track \cite{trec_filtering_track_2001} where given a question that often contained multiple aspects, participants had to filter a document stream with the binary judgment relevant or irrelevant. Participants also got human judgments to help update their scoring function making it different from most retrieval tasks where human judgments are not available.

Since then one of the most common types of complex query come from multi-hop tasks, where queries are specifically created to require information from multiple distinct documents. Some of the earliest multi-hop question answering (QA) datasets include QAngaroo \cite{qangaroo}, Complex Web Questions \cite{complex_web_questions}, and HotPotQA \cite{hotpotqa}. More recent multi-hop QA datasets have considered information in both tables and text \cite{hybridqa_multihop_tabular_text_dataset}.

Another type of complex query that has gained recent popularity is the instruction-following query. These queries generally contain more constraints than traditional queries which are expressed as instructions. The idea of adding instructions to queries was first introduced by \citet{one_embedder_instruction_finetuned_embeddings} and \citet{task_aware_retrieval_with_instructions} though both focus more on adaptability of retrieval models to different tasks and thus use task-level instructions. FollowIR \cite{followir} was the first paper to introduce a densely annotated evaluation set with query-level instructions.

Tip-of-the-tongue queries which come from users who remember aspects of an entity but forget the name (it is on the tip-of-their-tongue) are also generally complex by their nature. The term tip-of-the-tongue in the Information Retrieval context was first coined by \citet{tip_of_the_tongue_original} which investigated tip-of-the-tongue queries for movie recommendation. Recently, tip-of-the-tongue has become a TREC track \cite{trec_tot_2023, trec_tot_2024}, which includes items from movies, landmarks, and celebrities.

Other works that combine multiple aspects have been created in various domains and formats. Some have been built to specifically include logical operations such as \textit{and} and \textit{or} such as QUEST \cite{quest} and RoMQA \cite{romqa}. Others have focused on constructing multi-aspect queries such as DORIS-MAE \cite{doris_mae}, which focuses on paragraph-length queries for scientific paper search.

\subsubsection{Collections}
Although many datasets are complex, there are substantially fewer collections that primarily feature complex retrieval tasks. Generally, the collections that do can be split into specialized and non-specialized, where specialized collections focus on a specific topic whereas non-specialized cover multiple topics.

For specialized evaluations, there are a few collections which focus on topics that generally have multiple constraints. These include CoIR \cite{coir_code_benchmark_ir} for code-related retrieval tasks, MIRB \cite{mirb_mathmatical_benchmark_ir} for mathematical information retrieval, and R2MED \cite{r2med_reasoning_driven_benchmark_medical_ir} for reasoning-heavy medical retrieval. Each of these collections includes several tasks with the majority featuring complex queries. Though these benchmarks are useful to evaluate models in these specific domains, the lack of breadth limits the generalization of findings on any specific dataset. Although multiple specialized collections could be used to evaluate a system, this can require a significant effort and can lead to data variants that make direct comparisons difficult between methods. 

To remedy the problems with specialized collections, some non-specialized collections have been proposed. One such collection is BIRCO \cite{birco} which combined five complex tasks to form a unified collection. BIRCO specifically was constructed to evaluate LLM-based systems and as such they focus on making a small enough dataset to make LLM-based evaluations tractable, due to this design choice, BIRCO is designed as a reranking dataset with only a small number of documents provided for each query. Although they argue the difficulty is similar to full retrieval, the limited number of documents makes evaluating recall at larger cut-offs meaningless, which is highly relevant for complex retrieval tasks where models often struggle with recall and where strong LLM rankers are able to make up for precision errors. Additionally, the majority of tasks only have a single relevant query with an average of 8.4 relevant documents per query, this means evaluating recall with large number of relevant documents is not possible with BIRCO. The selected datasets have some additional shortcomings, WhatsThatBook \cite{whats_that_book_decomposing_complex_queries_for_tot_retrieval} a tip-of-the-tongue dataset for book retrieval uses Goodreads\footnote{goodreads.com} description for documents though these are often short and lack necessary details to satisfy the queries; RELIC \cite{relic_retrieving_evidence_for_literary_claims} uses a quotes from a literature analysis papers as documents and the original passages with the quote masked as the queries, though this task is complex, it is not a natural search task and is likely far out-of-domain for retrieval models. These datasets make BIRCO less reliable as an overall benchmark for complex retrieval. In contrast, we focus on high-quality and realistic datasets for complex retrieval tasks.

BRIGHT \cite{bright} is one of the most popular existing complex collections with a focus on reasoning-intensive retrieval tasks. BRIGHT has three main task types: (1) given a Stack Exchange question retrieve a relevant document (2) given a math or coding problem retrieve similar types of problems (3) given a math or coding question retrieve a relevant theorem or relevant code documentation. BRIGHT is a well-formulated and realistic dataset, but it has some limitations. First, the query types are limited to either Stack Exchange questions or code/math questions. Though these are complex, they are only a subset of the types of meaningful queries that we would expect users to ask of advanced information access systems. Second, for the Stack Exchange set, the number of documents for each subject area is low with an average around ~500. Though a chunked version is provided which increases the number of documents, the low number of source documents makes the collection far less challenging in the full document scenario and means the chunked passages do not show the diversity present in real-world large-scale collections. In this work, we focus on collecting a large set of diverse complex tasks which include several large collections with long documents and put an emphasis on tasks that cover various additional aspects beyond complexity. Additionally, we provide some additional benefits over BRIGHT, we provide a large portion of our dataset in a unified markdown format enabling research on how to handle semi-structured information. We use this structure to construct contextualized chunks, when possible, which is likely what would be used for real-world complex retrieval tasks if quality is desired, as preserving hierarchical context (like headers) within chunks helps to maintain semantic integrity and provides more complete information to the retrieval model, leading to more accurate relevance assessments \cite{advanced_chunking_strategies}. Furthermore, BRIGHT provides no official development/validation set which makes it: (1) hard to tune retrieval systems (2) makes it difficult to use modern retrieval strategies which often take advantage of few-shot prompting approaches \cite{dai2023promptagator, query_to_doc}.

\section{The Benchmark: \dataset}
One of our core contributions is our benchmark, which was constructed with the goal of producing a diverse and realistic collection of tasks to gauge how Information Retrieval (IR) systems generalize to complex retrieval tasks. Importantly, when we say complex we mean \textit{consisting of many parts} not difficult, though our results suggest that many of our tasks are also difficult for current retrieval systems. Our focus on complex retrieval is motivated by the desire to support more sophisticated information needs, which often require understanding queries with multiple distinct components and requirements. We believe that such information needs are quite common, though anecdotally most current retrieval systems struggle to accurately recall relevant information for these tasks. The need for traditional IR systems
to adapt to these information needs is especially relevant today given the rapid increase in LLM usage, which offers more flexibility in the types of request that can be answered. 
To promote the development of general retrieval systems capable of handling a wide range of complex tasks, we create the \textbf{C}omplex \textbf{R}etrieval \textbf{U}nified \textbf{M}ulti-task \textbf{B}enchmark (\textbf{\dataset}) which contains eight realistic and unique tasks. In this section, we discuss each dataset in \dataset in detail, including what makes them complex. In addition, we provide details on our method for standardizing document data, our approach to construct contextualized chunks, and how we constructed a validation set for each dataset.

\subsection{Datasets} \label{sec:overview_datasets}
\textbf{\paperretrieval} Derived from DORIS-MAE \cite{doris_mae} a test collection that contains 100 paragraph-length multi-aspect queries that describe information needs when searching for scientific papers. Each query is human-written and motivated by analyzing a set of one or more initial papers and reconstructing the early thought processes of the original authors. The corpus is made up of paper titles and abstracts. To facilitate data labeling, the authors break each query into various aspects and these aspects further into sub-aspects when appropriate. The authors then use GPT-3.5 to judge whether a paper meets an aspect or sub-aspect. The judgments on the aspects and sub-aspects can then be used to get an overall relevance score. This dataset is complex by design as it is constructed specifically to contain multiple aspects.

We remove all queries that do not have at least one document that satisfies all aspects (at least partially). This results in 79 queries. We provide two query relevance labels, one intended for binary metrics considers a paper relevant if all aspects are met, the other set is a sum of points where a document receives 1 point for each aspect that it mostly fulfilled and 2 points for each aspect it completely fulfilled.

\textbf{SetOps} Derived from QUEST \cite{quest} an entity retrieval dataset where queries contain implicit set relationships such as \textit{or}, \textit{and}, and \textit{not}. The dataset is constructed using Wikipedia categories to find relevant entities for a simulated query based on a set of templates such as \textit{"\_ or \_"} and \textit{"\_ but not \_"}. This query is then rewritten by crowd-source workers and scored for its naturalness and fluency. The documents used to represent each entity are the corresponding Wikipedia pages for a category. Although the known entities in a category and the query templates allow for an initial set of relevant documents, the information required by a query may not be present in the text of the Wikipedia page. Thus, the authors have crowd-workers assess whether a candidate page is relevant. This dataset is complex, as the queries often contain multiple requirements combined with set-based operations. Additionally, many of the "single" requirements contain multiple aspects on their own such as \textit{German spy comedy films}.

In this work, we only consider queries from the test collection in the book and film domains, as these are the only ones with human-relevance judgments. We further filter the queries, removing those that are not judged as having high naturalness and fluency. We consider an entity relevant if the average rater score is above 1.5 points where 1 point is given for likely relevant ratings and 2 points are given for definitely relevant ratings.

We provide both a full document and chunked document (passage) collection. The full document collection is just the Wikipedia page formatted in standard markdown. While the passage collection is a set of passages formatted using the contextualized chunking strategy described in Section \ref{FormattingChunkingSection}.

\textbf{Legal QA}
Derived from Reasoning Focused Legal Question Answering \cite{zheng2025reasoningfocused} a question answering dataset focusing on questions in the legal domain that require reasoning. We use only the Housing QA task which focuses on retrieving relevant legal statutes given a question about housing policy for a specific state. For this dataset, we construct full documents using the chunks provided by the original dataset and add headings using metadata from the original dataset. We also produce a chunked (passage) version following the chunking method from Section \ref{FormattingChunkingSection}. This dataset is complex as there are always at least two distinct features to consider, first is the actual legal question and second is the state that is being asked about. Furthermore, as the questions are reasoning heavy, often there are multiple considerations, for instance, in the example shown in Table \ref{tab:example_legal_qa} the query \textit{Are eviction cases first heard in high court? In the state of Tennessee} requires a relevant document to address both where eviction cases are first heard and whether that is a high court or not.

As the full documents are the complete list of statutes for a state, they are very long, and various chapters do not share topical relevance. For this reason, the documents released are a split version of the original documents. We split them so that each chunk contains content that shares the first three levels of headings (e.g. \textit{Laws > Ordinances > Housing > Construction} and \textit{Laws > Ordinances > Housing > Permitting} would be in the same chunk). We found that three levels generally aligned with topical boundaries well.

\textbf{Tip-of-the-Tongue}
Derived from TREC Tip-of-the-Tongue 2023 track \cite{trec_tot_2023} this dataset contains tip-of-the-tongue (TOT) queries for movies and TV shows with Wikipedia pages as documents. TOT queries are queries where a user knows of an item, but cannot remember the exact identifier such as title in the case of a movie or TV show. As such, users tend to add as many details as they can remember, which often results in long queries with many details at varying levels of relevance and certainty. For instance, the user might remember a certain interior vividly, but only provides a guess of when the movie might have been produced. As the entity identifier is not known and people's memories are often hazy, these queries tend to share minimal keywords with the relevant query. A natural consequence of this is also that queries have many aspects making them complex.

We process the original collection provided by TREC into our standardized markdown format. The original corpus is provided in MediaWiki format, Wikipedia's editing format, which includes templates that cannot be easily converted to markdown, since the exact way a template is rendered depends on the template's internal logic. To address this, we use WikiMedia's template expansion API to convert all templates to rendered HTML and then convert this to markdown. This results in a clean markdown collection that includes the relevant information from the templates, which include all Wikipedia information boxes and many tables. After creating a standardized markdown document, we use the chunking strategy outlined in Section \ref{FormattingChunkingSection}. We assume that all chunks derived from a relevant document are also relevant for our passage labels, though we recommend, and used while benchmarking, a MaxP strategy \cite{max_p_original} with the full document labels to not penalize models for failing to retrieve chunks that might not have relevant information.

\textbf{Theorem Retrieval} Derived from BRIGHT's theorem retrieval dataset \cite{bright} this dataset has math problems as queries and math theorems as documents. The task is to retrieve theorems that are relevant to solving the math problem. We largely leave the dataset unchanged as the documents are short enough that they do not need to be chunked and there would be minimal benefit to adding markdown formatting given the minimal structure. As such, we only split the original test set into a validation and test split. This dataset is complex, as the math problems generally have multiple constraints and considerations which are necessary to evaluate when finding the relevant theorem.

\textbf{\stackexchange} This task is derived from BRIGHT's questions from Stack Exchange which cover a range of topics including Economics, Biology, Robotics, Programming (Stack Overflow), Sustainable Living, Earth Science, and Psychology. The original BRIGHT data was created by taking questions from various Stack Exchange communities and using links from relevant answers to find relevant web pages. Then additional searches using Google were conducted to find additional relevant web pages or hard-negative web pages. This dataset is complex as generally the Stack Exchange questions include multiple components, such as an example and a question based on that example or multiple related questions.

To create a unified markdown collection, we used the original links collected by the BRIGHT authors to download the raw HTML for the web pages. We then used the boilerplate removal tool Trafilatura \cite{trafilatura_web_scraping_library} to clean the HTML and convert it to markdown. For Wikipedia pages, which made up a sizable subset of the collection, we directly downloaded the Wikipedia source using the WikiMedia API and converted this to markdown. Due to errors in accessing and converting the web pages as well as some missing document links, we were unable to completely recreate the BRIGHT corpus ourselves. To make up the difference, we used BRIGHT's original documents and converted them to our standard markdown to the best of our ability.

Aligning the relevance judgments for the full documents was straightforward, but as BRIGHT provides chunk-level relevance judgments, we also wanted chunk-level relevance for our chunked collection. We found this difficult because if their chunk was split over two of our chunks it is not clear which chunk contains the relevant information without additional labeling. To remedy this problem, we only use queries from BRIGHT where all the original relevant chunks are fully encapsulated in one of our chunks. We used sub-string coverage to automate this process with manual checks for lower-overlap sub-string chunks to ensure the chunk was valid. This reduced the number of queries, but ensured valid relevance labels. 

To increase difficulty and make evaluation simpler, we combined all Stack Exchange topics into a single collection and corpus. We believe that this has a very minimal chance of adding unlabeled positives given the significant differences in the topics.

\textbf{Code Retrieval} Derived from the dataset APPS \cite{apps_measuring_coding_challenge_competence} which was created to measure the ability of generative models to solve code problems. APPS contains 10k code problems such as those from LeetCode\footnote{leetcode.com} as well as several reference solutions. To create a retrieval dataset, we create a corpus made up of the reference solutions and use the code problems as queries. We find that the solutions have minimal keyword overlap with the problems in large part because these problems are often solved quickly with short variable names and minimal documentation. This makes the dataset a good measure of a retrieval model's ability to reason even with very minimal term overlap. This dataset is complex, as like Theorem Retrieval the code problems have multiple constraints and components which need to be considered to identify the correct solution document.

As the documents are fairly short and have no clear way to chunk them while remaining comprehensible, we do not do any document chunking. For the queries, we only include the natural language problem description and input-output format without any provided examples.

\textbf{Clinical Trial} Derived from the TREC Clinical Trial 2021 \cite{roberts2021overviewct} and 2022 \cite{roberts2022overviewct} Tracks this dataset uses a patient's medical history expressed in natural language as the query and tries to retrieve relevant clinical trials. For this dataset, we converted the original XML files to markdown and used the chunking approach described in Section\ref{FormattingChunkingSection}. We assume that all chunks derived from a relevant document are also relevant, but as recommended for Tip-of-the-tongue we suggest using MaxP and the full-document labels for evaluation. This dataset is complex because the patient history used has many aspects about the patient. Additionally, relevant documents can require a certain age range which provides a useful measure of whether retrieval models can adhere to numeric constraints.

\subsection{Formatting and Chunking Strategy} \label{FormattingChunkingSection}
\begin{figure}[htbp]
\centering
\resizebox{0.9\linewidth}{!}{
\begin{tikzpicture}[
    doc/.style={
        rectangle,
        draw=orange!40!black, %
        fill=orange!10,
        rounded corners,
        minimum height=1.5cm,
        minimum width=10cm,
        text width=9.5cm,
        align=left,
        font=\scriptsize
    },
    header/.style={font=\scriptsize\bfseries},
    paragraph/.style={font=\scriptsize},
    chunk/.style={
        rectangle,
        draw=red!40!black, %
        fill=red!10,
        rounded corners,
        minimum height=1.5cm,
        minimum width=4.5cm,
        text width=4cm,
        align=left,
        font=\tiny
    },
    arrow/.style={
        ->,
        line width=1pt, %
        color=black!80
    }
]

\node[doc, label={[yshift=0.2cm]above:Original Markdown Document}] (original_doc) at (0,0) {
    \texttt{\# Main Title} \\
    \texttt{\#\# Section 1} \\
    \texttt{This is the first paragraph of section 1. It contains some interesting information that spans multiple sentences.} \\
    \texttt{\#\#\# Subsection 1.1} \\
    \texttt{This is a paragraph in subsection 1.1. It provides more detailed information. Let's assume this paragraph is long enough to be its own chunk, or combined with the next one if it's short.} \\
    \texttt{Another paragraph in subsection 1.1, continuing the discussion with further details and examples.} \\
    \texttt{\#\# Section 2} \\
    \texttt{This is the first paragraph of section 2. It introduces a new topic.}
};

\def\yChunksTop{-2.8cm} %

\node[chunk, anchor=north] (chunk1) at (-5cm, \yChunksTop) {
    \texttt{\# Main Title} \\
    \texttt{\#\# Section 1} \\
    \texttt{This is the first paragraph of section 1. It contains some interesting information that spans multiple sentences.}
};

\node[chunk, anchor=north, label={[yshift=-0.5cm]below:Generated Chunks}] (chunk2) at (0cm, \yChunksTop) {
    \texttt{\# Main Title} \\
    \texttt{\#\# Section 1} \\
    \texttt{\#\#\# Subsection 1.1} \\
    \texttt{This is a paragraph in subsection 1.1. It provides more detailed information. Let's assume this paragraph is long enough to be its own chunk, or combined with the next one if it's short.}
};

\node[chunk, anchor=north] (chunk3) at (5cm, \yChunksTop) {
    \texttt{\# Main Title} \\
    \texttt{\#\# Section 1} \\
    \texttt{\#\#\# Subsection 1.1} \\
    \texttt{Another paragraph in subsection 1.1, continuing the discussion with further details and examples.} \\
    \texttt{\#\# Section 2} \\ %
    \texttt{This is the first paragraph of section 2. It introduces a new topic.}
};

\draw[arrow] ($(original_doc.south west) + (1cm, 0cm)$) to (chunk1.north);
\draw[arrow] (original_doc.south) to (chunk2.north);
\draw[arrow] ($(original_doc.south east) + (-1cm, 0cm)$) to (chunk3.north);

\end{tikzpicture}}
\caption{Example of our contextualized chunking strategy. The original markdown document content is segmented into chunks and each chunk prepended with its hierarchical header path. This approach preserves structural context for the retrieval models.}
\label{fig:chunking_example}
\end{figure}
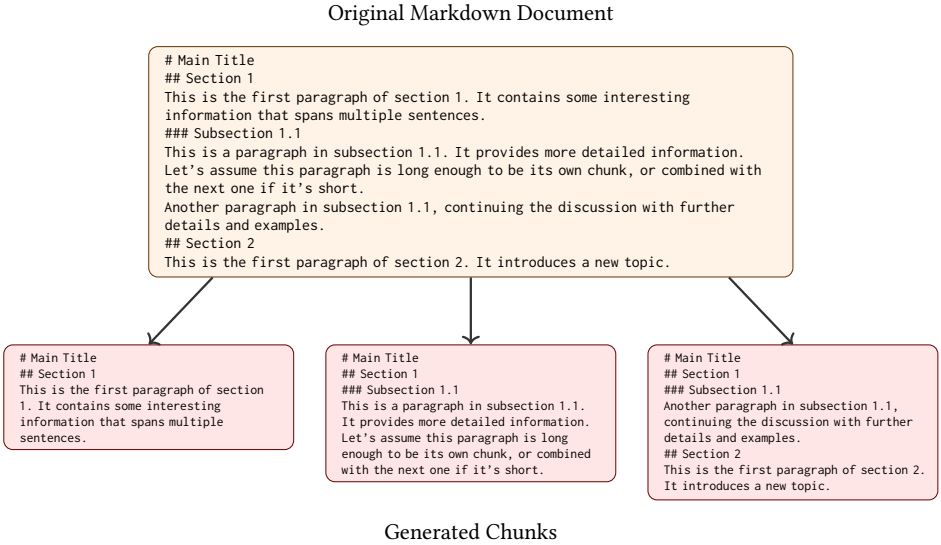

For all datasets that had original documents that had natural divisions (e.g. not a single code block) we formatted the documents in a uniform structured way using markdown. We provide the complete markdown document to enable long-context evaluation with structured documents. As many retrieval models are unable to process longer documents (or unable to represent them well), we also provide a passage version which features standardized chunks. For the chunked version we take advantage of the markdown structure to include all the relevant headings for a paragraph or other content block, see Figure \ref{fig:chunking_example} for an example of what this looks like. Our chunking strategy continues to add content to a chunk until the total number of BERT \cite{bert} tokens in the chunk would exceed 512 if the content was added. If a piece of content is more than 512 tokens on its own, it will be added as an individual chunk without truncation. 

We believe that our formatting and chunking strategy is unique in that it provides a unified structured format for documents. This enables future evaluation of retrieval models that use document structure and a truer understanding of retrieval model performance on our chunked datasets. This is in contrast to several existing datasets, which often use new-line chunking without headings for context, which can result in incomplete information for retrieval, which results in worse performance \cite{advanced_chunking_strategies}.

\subsection{Validation Data}
To enable validation of trained models and enable the use of few-shot techniques that have become common practice in LLM augmented retrieval \cite{izacard2022atlas} we include a validation/dev set for each dataset. Validation data is taken from the same pool as the evaluation data though it is of course not included in the final evaluation data. We take 10\% or 100 instances of the data, whichever is smaller, as the validation set for each task. 

This is a notable difference from other common retrieval benchmarks such as BEIR \cite{beir} and BRIGHT \cite{bright}, which do not include development sets for each of their tasks. This lack of development set limits the types of methods that can be evaluated e.g. few-shot prompt-based methods \cite{dai2023promptagator, query_to_doc}.

\begin{table*}[]
\centering
\caption{Overview of the instructions used for each dataset. }
\label{tab:dataset_instructions}
\resizebox{0.95\linewidth}{!}{
\begin{tabular}{l !{\color{darkgray}\vrule} l}
\toprule
Dataset Name              & Instruction  \\
\midrule
Tip-of-the-tongue         & Given a description of an entity, retrieve the entity that is described \\
\stackexchange             & Given a question, find relevant documents that can help answer it \\
\paperretrieval          & Given a query, find relevant scientific papers which satisfy the requirements  \\
Set-Ops                   & Given a query, find entity pages which satisfy it \\
Clinical Trial            & Given a patient's medical history, find clinical trials that are relevant to them \\
Legal QA                  & Given a query, find relevant legal documents which satisfy the requirements  \\
Theorem Retrieval         & Given a math question, find theorems that are relevant to it \\
Code Retrieval            & Given a code question, find code snippets that would answer it \\
\bottomrule
\end{tabular}
}
\end{table*}

\begin{table*}[]
\centering
\caption{Overview of the instructions used while retrieving using the various query rewriting approaches. These instructions were only used for retrieval models that were trained with instructions.}
\label{tab:rewriting_retrieval_instructions}
\begin{tabular}{l !{\color{darkgray}\vrule} l}
\toprule
Rewriter Name              & Instruction  \\
\midrule
Query-to-Answer             & Given an answer to a question find supporting/relevant items \\
Query-to-Doc                & Given an item find similar items \\
Query-as-Reasoning-Trace    & \makecell[lt]{Given a question, information about what should be included in \\ the answer, and a draft answer find supporting/relevant items}  \\
\bottomrule
\end{tabular}
\end{table*}

\begin{table*}[]
\centering
\caption{The format used for each retrieval model when indexing/encoding the documents and encoding/searching the queries. All formats were from the original author's recommendations when available. If models were not trained with instructions, the query and document were provided without any further information.}
\label{tab:instruction_formatting_per_model}
\begin{tabular}{l !{\color{darkgray}\vrule} l !{\color{darkgray}\vrule} l}
\toprule
Model Name       & Query Formatting & Document Formatting \\
\midrule
BM25             & \texttt{<query>}                                                 & \texttt{<document>} \\
Snowflake        & query: \texttt{<instruction>}\verb|\n|\texttt{<query>}           & \texttt{<document>} \\
GTE Qwen 1.5B    & Instruct: \texttt{<instruction>}\verb|\n|Query: \texttt{<query>} & \texttt{<document>} \\
GTE Qwen 7B      & Instruct: \texttt{<instruction>}\verb|\n|Query: \texttt{<query>} & \texttt{<document>} \\
Lion SB 1B       & \texttt{<query>}                                                 & \texttt{<document>} \\
Lion SB 7B       & \texttt{<query>}                                                 & \texttt{<document>} \\
Promptriever     & query:  \texttt{<query>} \texttt{<instruction>}                  & passage:  \texttt{<document>} \\
Lion DS 1B       & \texttt{<query>}                                                 & \texttt{<document>} \\
Lion DS 8B       & \texttt{<query>}                                                 & \texttt{<document>} \\
\bottomrule
\end{tabular}
\end{table*}

\definecolor{lightgray}{gray}{0.80} %

\section{Experimental Setup}
\label{sec:experimental_setup}
This section details the setup for our experiments.
\subsection{Evaluation Metrics}
We use nDCG@10 \cite{ndcg}, Recall@100 (R@100), and Recall@1000 (R@1000) as our primary evaluation metrics. We selected nDCG@10 as it is a common measure for retrieval datasets including BEIR, BRIGHT, and the TREC DL Tracks which makes it easier to compare the results from \dataset with existing datasets. Additionally, nDCG@10 captures the ability of models to directly satisfy users in a search engine setting where position is important and where being on the first page of results is significant for user satisfaction. Recall@100 and Recall@1000 are included because they demonstrate the upper bound of these models to retrieve relevant information. For instance, retrieving a relevant document in the top 100 is far simpler than putting it first. The same is true for $k=1000$, where $k$ is the recall cutoff, but to a more extreme level. The recall metrics also give insight about the potential gain that a reranker could provide. If R@1000 is low, this indicates that even with a perfect reranker there are only minimal gains to make and thus highlights the importance of capable first-stage retrieval models.

\subsection{Benchmarking Models}
To get a comprehensive picture of how existing neural retrieval models perform on complex retrieval tasks, we have assembled a diverse and representative collection of models. To achieve this goal, we considered several criteria, concretely we wanted: (1) models that were high-performing based on existing benchmarks (2) a range of model sizes (3) models trained on various types of data (4) models that used different retrieval paradigms. We also tried to pick models in a way that allowed for meaningful comparisons between models.

\subsubsection{Retrieval Models}
For general purpose strong retrieval models we use the retrieval subset of the Massive Text Embedding Benchmark (MTEB) leaderboard \cite{mteb} to select a few models across a variety of model sizes. The first is \textbf{Snowflake Arctic Embed L V2.0} (abbreviated Snowflake) \cite{snowflake_arctic_embed_l_v2} which is a dense retrieval model with  303M parameter (not including embedding parameters) and an embedding dimension of 1024. It is trained on a large collection of data including Stack Exchange, S2ORC, and Wikipedia \cite{yu2024arcticembed} and has one of the strongest scores on MTEB for a sub-0.5B model at the time of writing. The second model is \textbf{GTE Qwen2 1.5B Instruct} (abbreviated as GTE Qwen 1.5B) \cite{li2023towardsgte} a dense retrieval model based on Qwen2 1.5B. It has an embedding size of 1536 and uses bi-directional attention. Like other GTE models, GTE Qwen 1.5B is trained on a corpus of 800 million weakly supervised pairs from multiple domains including question-answering, code, social media, and link-text-webpage pairs. It is then fine-tuned on retrieval datasets including MS MARCO \cite{msmarco} and NQ \cite{natural_questions}. For its size, GTE Qwen is one of the best performing models on MTEB. The third model is \textbf{GTE Qwen2 7B Instruct} (abbreviated as GTE Qwen 7B) \cite{li2023towardsgte} is identical to GTE Qwen 1.5B except that it uses the larger Qwen2 7B as the base model and has an embedding dimension of 4096. It is one of the best performing embedding models on MTEB.

To get a sense of how models trained on the popular training dataset MSMARCO fare, we included several models which are only trained on MSMARCO (or derived datasets), but with a variety in the model architecture. To investigate how sparse retrieval models compare to dense ones we use \textbf{Lion-SB} and \textbf{Lion-DS} \cite{zeng2025scalingsparsedenseretrieval} which are Llama 3 \cite{llama3herd2024} based retrieval models that use sparse embeddings in the case of Lion-SB and dense embeddings in the case of Lion-DS. To investigate how size may impact sparse and dense retrieval we use Lion models in two sizes: 1B and 8B. All of the Lion models are trained with only MSMARCO data with a fixed compute budget, meaning that large models are trained for fewer epochs. To investigate how prompting might help models generalize we use \textbf{Promptriever Llama 3 8B Instruct} (abbreviated as Promptriever) \cite{promptriever}. This model was trained with a modified version of MSMARCO which added additional instruction and negative documents which makes it capable of responding to instructions to modify the inference-time behavior.

We also include the term-based model \textbf{BM25} \cite{bm25} as a reference.

\subsubsection{Query Rewriting and Expansion Approaches} \label{sec:QueryRewriting}
Using LLMs to rewrite queries has become a popular method to improve retrieval performance \cite{bright, query_to_doc, query_expansion_by_prompting}. For completeness, we include several LLM based query rewriting techniques to investigate how these impact model performance on complex tasks. As chain-of-thought (CoT) \cite{wei2022chainofthought} prompts are generally known to produce better outcomes \cite{wei2022chainofthought, wang2022selfconsistency, query_expansion_by_prompting}, which was verified by our early experiments, we use CoT whenever possible and do not provide zero-shot results.

We experiment with three query rewriting techniques:
\begin{itemize}
    \item \textbf{Query-to-Answer CoT} Given a query, we prompt the LLM to directly answer the question after an extended "thinking" period. The thinking information is not used.
    \item \textbf{Query-to-Document CoT} Given a query, we prompt the LLM to provide a document which provides an answer to the query after an extended "thinking" period. The query instruction provides the expected document type (e.g. a piece of code, web page, etc.). The thinking trace is not used for retrieval only the final document.
    \item \textbf{Query as Reasoning Trace} Given a query, the model is asked to understand the query, think about it, and produce a complete answer. Unlike previous approaches, the entire reasoning process is used as part of the query. This is similar to the approach used in BRIGHT \cite{bright}. 
\end{itemize}

The prompts for each rewriting technique can be seen in Appendix \ref{sec:appendix_prompts}.

\subsection{Retrieval Model Implementation} \label{sec:retrieval_model_implementation}
For all neural retrieval models, we used PyTorch \cite{pytorch} and Huggingface Transformers \cite{huggingface_transformers} for the implementations. For the Snowflake Arctic Embed L V2.0 we used SentenceTransformers \cite{sentence_transformers}. For the other models, we followed the exact usage suggested by the model authors and directly used the original code when necessary. For prompts, we followed the published prompt format for all models that included suggested prompt formats. For models that do not support prompting, we include only the query text. The prompt templates for all baseline models can be seen in Table \ref{tab:instruction_formatting_per_model}.

For each task, we include a specific instruction for the models that support instructions. The instructions for each task can be found in Table \ref{tab:dataset_instructions}. For rewritten queries, we used a shared instruction between tasks that was unique to the rewriting approach. These instructions can be seen in Table \ref{tab:rewriting_retrieval_instructions}.

For dense retrieval models, we used an exact nearest neighbor (i.e. flat index) implemented in PyTorch. For sparse retrieval models, we used an adapted version of the Numba optimized version released by \citet{splade_v1}. For BM25, we used the implementation from the Pyserini library \cite{pyserini} with default parameters. 

The results on the chunked-document (passage) version of \dataset (the main results in this paper) were found by retrieving 2000 documents for each query. For datasets where chunking was employed and where the original dataset only included document-level relevance labels -- namely Tip-of-the-tongue, Clinical Trial, and SetOps -- we used MaxP \cite{max_p_original} to produce the final retrieval list for all models. MaxP selects the maximum passage (i.e. chunk) from a parent document and only keeps this maximum passage. We chose this approach as expecting models to retrieve all chunks from a relevant document would likely add significant amounts of noise given that many chunks from a relevant document may not be directly relevant to the query.

\subsection{LLM for Query Rewriting}
For all query rewriting experiments, we use Gemma-3 27B \cite{gemma_3_technical_report} as the rewriting model. We selected this model because: (1) it is an open-weight model allowing full reproducibility (2) we found it to follow instructions better than other open-weight models (3) it has strong performance for its size (4) it is substantially larger than the largest retrieval models we evaluated. We allowed a maximum of 8,192 tokens to be generated and used a temperature of 0 for query rewriting.

\section{Results} \label{sec:results}
In this section, we discuss the results of our baseline models on \dataset. The main focus is on the chunked document (passage) collection, which uses chunked documents for tasks with longer original documents. This is because it is well known that chunking produces better results, and likely the passage version is more informative about general retrieval model performance and broader trends. In addition, we find that general trends between the full-document results and chunked-document results are similar. We start by looking at the overall performance and trends with a focus on what the top-performing models seem to have, which the others do not. We then do a deep dive on each dataset to understand what characteristics of the dataset may explain the results across our baselines. Next, we explore the impact of sparse versus dense neural models on \dataset and what that might suggest about the strengths and weaknesses of both approaches. Following this, we discuss the impact of query rewriting on retrieval performance on \dataset. Finally, we investigate the impact of informative and task-specific instructions on retrieval performance for the top performing models.

\subsection{Overall Performance}
\begin{table*}[]
\centering
\caption{Overview of baseline model performance for the chunked document (passage) version of \dataset. We use three metrics: normalized discounted cumulative gain at 10 (nDCG@10), recall at 100 (R@100), and recall at 1000 (R@1000). The last row shows the average (mean) value of each metric across all the tasks.}
\label{tab:main_results_table}
\resizebox{0.88\linewidth}{!}{
\begin{tabularx}{0.98\linewidth}{l!{\color{lightgray}\vrule}C !{\color{lightgray}\vrule} C !{\color{lightgray}\vrule} C !{\color{lightgray}\vrule} C !{\color{lightgray}\vrule} C !{\color{lightgray}\vrule} C !{\color{lightgray}\vrule} C !{\color{lightgray}\vrule} C !{\color{lightgray}\vrule} C}
\toprule
Metric / Dataset & \rotcol{BM25} & \rotcol{Snowflake} & \rotcol{GTE Qwen 1.5B} & \rotcol{GTE Qwen 7B} & \rotcol{Lion SB 1B} & \rotcol{Lion SB 8B} & \rotcol{Promptriever} & \rotcol{Lion DS 1B} & \rotcol{Lion DS 8B} \\
\midrule
\multicolumn{10}{@{}p{\dimexpr\linewidth-2\tabcolsep}}{\textbf{Tip-of-the-tongue}} \\
\hspace{1.5em} nDCG@10 & 0.011 & 0.092 & 0.156 & 0.253 & 0.096 & 0.193 & 0.305 & 0.040 & 0.104 \\
\hspace{1.5em} R@100 & 0.052 & 0.326 & 0.489 & 0.674 & 0.311 & 0.496 & 0.578 & 0.193 & 0.193 \\
\hspace{1.5em} R@1000 & 0.141 & 0.519 & 0.733 & 0.867 & 0.615 & 0.741 & 0.844 & 0.415 & 0.400 \\
\midrule
\multicolumn{10}{@{}p{\dimexpr\linewidth-2\tabcolsep}}{\textbf{\stackexchange}} \\
\hspace{1.5em} nDCG@10 & 0.088 & 0.273 & 0.300 & 0.327 & 0.180 & 0.197 & 0.262 & 0.223 & 0.185 \\
\hspace{1.5em} R@100 & 0.450 & 0.740 & 0.773 & 0.836 & 0.678 & 0.696 & 0.730 & 0.649 & 0.578 \\
\hspace{1.5em} R@1000 & 0.752 & 0.949 & 0.977 & 0.961 & 0.898 & 0.902 & 0.940 & 0.895 & 0.859 \\
\midrule
\multicolumn{10}{@{}p{\dimexpr\linewidth-2\tabcolsep}}{\textbf{\paperretrieval}} \\
\hspace{1.5em} nDCG@10 & 0.376 & 0.338 & 0.429 & 0.391 & 0.414 & 0.424 & 0.421 & 0.429 & 0.457 \\
\hspace{1.5em} R@100 & 0.449 & 0.427 & 0.532 & 0.486 & 0.488 & 0.492 & 0.536 & 0.450 & 0.517 \\
\hspace{1.5em} R@1000 & 0.738 & 0.763 & 0.840 & 0.802 & 0.759 & 0.770 & 0.802 & 0.774 & 0.820 \\
\midrule
\multicolumn{10}{@{}p{\dimexpr\linewidth-2\tabcolsep}}{\textbf{SetOps}} \\
\hspace{1.5em} nDCG@10 & 0.197 & 0.096 & 0.156 & 0.227 & 0.217 & 0.231 & 0.194 & 0.156 & 0.156 \\
\hspace{1.5em} R@100 & 0.413 & 0.253 & 0.364 & 0.484 & 0.447 & 0.484 & 0.430 & 0.316 & 0.351 \\
\hspace{1.5em} R@1000 & 0.644 & 0.511 & 0.650 & 0.771 & 0.717 & 0.740 & 0.710 & 0.555 & 0.590 \\
\midrule
\multicolumn{10}{@{}p{\dimexpr\linewidth-2\tabcolsep}}{\textbf{Clinical Trial}} \\
\hspace{1.5em} nDCG@10 & 0.224 & 0.294 & 0.352 & 0.370 & 0.179 & 0.254 & 0.468 & 0.291 & 0.301 \\
\hspace{1.5em} R@100 & 0.102 & 0.153 & 0.216 & 0.258 & 0.084 & 0.129 & 0.283 & 0.150 & 0.145 \\
\hspace{1.5em} R@1000 & 0.276 & 0.373 & 0.532 & 0.625 & 0.271 & 0.333 & 0.620 & 0.369 & 0.364 \\
\midrule
\multicolumn{10}{@{}p{\dimexpr\linewidth-2\tabcolsep}}{\textbf{Legal QA}} \\
\hspace{1.5em} nDCG@10 & 0.064 & 0.181 & 0.255 & 0.428 & 0.223 & 0.161 & 0.300 & 0.162 & 0.116 \\
\hspace{1.5em} R@100 & 0.257 & 0.561 & 0.680 & 0.821 & 0.598 & 0.504 & 0.703 & 0.506 & 0.425 \\
\hspace{1.5em} R@1000 & 0.520 & 0.775 & 0.835 & 0.863 & 0.796 & 0.751 & 0.834 & 0.751 & 0.687 \\
\midrule
\multicolumn{10}{@{}p{\dimexpr\linewidth-2\tabcolsep}}{\textbf{Theorem Retrieval}} \\
\hspace{1.5em} nDCG@10 & 0.021 & 0.078 & 0.177 & 0.374 & 0.044 & 0.080 & 0.204 & 0.046 & 0.033 \\
\hspace{1.5em} R@100 & 0.114 & 0.325 & 0.583 & 0.723 & 0.248 & 0.255 & 0.511 & 0.173 & 0.124 \\
\hspace{1.5em} R@1000 & 0.399 & 0.647 & 0.899 & 0.921 & 0.553 & 0.524 & 0.764 & 0.367 & 0.370 \\
\midrule
\multicolumn{10}{@{}p{\dimexpr\linewidth-2\tabcolsep}}{\textbf{Code Retrieval}} \\
\hspace{1.5em} nDCG@10 & 0.041 & 0.100 & 0.302 & 0.395 & 0.072 & 0.113 & 0.390 & 0.042 & 0.077 \\
\hspace{1.5em} R@100 & 0.046 & 0.082 & 0.293 & 0.417 & 0.092 & 0.128 & 0.402 & 0.042 & 0.074 \\
\hspace{1.5em} R@1000 & 0.108 & 0.184 & 0.529 & 0.666 & 0.247 & 0.305 & 0.658 & 0.111 & 0.186 \\
\midrule
\multicolumn{10}{@{}p{\dimexpr\linewidth-2\tabcolsep}}{\textbf{Average}} \\
\hspace{1.5em} nDCG@10 & 0.128 & 0.182 & 0.266 & 0.346 & 0.178 & 0.207 & 0.318 & 0.174 & 0.179 \\
\hspace{1.5em} R@100 & 0.235 & 0.358 & 0.491 & 0.587 & 0.368 & 0.398 & 0.522 & 0.310 & 0.301 \\
\hspace{1.5em} R@1000 & 0.447 & 0.590 & 0.749 & 0.809 & 0.607 & 0.633 & 0.772 & 0.530 & 0.534 \\
\bottomrule
\end{tabularx}}
\end{table*}

The main results can be seen in Table \ref{tab:main_results_table}. A general trend is that model performance is substantially lower than on common retrieval evaluation datasets such as BEIR \cite{beir} and TREC DL \cite{craswell2020overview}, suggesting that additional developments are needed to produce high-performance retrieval models on complex tasks. Looking at the models that perform best, we find that the clear standouts are GTE Qwen 7B, Promptriever, and GTE Qwen 1.5B. For their size, the Qwen models lead models in a similar range. This is likely due to the substantial text-pair training data that is used during the first training stage for both models. As these data are diverse and vast they likely increases the models capacity to represent a broader range of concepts, which could explain the improvement in domains far from traditional training datasets. 

Promptriever's high performance is interesting because unlike the GTE models it is trained on a narrow retrieval dataset based on MSMARCO but with instructions added to queries and hard negative passages created to encourage instruction following. Compared to Lion DS 8B, which uses the same Llama-3 8B architecture and MSMARCO training dataset, Promptriever performs substantially better with close to double the nDCG@10 and a sizable increase in both recall metrics. There are a few differences beyond the training data for each model that might play a role in this disparity. One substantial difference is that although both use the same architecture, Promptriever uses the instruction-tuned version of Llama-3 8B while Lion DS uses the base version. This could help Promptriever as during instruction tuning the model may be exposed to examples that are more similar to the complex retrieval tasks in \dataset such as answering code questions. The knowledge gained from instruction tuning might be transferred to the retrieval model and thus improve performance. There are other differences between Promptriever and Lion DS, for one Lion uses knowledge distillation loss and contrastive loss while Promptriever only uses contrastive loss. The impact of knowledge distillation loss is demonstrated in the original Lion paper where the authors find that compared to contrastive loss, the addition of knowledge distillation reduces the model's performance on out-of-domain tasks. There are two other notable differences between Promptriever and Lion DS 8B: Lion uses bi-directional attention while Promptriever uses causal attention. It seems unlikely that the bidirectional attention would result in a decrease in performance, though perhaps it would require more time to adapt to, and thus results in worse performance. Although these factors likely play some role, the introduction of instructions seems like a very important difference, especially as \dataset contains such varied datasets. Having the ability to modify the search-time behavior would probably confer a substantial advantage. Our analysis of the importance of task-specific instruction in Section \ref{sec:impact_of_instructions} shows that instructions are important for performance, but they do not explain Promptriever's substantial performance gain over Lion DS 8B. It might be that the training data helps Promptriever beyond allowing it to follow instructions, as in addition to adding instructions, the training data include new hard negatives which are only negative due to an added requirement from the instructions. This requires the model to learn to judge relevance in a way that might be more subtle than what is present in the original MSMARCO dataset and make Promptriever better suited for complex retrieval tasks. Note that the GTE Qwen models also use the instruction-tuned versions of Qwen as base models and are also trained with instructions, both of which likely play a role in their success. Though like Lion they use bi-directional attention.

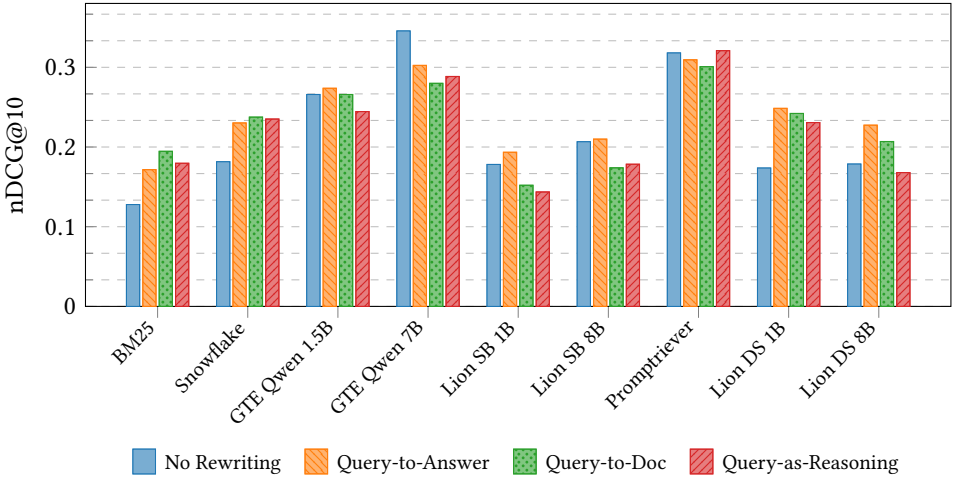
\begin{figure}
\centering

\begin{tikzpicture}

\begin{axis}[
    height=0.4\textwidth,
    width=\textwidth,
    ybar=1.2pt,
    ymajorgrids=true,
    yminorgrids=true,
    grid style={dashed, gray!50},
    x=1.2cm,
    xtick pos=left,
    ytick pos=left,
    minor y tick num=2,
    bar width=5pt, %
    ylabel={nDCG@10},
    ymin=0, %
    symbolic x coords={
        BM25, 
        Snowflake,
        GTE Qwen 1.5B,
        GTE Qwen 7B,
        Lion SB 1B, 
        Lion SB 8B, 
        Promptriever, 
        Lion DS 1B, 
        Lion DS 8B,
    },
    xtick=data, %
    xticklabel style={
        rotate=45, 
        anchor=east,
        font=\footnotesize,
    },
    legend style={
        at={(0.5, -0.45)}, %
        anchor=north,     %
        legend columns=-1,
        font=\footnotesize,
        /tikz/every even column/.append style={column sep=0.24cm},
        /tikz/every odd column/.append style={column sep=0.05cm},
        draw=white,
    },
    legend image code/.code={
        \fill[fill] (0cm,-0.125cm) rectangle (0.3cm,0.2cm);
    },
]

\addplot[mybarstyle={colorMain}{}] coordinates {({BM25}, 0.1278) ({GTE Qwen 1.5B}, 0.2659) ({GTE Qwen 7B}, 0.3456) ({Lion DS 1B}, 0.1736) ({Lion DS 8B}, 0.1786) ({Lion SB 1B}, 0.1781) ({Lion SB 8B}, 0.2066) ({Promptriever}, 0.3180) ({Snowflake}, 0.1815)};
\addplot[mybarstyle={colorQTA}{north west lines}] coordinates {({BM25}, 0.1716) ({GTE Qwen 1.5B}, 0.2737) ({GTE Qwen 7B}, 0.3024) ({Lion DS 1B}, 0.2485) ({Lion DS 8B}, 0.2274) ({Lion SB 1B}, 0.1934) ({Lion SB 8B}, 0.2099) ({Promptriever}, 0.3094) ({Snowflake}, 0.2302)};
\addplot[mybarstyle={colorQTD}{crosshatch dots}] coordinates {({BM25}, 0.1946) ({GTE Qwen 1.5B}, 0.2659) ({GTE Qwen 7B}, 0.2799) ({Lion DS 1B}, 0.2421) ({Lion DS 8B}, 0.2069) ({Lion SB 1B}, 0.1520) ({Lion SB 8B}, 0.1739) ({Promptriever}, 0.3008) ({Snowflake}, 0.2375)};
\addplot[mybarstyle={colorQAR}{north east lines}] coordinates {({BM25}, 0.1796) ({GTE Qwen 1.5B}, 0.2445) ({GTE Qwen 7B}, 0.2885) ({Lion DS 1B}, 0.2307) ({Lion DS 8B}, 0.1679) ({Lion SB 1B}, 0.1436) ({Lion SB 8B}, 0.1784) ({Promptriever}, 0.3209) ({Snowflake}, 0.2351)};

\legend{
    No Rewriting, 
    Query-to-Answer, 
    Query-to-Doc, 
    Query-as-Reasoning
}

\end{axis}
\end{tikzpicture}
\vspace{-25pt}
\caption{Plot of average nDCG@10 on the chunked-document version of \dataset with various query rewriting techniques.}

\label{fig:query_rewriting_short_ndcg}

\end{figure}

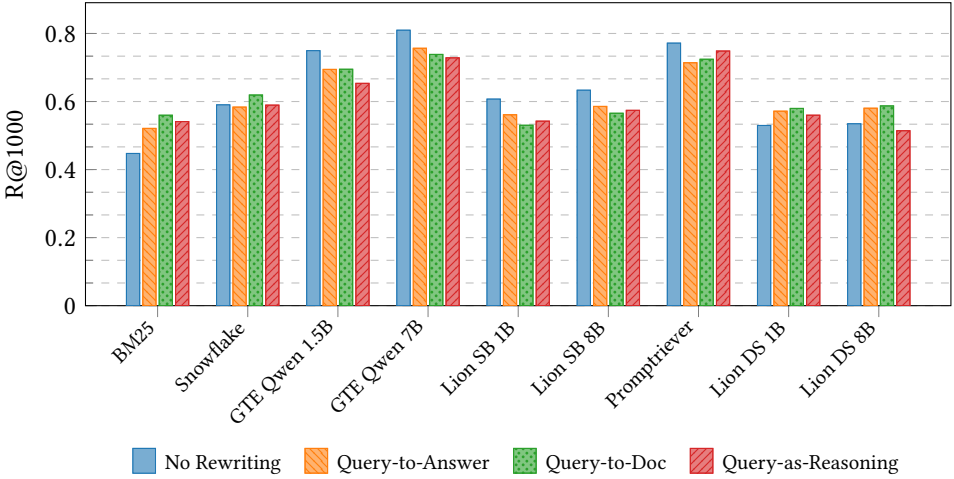
\begin{figure}
\centering

\begin{tikzpicture}

\begin{axis}[
    height=0.4\textwidth,
    width=\textwidth,
    ybar=1.2pt,
    ymajorgrids=true,
    yminorgrids=true,
    grid style={dashed, gray!50},
    x=1.2cm,
    xtick pos=left,
    ytick pos=left,
    minor y tick num=2,
    bar width=5pt, %
    ylabel={R@1000},
    ymin=0, %
    symbolic x coords={
        BM25, 
        Snowflake,
        GTE Qwen 1.5B,
        GTE Qwen 7B,
        Lion SB 1B, 
        Lion SB 8B, 
        Promptriever, 
        Lion DS 1B, 
        Lion DS 8B,
    },
    xtick=data, %
    xticklabel style={
        rotate=45, 
        anchor=east,
        font=\footnotesize,
    },
    legend style={
        at={(0.5, -0.45)}, %
        anchor=north,     %
        legend columns=-1,
        font=\footnotesize,
        /tikz/every even column/.append style={column sep=0.24cm},
        /tikz/every odd column/.append style={column sep=0.05cm},
        draw=white,
    },
    legend image code/.code={
        \fill[fill] (0cm,-0.125cm) rectangle (0.3cm,0.2cm);
    },
]

\addplot[mybarstyle={colorMain}{}] coordinates {({BM25}, 0.4473) ({GTE Qwen 1.5B}, 0.7494) ({GTE Qwen 7B}, 0.8095) ({Lion DS 1B}, 0.5296) ({Lion DS 8B}, 0.5345) ({Lion SB 1B}, 0.6070) ({Lion SB 8B}, 0.6332) ({Promptriever}, 0.7715) ({Snowflake}, 0.5901)};
\addplot[mybarstyle={colorQTA}{north west lines}] coordinates {({BM25}, 0.5210) ({GTE Qwen 1.5B}, 0.6944) ({GTE Qwen 7B}, 0.7565) ({Lion DS 1B}, 0.5716) ({Lion DS 8B}, 0.5803) ({Lion SB 1B}, 0.5611) ({Lion SB 8B}, 0.5856) ({Promptriever}, 0.7140) ({Snowflake}, 0.5836)};
\addplot[mybarstyle={colorQTD}{crosshatch dots}] coordinates {({BM25}, 0.5597) ({GTE Qwen 1.5B}, 0.6948) ({GTE Qwen 7B}, 0.7384) ({Lion DS 1B}, 0.5797) ({Lion DS 8B}, 0.5871) ({Lion SB 1B}, 0.5302) ({Lion SB 8B}, 0.5655) ({Promptriever}, 0.7242) ({Snowflake}, 0.6191)};
\addplot[mybarstyle={colorQAR}{north east lines}] coordinates {({BM25}, 0.5411) ({GTE Qwen 1.5B}, 0.6534) ({GTE Qwen 7B}, 0.7284) ({Lion DS 1B}, 0.5598) ({Lion DS 8B}, 0.5142) ({Lion SB 1B}, 0.5423) ({Lion SB 8B}, 0.5741) ({Promptriever}, 0.7484) ({Snowflake}, 0.5894)};

\legend{
    No Rewriting, 
    Query-to-Answer, 
    Query-to-Doc, 
    Query-as-Reasoning
}

\end{axis}
\end{tikzpicture}
\vspace{-25pt}
\caption{Plot of average R@1000 on the chunked-document version of \dataset with various query rewriting techniques.}

\label{fig:query_rewriting_short_recall}

\end{figure}

The takeaway from the best performing models is that training data, unsurprisingly, plays a significant role in the ability of retrieval models to perform well in varied and complex retrieval tasks. Interestingly, Promptriever shows that a vast diverse set of training data may not be necessary for good performance and that instruction-tuning alone can have a substantial impact on the capability of retrieval models. With that said, the performance of GTE Qwen suggests there are still benefits to training on a large collection of diverse text pairs. Another takeaway is that larger models generally do better, with the two best performing models having seven billion or more parameters. In support of this conclusion, directly comparing GTE Qwen 1.5B and GTE Qwen 7B we can see a notable increase in the average performance of the larger model; though the smaller Qwen does do better on \paperretrieval.

\subsection{Performance per Task} \label{sec:perfromance_per_task}
To understand how the unique features of each task within \dataset impact the performance of the models, in this section, we breakdown the model results by task with a focus on what might cause the models to succeed or fail on a given task.

\subsubsection{Tip-of-the-tongue}
The low nDCG and recall values of BM25 suggest that there is very little term overlap which is supported by the properties of tip-of-the-tongue queries where people try to recall features from memory without remembering key identifiers such as the title or actors names. There is substantial variance in both nDCG and recall across the various models. The general ranking is similar to the average with the notable exception that Promptriever has the highest nDCG value.

The best model in terms of R@1000, GTE Qwen 7B, is able to achieve a respectable 0.867 which is close to what state-of-the-art models achieve on datasets like TREC DL. The high recall contrasted with the low nDCG suggests that models struggle to find the correct item, but might be able to find reasonable candidates quite well. This highlights an important aspect of retrieval which is how plausible candidates scale as the granularity of relevance decreases. In this case, the model may not be able to exactly match the qualities of the query with the document, but might be able to find broader features which match such as genre, general time period, etc. By leveraging these broader features the candidate space might decrease significantly which explains the high recall as only a small subset of the movies in the corpus fall into the intersection of these broad subsets. 

\subsubsection{\stackexchange}
The nDCG values are fairly low, and BM25 is especially low, suggesting that term matching is not able to easily distinguish relevant documents from irrelevant ones. The variance of nDCG values is not as high as other datasets, but still reasonably large. The variance of R@100 is substantial while R@1000 is fairly compressed. The high recall values suggest models are able to find a reasonable set of candidate documents, but struggle to distinguish the subtle differences between relevant and hard negative documents. 

The low nDCG and high recall likely indicates a similar phenomenon as in Tip-of-the-tongue where it is pretty easy to retrieve candidates which share broad features with relevant documents such as topic, but it is difficult for the models to differentiate relevant documents within this candidate pool. The construction of the dataset may make finding the broad pool of topically relevant candidates easier, as each query is meant to be distinct enough from other queries to ensure there would be no overlap in relevant documents, which likely results in minimal overlap in document topics.

\subsubsection{\paperretrieval}

This dataset is characterized by long multi-aspect queries with specialized terminology on scientific topics. Performance is relatively high across the board compared to other datasets, with even BM25 achieving a respectable nDCG@10 of 0.376. This suggests that the aspects in the queries often contain keywords present in the relevant paper titles and abstracts. However, the top neural models still provide a significant boost, with Lion DS 8B achieving the highest nDCG@10 of 0.457. This is one of the few datasets where the dense Lion models outperform their sparse counterparts, lending credence to the idea explored in Section \ref{sec_comparing_dense_and_sparse} that dense embeddings are better at capturing the meaning of rare specialized terminology. The instruction-tuned models like Promptriever and GTE Qwen also perform strongly, though worse than Lion DS 8B which makes this the only dataset where this is true. This may be because the paragraph length query and abstract are similar to the passages in MSMARCO which allows the Lion models to be closer to their "in-domain" performance.

\subsubsection{Set-Ops}
The results of this dataset are unique in that BM25 outperforms several models in both nDCG and recall. This suggests that term matching is an important feature which is not surprising, as many of the queries contain entity names such as \textit{books by Leo Tolstoy written in 1863}. This would clearly explain why BM25 is able to do well on the recall metrics, but given that the queries also require set-based operation logical such as \textit{and}, \textit{or}, and \textit{not} it is surprising that BM25 is able to do so well. This result might indicate that the neural retrieval models that do worse than BM25 fail in two ways: (1) they fail to represent the precise information in the queries and documents and (2) they fail to understand the set-based operations. Both seem plausible with prior work on the ability of neural retrieval models to handle negations and other set-based operations \cite{negation_in_neural_information_retrieval, reproducing_nevir, quest} show that unless models are specifically trained they will often fail.

Additional evidence for the importance of term matching can be seen in the performance of Lion SB 8B which takes second place beating out the overall silver medalist Promptriever. This is unusual as generally Promptriever is significantly better than the Lion models. The stronger performance from Lion may also be due to the similarity between the queries from SetOps and the MSMARCO queries used for training. Unlike many of the other tasks, the SetOps queries are fairly short and have an implicit information need that is similar to the queries found in MSMARCO.

Overall, the performance is below average even for the best models. This suggests that models struggle with set-based logic in queries which is supported by existing work \cite{quest, negation_in_neural_information_retrieval} and that models may also struggle to represent the entity-dense information from both the queries and documents.

\subsubsection{Clinical Trial}
In this task, models must match a patient's medical history to relevant clinical trial descriptions, both of which are rich in specialized medical terminology. The results show that this is a challenging task, with low recall scores across all models; even the best model, GTE Qwen 7B, only achieves an R@1000 of 0.625. This indicates difficulty in even identifying a broad set of candidate trials from the large corpus. It is worth noting that this is one of the larger datasets with around 900k chunks and has by far the highest number of positive documents per query at 391.5. It may be the case that having more positive documents makes recalling them all effectively more difficult, as they may be more spread out in embedding space. Interestingly, although models struggle with recall the best nDCG@10 (0.468) achieved by Promptriever is well above the average for the other datasets. This may be because distinguishing relevant clinical trials requires paying attention to small details which may align well with the generated hard negatives that Promptriever is trained with. Although Promptriever is unique in how high its nDCG is, it is not unique in having a higher nDCG than the benchmark average. Several models score above their average, suggesting that high precision is a general trend. This is interesting as in other datasets such as \stackexchange and Tip-of-the-tongue, we see the opposite, models do well on recall but poorly on nDCG. 
This might be explained by revisiting the number of positive documents and how they might be distributed in embedding space. Consider that there are likely some kinds of clusters in embedding space with similar documents grouped together, when there are more positives these clusters have a higher likelihood of contain several positives for a given query. For high precision at a low k, such as 10, the model may only need to place the query embedding close to one of these positive clusters. Conversely, to get high recall the model would need to place multiple embeddings to be close to multiple clusters, which, as these are all single-embedding models, is impossible.  
Another observation is that similar to the \paperretrieval dataset, this is another domain where dense models (like Lion DS) show a competitive advantage over their sparse counterparts, likely due to their ability to better handle the complex, domain-specific vocabulary. This may be due to tokenization which may split medical terms into several sub-word tokens, and thus reduce the precision the model can express. This may be exacerbated by the property of medical terminology to share prefixes and suffixes (e.g. \textit{Hemophilia} and \textit{Hemoglobin}).

\subsubsection{Legal QA}
The Legal QA task requires matching a natural language question about a housing issue to a specific legal statute. Overall, the results seem to be above average especially in terms of recall. The low nDCG for BM25 suggests that term matching is not enough to distinguish relevant from irrelevant documents. This could be because questions are state specific, so relevant documents might exist for various states, not just the one specified in the question. It might also be because the questions require reasoning and thus cannot be solved with keyword matching alone.
The model performances are inline with the average model rankings for the most part with the GTE Qwen models and Promptriever leading the pack for both recall and nDCG. Overall, the results suggest strong models are pretty good at finding candidate documents and are better at distinguishing them from irrelevant documents than on some of the other datasets. This seems reasonable given the characteristics of the dataset. Compared to many of the other datasets the number of aspects is pretty small and one of them, the state being inquired about, is very simple to match with the documents. Additionally, the format is similar to those seen in common training datasets like MSMARCO and NQ \cite{msmarco, natural_questions}. Further, the terminology is likely less specialized than in \paperretrieval and Clinical Trial which likely further improves model performance.

\subsubsection{Theorem Retrieval}
This dataset represents a reasoning-intensive retrieval task where models must connect a math problem (query) to a theorem definition (document) that could be used to solve it. This requires a deep level of mathematical reasoning and also an ability to understand the various constraints introduced in the problem definition. Performance is low for most models, indicating extreme difficulty. However, there is a clear separation at the top, with GTE Qwen 7B achieving a significantly higher nDCG@10 (0.374) than any other model. This large gap demonstrates the importance of model size for complex retrieval tasks as well as the diversity of training data. There is also a significant difference in recall between the Qwen models and all other models. The sizable gap likely also indicates that the training data for these models include similar retrieval tasks.
The gap between the sparse and dense Lion models is fairly surprising given one might expect it is difficult to represent mathematical information as a bag-of-tokens and a more semantically-rich, dense embedding would likely do a better job. It may be because the task is not pure math, there are descriptions of each theorem to use for matching. It might also be the case that the sparse model is able to use some shared tokens to convey the meaning without trying to represent the math equations explicitly. 
Overall, the results underscore the importance of training data, especially for recall. The strong recall performance of the best models shows that even on reasoning-intensive complex tasks existing neural retrieval models can find a strong set of candidate documents even if the fine-grain ranking needs work. 

\subsubsection{Code Retrieval}
For the Code Retrieval task, models must retrieve code solutions given a natural language problem description. As noted, there is often minimal keyword overlap between the problem and the solution, making this a challenging complex retrieval problem; which is supported by the low BM25 scores for both recall and nDCG. Again, we see that the Qwen models and Promptriever dominate and even outperform their average nDCG. In contrast, recall is below average for every model. The low recall and high nDCG mirrors the trends from Clinical Trial which suggests models may be able to find some of the correct answers, but struggle to recall the potentially diverse set of relevant answers. Though the top three models do above average in terms of nDCG this is not true of the other models which underperform their averages often significantly. The large range of nDCG might indicate that whatever is different about the top performing models is well suited for code-based retrieval.
On this topic, it is notable that the best three models are also those which use instruction tuned LLMs as base models. This could indicate that the additional instruction data used on these models makes them more capable of understanding code. As instruction-tuning often includes a specific focus on code it seems reasonable that these models are better at understanding and thus retrieving code. Additionally, the GTE models are specifically trained on code-text pairs which likely explains the slight performance edge over Promptriever though it is not as significant as in some of the other tasks such as Theorem Retrieval or \stackexchange.

\subsection{Full Document Results}
The results of the baseline models for the full-document variant of \dataset are presented in Table \ref{tab:auto_main_full_doc}. As mentioned previously, we believe the passage version of \dataset is more realistic and better reflects the general trends and capabilities of retrieval model performance. Thus, we only include a limited analysis of the results on the full document collection with a focus on the change in model behavior between the chunked and full document versions.

Before analyzing the results, it is worth noting that some datasets only have one corpus as the original documents do not require chunking. These datasets---\paperretrieval, Theorem Retrieval, and Code Retrieval---are included in both result tables with identical results, as they can be used in either collection without modification. Since their results are unchanged from the passage version, we focus our analysis on the datasets that differ between the two collections.

The remaining datasets show general trends similar to the passage version in terms of model rankings and relative performance, with the top-performing models (GTE Qwen 7B, Promptriever, and GTE Qwen 1.5B) maintaining their lead. However, there are notable differences in absolute performance levels that provide insights into the trade-offs between chunking and full-document retrieval.

For the datasets that had per-chunk labels in the passage variant---Legal QA and \stackexchange---we observe interesting and contrasting patterns. \stackexchange shows a substantial improvement in the full document version, with nDCG@10 increasing from a range of 0.088-0.327 in the short version to 0.129-0.561 in the long version. The recall improvements are even more dramatic, with R@1000 jumping from 0.752-0.977 to 0.869-1.000, with several models achieving perfect recall. This substantial improvement can be attributed to the significantly smaller document collection size (5k versus 41k documents), which makes finding relevant documents considerably easier as there are fewer topically similar but irrelevant documents to distinguish between.

Legal QA presents a more varied picture with mixed results across different models. The top-performing models---GTE Qwen 7B, Promptriever, and GTE Qwen 1.5B---generally show improvements in the full document version, as does BM25 and Lion SB 1B. However, other neural models including Lion DS 1B, Lion DS 8B, Snowflake, and Lion SB 8B exhibit notable performance decreases. This suggests that while some models can effectively leverage the additional context and reduced corpus size, others struggle with the longer sequences that must be processed and represented.

The remaining datasets---Tip-of-the-tongue, SetOps, and Clinical Trial---allow for the most direct comparison between the passage and document versions of \dataset, as they only have document-level labels and were evaluated using MaxP aggregation on the chunked version. This comparison essentially examines whether chunking or using full documents produces better retrieval performance.

For SetOps, we observe that nearly all neural models perform worse on the document collection compared to the chunked version. This aligns with the common wisdom that chunking is beneficial. The queries in SetOps are often fact-based and entity-rich, requiring the model to find a specific piece of information within a document. Chunking helps by isolating these facts into smaller, more focused passages. A single embedding for a long document, such as a full Wikipedia page, may dilute or "average out" these crucial details, making it harder for the model to perform precise matching.

On Tip-of-the-tongue, the results are mixed. An interesting trend is that for all small and large model pairs---Qwen 1.5B and Qwen 7B, Lion SB 1B and Lion SB 8B, and Lion DS 1B and Lion DS 8B---the large models do worse in both nDCG and recall on the full document collections than on the passage collection. In contrast, the smaller models see an increase in both nDCG and recall. Looking beyond the pairs of models, the other unpaired models show a similar trend, with the larger Promptriever performing worse and the smaller Snowflake performing better. BM25 seems to be included in the smaller model camp and sees better performance. This trend is pretty odd given that generally larger models have both larger embedding sizes and are better at understanding long-context information. One possible interpretation is that chunking is better, but a capable model is required to fully take advantage of the contextualized chunks. If the smaller models are not able to understand the contextualized chunking format, it may add noise to the chunked version, which is less prominent, and a smaller ratio of the overall text, in the full document version. Another potential reason for this odd behavior is that larger models are able to fill in missing context with their parametric knowledge in a way that a smaller model may not be able to. If this were the case for the chunked collection, they may be able to better represent chunks than the smaller models. As for why the larger models would see a lower performance on the full document collection, it could be that they have to do more compression, which hurts the representation quality, than they need to do for the chunked collection. Although these theories might explain this phenomenon, future experiments are needed to fully understand it.

Clinical Trial does not show the same inverse scaling; instead nearly all models see a boost to nDCG with the full documents, with the notable exception of the best performing models in both collections, Promptriever which sees a lower nDCG. Promptriever also has a decrease in recall though it is not the only model to have lower recall. Qwen 7B and Lion SB 8B also have lower recall with the full document collection, while the rest of the models have some improvement. This might suggest a weaker recall-only version of the inverse performance trend from Tip-of-the-tongue expect that Lion DS 8B sees a large improvement in recall. The variation in performance makes it hard to draw a solid conclusion from this task. One clear takeaway from both Clinical Trial and some of the other tasks is that better performing models in the passage collection, especially large ones, tend to see performance degrade with full documents.  

There is no clear message from the comparison between the full documents and chunked documents. In some cases, chunking leads to clear improvements, while in other cases there is substantial degradation. While it may be useful to some extent to compare \stackexchange and Legal QA the difference in labels makes the comparison more muddied than the other datasets which have document-level labels. Even with this benefit, the picture from the tasks with document-level label is still murky with a range of outcomes across models and tasks. One observation is that the chunked versions of these collections always contain the highest nDCG and recall which indicates that chunking does seem to be optimal for the best performing models. This aligns with prior work that shows that chunking, generally, improves retrieval performance. The gain that many models show seems to contradict these findings and requires more investigation to fully understand. We leave this to future work.

\subsection{Comparison of Learned Dense and Sparse Retrieval Models} \label{sec_comparing_dense_and_sparse}
Beyond exploring the top-performing models, our experiments reveal some other interesting findings. Notably, comparing the sparse and dense Lion models, we find that there is a substantial difference between the two. For instance, Lion SB 1B outperforms Lion DS 8B on both recall metric and is only 0.01 points worse on nDCG despite being less than a quarter of the size. The aggregated results suggest that when trained in the same way, sparse retrieval models have a significant advantage over their dense counterparts. 

Inspecting the individual task performance, we find that the sparse models perform better in the majority of the datasets often by substantial amounts; for example in Tip-of-the-tongue, the sparse models have nDCG@10 values almost double the dense version with large increases in recall. The only datasets where the dense models performed better are the Clinical Trial and \paperretrieval datasets. These datasets likely represent the datasets with the most specialized terminology, except perhaps Theorem retrieval, which might indicate that sparse models struggle to represent rarer terminology.

\subsection{Impact of Query Rewriting}
The results of the average nDCG and R@1000 with query rewriting techniques can be seen in Figure \ref{fig:query_rewriting_short_ndcg} and Figure \ref{fig:query_rewriting_short_recall}, respectively. The complete result table for each rewriting technique can be viewed in Table \ref{tab:auto_query_to_answer_cot} (Query-to-Answer), Table \ref{tab:auto_query_to_doc_cot} (Query-to-Doc), and Table \ref{tab:auto_query_as_reasoning_steps} (Query-as-Reasoning-Steps). The results show that rewriting techniques tend to help weaker models while hurting stronger models. For the best model, GTE Qwen 7B, all rewriting techniques harm nDCG@10 and R@1000. Rewriting impacts recall and nDCG differently, only the bottom four models see any improvement on R@1000 with any of the techniques while for nDCG@10 all but the strongest model has some improvement or stay the same. Looking only at the models that see improved nDCG there are differences between the weaker and stronger models. Weaker models see gains with more of the rewriting techniques on average, while stronger models tend to only have one technique which helps. The gains from rewriting techniques are also larger for weaker models.

Of the rewriting techniques, Query-to-Answer is the approach which improves nDCG@10 for the most models with it improving performance for five out of the eight models which saw improvements from rewriting. The Query-to-Doc is second bringing the improvement to two models and Query-as-Reasoning is last with improvements to only one model. Looking at R@1000 all four models that saw improvement got the most improvement from Query-to-Doc. These results are reasonable as directly providing an answer, when the rewriting model is correct and there are a limited number of correct answers, would likely lead to high overlap with the relevant document, thus resulting in higher nDCG@10. Query-to-Doc on the other hand may introduce more noise in the form of context for the answer, formatting, background, etc. which hurts precision, but which adds useful key-words and concepts increasing recall. The minimal or lack of gains in recall for strong models might indicate that they already have enough general knowledge to match broad categories and topics which enables strong recall. Rewriting may hurt this ability by adding superfluous information or incorrect information which dilutes the original query intent. Additionally, the rewritten query may modify the distribution of the queries making them less like the query distributions the models were trained on which may contribute in the degraded performance.

\begin{table*}[]
\centering
\caption{Overview of baseline model performance for the full-document version of \dataset. We use three metrics: normalized discounted cumulative gain at 10 (nDCG@10), recall at 100 (R@100), and recall at 1000 (R@1000). The last row shows the average (mean) value of each metric across all the tasks.}
\label{tab:auto_main_full_doc}
\resizebox{0.88\linewidth}{!}{
\begin{tabularx}{0.98\linewidth}{l!{\color{lightgray}\vrule}C !{\color{lightgray}\vrule} C !{\color{lightgray}\vrule} C !{\color{lightgray}\vrule} C !{\color{lightgray}\vrule} C !{\color{lightgray}\vrule} C !{\color{lightgray}\vrule} C !{\color{lightgray}\vrule} C !{\color{lightgray}\vrule} C}
\toprule
Metric / Dataset & \rotcol{BM25} & \rotcol{Snowflake} & \rotcol{GTE Qwen 1.5B} & \rotcol{GTE Qwen 7B} & \rotcol{Lion SB 1B} & \rotcol{Lion SB 8B} & \rotcol{Promptriever} & \rotcol{Lion DS 1B} & \rotcol{Lion DS 8B} \\
\midrule
\multicolumn{10}{@{}p{\dimexpr\linewidth-2\tabcolsep}}{\textbf{Tip-of-the-tongue}} \\
\hspace{1.5em} nDCG@10 & 0.042 & 0.110 & 0.212 & 0.237 & 0.140 & 0.075 & 0.225 & 0.043 & 0.045 \\
\hspace{1.5em} R@100 & 0.148 & 0.348 & 0.519 & 0.600 & 0.378 & 0.215 & 0.593 & 0.222 & 0.141 \\
\hspace{1.5em} R@1000 & 0.289 & 0.556 & 0.793 & 0.830 & 0.652 & 0.393 & 0.837 & 0.437 & 0.363 \\
\midrule
\multicolumn{10}{@{}p{\dimexpr\linewidth-2\tabcolsep}}{\textbf{\stackexchange}} \\
\hspace{1.5em} nDCG@10 & 0.129 & 0.435 & 0.447 & 0.561 & 0.306 & 0.312 & 0.442 & 0.357 & 0.249 \\
\hspace{1.5em} R@100 & 0.505 & 0.944 & 0.953 & 0.981 & 0.850 & 0.879 & 0.944 & 0.855 & 0.729 \\
\hspace{1.5em} R@1000 & 0.869 & 0.991 & 1.000 & 1.000 & 0.967 & 0.977 & 0.995 & 1.000 & 0.939 \\
\midrule
\multicolumn{10}{@{}p{\dimexpr\linewidth-2\tabcolsep}}{\textbf{\paperretrieval}} \\
\hspace{1.5em} nDCG@10 & 0.376 & 0.338 & 0.429 & 0.391 & 0.414 & 0.424 & 0.421 & 0.429 & 0.457 \\
\hspace{1.5em} R@100 & 0.449 & 0.427 & 0.532 & 0.486 & 0.488 & 0.492 & 0.536 & 0.450 & 0.517 \\
\hspace{1.5em} R@1000 & 0.738 & 0.763 & 0.840 & 0.802 & 0.759 & 0.770 & 0.802 & 0.774 & 0.820 \\
\midrule
\multicolumn{10}{@{}p{\dimexpr\linewidth-2\tabcolsep}}{\textbf{SetOps}} \\
\hspace{1.5em} nDCG@10 & 0.202 & 0.092 & 0.152 & 0.215 & 0.197 & 0.204 & 0.198 & 0.140 & 0.142 \\
\hspace{1.5em} R@100 & 0.408 & 0.254 & 0.352 & 0.457 & 0.432 & 0.390 & 0.432 & 0.279 & 0.295 \\
\hspace{1.5em} R@1000 & 0.655 & 0.515 & 0.623 & 0.722 & 0.703 & 0.634 & 0.710 & 0.509 & 0.505 \\
\midrule
\multicolumn{10}{@{}p{\dimexpr\linewidth-2\tabcolsep}}{\textbf{Clinical Trial}} \\
\hspace{1.5em} nDCG@10 & 0.260 & 0.343 & 0.371 & 0.403 & 0.233 & 0.278 & 0.428 & 0.331 & 0.327 \\
\hspace{1.5em} R@100 & 0.119 & 0.178 & 0.237 & 0.279 & 0.104 & 0.132 & 0.264 & 0.164 & 0.161 \\
\hspace{1.5em} R@1000 & 0.298 & 0.421 & 0.569 & 0.620 & 0.280 & 0.332 & 0.577 & 0.413 & 0.394 \\
\midrule
\multicolumn{10}{@{}p{\dimexpr\linewidth-2\tabcolsep}}{\textbf{Legal QA}} \\
\hspace{1.5em} nDCG@10 & 0.279 & 0.087 & 0.322 & 0.457 & 0.281 & 0.101 & 0.330 & 0.094 & 0.036 \\
\hspace{1.5em} R@100 & 0.599 & 0.442 & 0.689 & 0.766 & 0.653 & 0.345 & 0.690 & 0.438 & 0.208 \\
\hspace{1.5em} R@1000 & 0.807 & 0.699 & 0.847 & 0.872 & 0.827 & 0.591 & 0.864 & 0.689 & 0.443 \\
\midrule
\multicolumn{10}{@{}p{\dimexpr\linewidth-2\tabcolsep}}{\textbf{Theorem Retrieval}} \\
\hspace{1.5em} nDCG@10 & 0.021 & 0.078 & 0.177 & 0.374 & 0.044 & 0.080 & 0.204 & 0.046 & 0.033 \\
\hspace{1.5em} R@100 & 0.114 & 0.325 & 0.583 & 0.723 & 0.248 & 0.255 & 0.511 & 0.173 & 0.124 \\
\hspace{1.5em} R@1000 & 0.399 & 0.647 & 0.899 & 0.921 & 0.553 & 0.524 & 0.764 & 0.367 & 0.370 \\
\midrule
\multicolumn{10}{@{}p{\dimexpr\linewidth-2\tabcolsep}}{\textbf{Code Retrieval}} \\
\hspace{1.5em} nDCG@10 & 0.041 & 0.100 & 0.302 & 0.395 & 0.072 & 0.113 & 0.390 & 0.042 & 0.077 \\
\hspace{1.5em} R@100 & 0.046 & 0.082 & 0.293 & 0.417 & 0.092 & 0.128 & 0.402 & 0.042 & 0.074 \\
\hspace{1.5em} R@1000 & 0.108 & 0.184 & 0.529 & 0.666 & 0.247 & 0.305 & 0.658 & 0.111 & 0.186 \\
\midrule
\multicolumn{10}{@{}p{\dimexpr\linewidth-2\tabcolsep}}{\textbf{Average}} \\
\hspace{1.5em} nDCG@10 & 0.169 & 0.198 & 0.302 & 0.379 & 0.211 & 0.216 & 0.330 & 0.185 & 0.189 \\
\hspace{1.5em} R@100 & 0.298 & 0.375 & 0.520 & 0.589 & 0.406 & 0.374 & 0.546 & 0.328 & 0.301 \\
\hspace{1.5em} R@1000 & 0.520 & 0.597 & 0.762 & 0.804 & 0.624 & 0.590 & 0.776 & 0.537 & 0.522 \\
\bottomrule
\end{tabularx}}
\end{table*}

\begin{table*}[]
\centering
\caption{
Overview of model performance using the \textit{Query-to-Answer CoT} query rewriting approach on the chunked-document version of \dataset. The \textit{Query-to-Answer CoT} approach uses a generated direct answer to the original query as the rewritten query. We use three metrics: normalized discounted cumulative gain at 10 (nDCG@10), recall at 100 (R@100), and recall at 1000 (R@1000). The last row shows the average (mean) value of each metric across all the tasks.}
\label{tab:auto_query_to_answer_cot}
\resizebox{0.88\linewidth}{!}{
\begin{tabularx}{0.98\linewidth}{l!{\color{lightgray}\vrule}C !{\color{lightgray}\vrule} C !{\color{lightgray}\vrule} C !{\color{lightgray}\vrule} C !{\color{lightgray}\vrule} C !{\color{lightgray}\vrule} C !{\color{lightgray}\vrule} C !{\color{lightgray}\vrule} C !{\color{lightgray}\vrule} C}
\toprule
Metric / Dataset & \rotcol{BM25} & \rotcol{Snowflake} & \rotcol{GTE Qwen 1.5B} & \rotcol{GTE Qwen 7B} & \rotcol{Lion SB 1B} & \rotcol{Lion SB 8B} & \rotcol{Promptriever} & \rotcol{Lion DS 1B} & \rotcol{Lion DS 8B} \\
\midrule
\multicolumn{10}{@{}p{\dimexpr\linewidth-2\tabcolsep}}{\textbf{Tip-of-the-tongue}} \\
\hspace{1.5em} nDCG@10 & 0.045 & 0.062 & 0.068 & 0.114 & 0.073 & 0.090 & 0.101 & 0.049 & 0.052 \\
\hspace{1.5em} R@100 & 0.059 & 0.133 & 0.193 & 0.341 & 0.185 & 0.222 & 0.274 & 0.081 & 0.119 \\
\hspace{1.5em} R@1000 & 0.119 & 0.252 & 0.444 & 0.726 & 0.356 & 0.415 & 0.467 & 0.230 & 0.267 \\
\midrule
\multicolumn{10}{@{}p{\dimexpr\linewidth-2\tabcolsep}}{\textbf{\stackexchange}} \\
\hspace{1.5em} nDCG@10 & 0.258 & 0.352 & 0.373 & 0.359 & 0.292 & 0.327 & 0.378 & 0.391 & 0.364 \\
\hspace{1.5em} R@100 & 0.644 & 0.799 & 0.805 & 0.815 & 0.675 & 0.698 & 0.786 & 0.779 & 0.772 \\
\hspace{1.5em} R@1000 & 0.867 & 0.952 & 0.970 & 0.960 & 0.927 & 0.923 & 0.955 & 0.926 & 0.926 \\
\midrule
\multicolumn{10}{@{}p{\dimexpr\linewidth-2\tabcolsep}}{\textbf{\paperretrieval}} \\
\hspace{1.5em} nDCG@10 & 0.205 & 0.254 & 0.295 & 0.280 & 0.169 & 0.150 & 0.325 & 0.250 & 0.225 \\
\hspace{1.5em} R@100 & 0.311 & 0.357 & 0.413 & 0.387 & 0.254 & 0.246 & 0.426 & 0.308 & 0.292 \\
\hspace{1.5em} R@1000 & 0.595 & 0.687 & 0.767 & 0.748 & 0.552 & 0.540 & 0.778 & 0.614 & 0.659 \\
\midrule
\multicolumn{10}{@{}p{\dimexpr\linewidth-2\tabcolsep}}{\textbf{SetOps}} \\
\hspace{1.5em} nDCG@10 & 0.080 & 0.098 & 0.154 & 0.163 & 0.127 & 0.134 & 0.167 & 0.142 & 0.120 \\
\hspace{1.5em} R@100 & 0.208 & 0.229 & 0.349 & 0.371 & 0.284 & 0.288 & 0.354 & 0.267 & 0.243 \\
\hspace{1.5em} R@1000 & 0.450 & 0.484 & 0.620 & 0.672 & 0.550 & 0.534 & 0.614 & 0.464 & 0.453 \\
\midrule
\multicolumn{10}{@{}p{\dimexpr\linewidth-2\tabcolsep}}{\textbf{Clinical Trial}} \\
\hspace{1.5em} nDCG@10 & 0.255 & 0.306 & 0.329 & 0.347 & 0.148 & 0.196 & 0.349 & 0.311 & 0.294 \\
\hspace{1.5em} R@100 & 0.183 & 0.188 & 0.225 & 0.245 & 0.061 & 0.096 & 0.253 & 0.191 & 0.165 \\
\hspace{1.5em} R@1000 & 0.464 & 0.384 & 0.519 & 0.549 & 0.176 & 0.230 & 0.539 & 0.407 & 0.362 \\
\midrule
\multicolumn{10}{@{}p{\dimexpr\linewidth-2\tabcolsep}}{\textbf{Legal QA}} \\
\hspace{1.5em} nDCG@10 & 0.133 & 0.307 & 0.306 & 0.413 & 0.266 & 0.232 & 0.358 & 0.302 & 0.215 \\
\hspace{1.5em} R@100 & 0.413 & 0.700 & 0.708 & 0.805 & 0.644 & 0.609 & 0.732 & 0.682 & 0.572 \\
\hspace{1.5em} R@1000 & 0.669 & 0.836 & 0.841 & 0.862 & 0.814 & 0.801 & 0.849 & 0.819 & 0.775 \\
\midrule
\multicolumn{10}{@{}p{\dimexpr\linewidth-2\tabcolsep}}{\textbf{Theorem Retrieval}} \\
\hspace{1.5em} nDCG@10 & 0.315 & 0.372 & 0.413 & 0.427 & 0.411 & 0.412 & 0.422 & 0.425 & 0.379 \\
\hspace{1.5em} R@100 & 0.584 & 0.717 & 0.739 & 0.774 & 0.669 & 0.690 & 0.725 & 0.696 & 0.636 \\
\hspace{1.5em} R@1000 & 0.790 & 0.831 & 0.892 & 0.905 & 0.877 & 0.888 & 0.849 & 0.850 & 0.851 \\
\midrule
\multicolumn{10}{@{}p{\dimexpr\linewidth-2\tabcolsep}}{\textbf{Code Retrieval}} \\
\hspace{1.5em} nDCG@10 & 0.082 & 0.091 & 0.252 & 0.316 & 0.061 & 0.138 & 0.375 & 0.118 & 0.170 \\
\hspace{1.5em} R@100 & 0.094 & 0.097 & 0.261 & 0.352 & 0.080 & 0.151 & 0.403 & 0.116 & 0.165 \\
\hspace{1.5em} R@1000 & 0.214 & 0.243 & 0.502 & 0.630 & 0.237 & 0.354 & 0.661 & 0.263 & 0.349 \\
\midrule
\multicolumn{10}{@{}p{\dimexpr\linewidth-2\tabcolsep}}{\textbf{Average}} \\
\hspace{1.5em} nDCG@10 & 0.172 & 0.230 & 0.274 & 0.302 & 0.193 & 0.210 & 0.309 & 0.248 & 0.227 \\
\hspace{1.5em} R@100 & 0.312 & 0.403 & 0.462 & 0.511 & 0.357 & 0.375 & 0.494 & 0.390 & 0.370 \\
\hspace{1.5em} R@1000 & 0.521 & 0.584 & 0.694 & 0.756 & 0.561 & 0.586 & 0.714 & 0.572 & 0.580 \\
\bottomrule
\end{tabularx}}
\end{table*}

\begin{table*}[]
\centering
\caption{
Overview of model performance using the \textit{Query-to-Doc CoT} query rewriting approach on the chunked-document version of \dataset. The \textit{Query-to-Doc CoT} approach uses a generated relevant document created based on the original query as the rewritten query. We use three metrics: normalized discounted cumulative gain at 10 (nDCG@10), recall at 100 (R@100), and recall at 1000 (R@1000). The last row shows the average (mean) value of each metric across all the tasks.
}
\label{tab:auto_query_to_doc_cot}
\resizebox{0.88\linewidth}{!}{
\begin{tabularx}{0.98\linewidth}{l!{\color{lightgray}\vrule}C !{\color{lightgray}\vrule} C !{\color{lightgray}\vrule} C !{\color{lightgray}\vrule} C !{\color{lightgray}\vrule} C !{\color{lightgray}\vrule} C !{\color{lightgray}\vrule} C !{\color{lightgray}\vrule} C !{\color{lightgray}\vrule} C}
\toprule
Metric / Dataset & \rotcol{BM25} & \rotcol{Snowflake} & \rotcol{GTE Qwen 1.5B} & \rotcol{GTE Qwen 7B} & \rotcol{Lion SB 1B} & \rotcol{Lion SB 8B} & \rotcol{Promptriever} & \rotcol{Lion DS 1B} & \rotcol{Lion DS 8B} \\
\midrule
\multicolumn{10}{@{}p{\dimexpr\linewidth-2\tabcolsep}}{\textbf{Tip-of-the-tongue}} \\
\hspace{1.5em} nDCG@10 & 0.051 & 0.061 & 0.060 & 0.084 & 0.041 & 0.067 & 0.100 & 0.061 & 0.079 \\
\hspace{1.5em} R@100 & 0.119 & 0.215 & 0.207 & 0.341 & 0.156 & 0.237 & 0.267 & 0.200 & 0.222 \\
\hspace{1.5em} R@1000 & 0.185 & 0.459 & 0.548 & 0.726 & 0.385 & 0.511 & 0.630 & 0.333 & 0.385 \\
\midrule
\multicolumn{10}{@{}p{\dimexpr\linewidth-2\tabcolsep}}{\textbf{\stackexchange}} \\
\hspace{1.5em} nDCG@10 & 0.258 & 0.371 & 0.387 & 0.336 & 0.241 & 0.291 & 0.333 & 0.367 & 0.304 \\
\hspace{1.5em} R@100 & 0.652 & 0.780 & 0.809 & 0.802 & 0.623 & 0.661 & 0.776 & 0.762 & 0.704 \\
\hspace{1.5em} R@1000 & 0.885 & 0.930 & 0.961 & 0.934 & 0.906 & 0.898 & 0.935 & 0.920 & 0.924 \\
\midrule
\multicolumn{10}{@{}p{\dimexpr\linewidth-2\tabcolsep}}{\textbf{\paperretrieval}} \\
\hspace{1.5em} nDCG@10 & 0.332 & 0.287 & 0.324 & 0.312 & 0.171 & 0.150 & 0.339 & 0.299 & 0.260 \\
\hspace{1.5em} R@100 & 0.388 & 0.411 & 0.434 & 0.412 & 0.266 & 0.262 & 0.411 & 0.383 & 0.328 \\
\hspace{1.5em} R@1000 & 0.750 & 0.715 & 0.747 & 0.750 & 0.538 & 0.553 & 0.759 & 0.651 & 0.679 \\
\midrule
\multicolumn{10}{@{}p{\dimexpr\linewidth-2\tabcolsep}}{\textbf{SetOps}} \\
\hspace{1.5em} nDCG@10 & 0.149 & 0.138 & 0.163 & 0.166 & 0.117 & 0.118 & 0.172 & 0.151 & 0.134 \\
\hspace{1.5em} R@100 & 0.279 & 0.266 & 0.327 & 0.340 & 0.259 & 0.228 & 0.338 & 0.269 & 0.248 \\
\hspace{1.5em} R@1000 & 0.505 & 0.528 & 0.596 & 0.616 & 0.490 & 0.447 & 0.594 & 0.493 & 0.454 \\
\midrule
\multicolumn{10}{@{}p{\dimexpr\linewidth-2\tabcolsep}}{\textbf{Clinical Trial}} \\
\hspace{1.5em} nDCG@10 & 0.284 & 0.250 & 0.305 & 0.298 & 0.100 & 0.151 & 0.282 & 0.268 & 0.250 \\
\hspace{1.5em} R@100 & 0.189 & 0.155 & 0.201 & 0.205 & 0.031 & 0.073 & 0.183 & 0.140 & 0.131 \\
\hspace{1.5em} R@1000 & 0.459 & 0.365 & 0.478 & 0.502 & 0.105 & 0.185 & 0.458 & 0.324 & 0.314 \\
\midrule
\multicolumn{10}{@{}p{\dimexpr\linewidth-2\tabcolsep}}{\textbf{Legal QA}} \\
\hspace{1.5em} nDCG@10 & 0.093 & 0.323 & 0.297 & 0.387 & 0.167 & 0.158 & 0.380 & 0.278 & 0.158 \\
\hspace{1.5em} R@100 & 0.355 & 0.706 & 0.693 & 0.791 & 0.501 & 0.506 & 0.740 & 0.644 & 0.494 \\
\hspace{1.5em} R@1000 & 0.645 & 0.830 & 0.833 & 0.863 & 0.739 & 0.751 & 0.846 & 0.811 & 0.746 \\
\midrule
\multicolumn{10}{@{}p{\dimexpr\linewidth-2\tabcolsep}}{\textbf{Theorem Retrieval}} \\
\hspace{1.5em} nDCG@10 & 0.308 & 0.383 & 0.365 & 0.367 & 0.326 & 0.329 & 0.442 & 0.395 & 0.310 \\
\hspace{1.5em} R@100 & 0.587 & 0.703 & 0.744 & 0.788 & 0.638 & 0.611 & 0.742 & 0.667 & 0.590 \\
\hspace{1.5em} R@1000 & 0.841 & 0.903 & 0.929 & 0.915 & 0.848 & 0.851 & 0.935 & 0.855 & 0.864 \\
\midrule
\multicolumn{10}{@{}p{\dimexpr\linewidth-2\tabcolsep}}{\textbf{Code Retrieval}} \\
\hspace{1.5em} nDCG@10 & 0.082 & 0.087 & 0.226 & 0.289 & 0.053 & 0.127 & 0.358 & 0.118 & 0.160 \\
\hspace{1.5em} R@100 & 0.091 & 0.092 & 0.231 & 0.327 & 0.077 & 0.142 & 0.382 & 0.113 & 0.156 \\
\hspace{1.5em} R@1000 & 0.208 & 0.223 & 0.466 & 0.601 & 0.231 & 0.328 & 0.637 & 0.251 & 0.331 \\
\midrule
\multicolumn{10}{@{}p{\dimexpr\linewidth-2\tabcolsep}}{\textbf{Average}} \\
\hspace{1.5em} nDCG@10 & 0.195 & 0.237 & 0.266 & 0.280 & 0.152 & 0.174 & 0.301 & 0.242 & 0.207 \\
\hspace{1.5em} R@100 & 0.333 & 0.416 & 0.456 & 0.501 & 0.319 & 0.340 & 0.480 & 0.397 & 0.359 \\
\hspace{1.5em} R@1000 & 0.560 & 0.619 & 0.695 & 0.738 & 0.530 & 0.566 & 0.724 & 0.580 & 0.587 \\
\bottomrule
\end{tabularx}}
\end{table*}

\begin{table*}[]
\centering
\caption{Overview of model performance using the \textit{Query-as-Reasoning-Steps} query rewriting approach on the chunked-document version of \dataset. The \textit{Query-as-Reasoning-Steps} approach uses a reasoning trace as the query which includes what the original query is asking, what a relevant document should contain, and an example relevant document. We use three metrics: normalized discounted cumulative gain at 10 (nDCG@10), recall at 100 (R@100), and recall at 1000 (R@1000). The last row shows the average (mean) value of each metric across all the tasks.}
\label{tab:auto_query_as_reasoning_steps}
\resizebox{0.88\linewidth}{!}{
\begin{tabularx}{0.98\linewidth}{l!{\color{lightgray}\vrule}C !{\color{lightgray}\vrule} C !{\color{lightgray}\vrule} C !{\color{lightgray}\vrule} C !{\color{lightgray}\vrule} C !{\color{lightgray}\vrule} C !{\color{lightgray}\vrule} C !{\color{lightgray}\vrule} C !{\color{lightgray}\vrule} C}
\toprule
Metric / Dataset & \rotcol{BM25} & \rotcol{Snowflake} & \rotcol{GTE Qwen 1.5B} & \rotcol{GTE Qwen 7B} & \rotcol{Lion SB 1B} & \rotcol{Lion SB 8B} & \rotcol{Promptriever} & \rotcol{Lion DS 1B} & \rotcol{Lion DS 8B} \\
\midrule
\multicolumn{10}{@{}p{\dimexpr\linewidth-2\tabcolsep}}{\textbf{Tip-of-the-tongue}} \\
\hspace{1.5em} nDCG@10 & 0.023 & 0.050 & 0.044 & 0.160 & 0.062 & 0.119 & 0.131 & 0.063 & 0.057 \\
\hspace{1.5em} R@100 & 0.081 & 0.148 & 0.148 & 0.519 & 0.170 & 0.333 & 0.400 & 0.148 & 0.126 \\
\hspace{1.5em} R@1000 & 0.170 & 0.296 & 0.333 & 0.793 & 0.467 & 0.578 & 0.644 & 0.304 & 0.252 \\
\midrule
\multicolumn{10}{@{}p{\dimexpr\linewidth-2\tabcolsep}}{\textbf{\stackexchange}} \\
\hspace{1.5em} nDCG@10 & 0.241 & 0.363 & 0.345 & 0.336 & 0.208 & 0.288 & 0.354 & 0.357 & 0.299 \\
\hspace{1.5em} R@100 & 0.624 & 0.801 & 0.816 & 0.768 & 0.651 & 0.687 & 0.799 & 0.779 & 0.710 \\
\hspace{1.5em} R@1000 & 0.901 & 0.942 & 0.972 & 0.944 & 0.910 & 0.903 & 0.940 & 0.937 & 0.914 \\
\midrule
\multicolumn{10}{@{}p{\dimexpr\linewidth-2\tabcolsep}}{\textbf{\paperretrieval}} \\
\hspace{1.5em} nDCG@10 & 0.330 & 0.285 & 0.296 & 0.304 & 0.168 & 0.161 & 0.368 & 0.295 & 0.241 \\
\hspace{1.5em} R@100 & 0.424 & 0.440 & 0.433 & 0.399 & 0.298 & 0.257 & 0.486 & 0.376 & 0.362 \\
\hspace{1.5em} R@1000 & 0.772 & 0.724 & 0.732 & 0.730 & 0.558 & 0.535 & 0.800 & 0.665 & 0.651 \\
\midrule
\multicolumn{10}{@{}p{\dimexpr\linewidth-2\tabcolsep}}{\textbf{SetOps}} \\
\hspace{1.5em} nDCG@10 & 0.120 & 0.045 & 0.059 & 0.098 & 0.123 & 0.131 & 0.148 & 0.109 & 0.083 \\
\hspace{1.5em} R@100 & 0.245 & 0.136 & 0.183 & 0.246 & 0.280 & 0.265 & 0.348 & 0.209 & 0.168 \\
\hspace{1.5em} R@1000 & 0.464 & 0.352 & 0.420 & 0.526 & 0.528 & 0.496 & 0.620 & 0.393 & 0.343 \\
\midrule
\multicolumn{10}{@{}p{\dimexpr\linewidth-2\tabcolsep}}{\textbf{Clinical Trial}} \\
\hspace{1.5em} nDCG@10 & 0.289 & 0.345 & 0.409 & 0.382 & 0.077 & 0.133 & 0.406 & 0.289 & 0.127 \\
\hspace{1.5em} R@100 & 0.161 & 0.175 & 0.250 & 0.252 & 0.022 & 0.055 & 0.264 & 0.119 & 0.041 \\
\hspace{1.5em} R@1000 & 0.377 & 0.378 & 0.560 & 0.558 & 0.076 & 0.147 & 0.559 & 0.294 & 0.130 \\
\midrule
\multicolumn{10}{@{}p{\dimexpr\linewidth-2\tabcolsep}}{\textbf{Legal QA}} \\
\hspace{1.5em} nDCG@10 & 0.089 & 0.338 & 0.286 & 0.408 & 0.181 & 0.187 & 0.389 & 0.333 & 0.186 \\
\hspace{1.5em} R@100 & 0.339 & 0.728 & 0.708 & 0.806 & 0.543 & 0.550 & 0.768 & 0.716 & 0.536 \\
\hspace{1.5em} R@1000 & 0.635 & 0.843 & 0.843 & 0.867 & 0.745 & 0.768 & 0.856 & 0.836 & 0.749 \\
\midrule
\multicolumn{10}{@{}p{\dimexpr\linewidth-2\tabcolsep}}{\textbf{Theorem Retrieval}} \\
\hspace{1.5em} nDCG@10 & 0.291 & 0.342 & 0.333 & 0.404 & 0.322 & 0.352 & 0.417 & 0.341 & 0.261 \\
\hspace{1.5em} R@100 & 0.670 & 0.724 & 0.786 & 0.777 & 0.659 & 0.680 & 0.725 & 0.645 & 0.599 \\
\hspace{1.5em} R@1000 & 0.854 & 0.937 & 0.986 & 0.930 & 0.880 & 0.924 & 0.931 & 0.860 & 0.825 \\
\midrule
\multicolumn{10}{@{}p{\dimexpr\linewidth-2\tabcolsep}}{\textbf{Code Retrieval}} \\
\hspace{1.5em} nDCG@10 & 0.054 & 0.113 & 0.184 & 0.216 & 0.008 & 0.056 & 0.354 & 0.059 & 0.089 \\
\hspace{1.5em} R@100 & 0.062 & 0.103 & 0.180 & 0.236 & 0.032 & 0.085 & 0.375 & 0.072 & 0.102 \\
\hspace{1.5em} R@1000 & 0.156 & 0.243 & 0.381 & 0.479 & 0.174 & 0.242 & 0.637 & 0.189 & 0.250 \\
\midrule
\multicolumn{10}{@{}p{\dimexpr\linewidth-2\tabcolsep}}{\textbf{Average}} \\
\hspace{1.5em} nDCG@10 & 0.180 & 0.235 & 0.244 & 0.289 & 0.144 & 0.178 & 0.321 & 0.231 & 0.168 \\
\hspace{1.5em} R@100 & 0.326 & 0.407 & 0.438 & 0.500 & 0.332 & 0.364 & 0.521 & 0.383 & 0.330 \\
\hspace{1.5em} R@1000 & 0.541 & 0.589 & 0.653 & 0.728 & 0.542 & 0.574 & 0.748 & 0.560 & 0.514 \\
\bottomrule
\end{tabularx}}
\end{table*}

\subsection{Impact of Instructions on Performance} \label{sec:impact_of_instructions}
Both of the top performing models, GTE Qwen 7B and Promptriever, are trained to follow instructions. As each task has instructions outlining the objective, it is likely that these models modify their query representation to better complete the task. To quantify the impact of instruction following on retrieval performance, we perform an experiment in which we replace the task-specific instructions with the generic instruction \textit{Given a query, find relevant documents}. This generic instruction provides no insight about the task and thus can help to gauge the importance of task-specific instructions in the performance of the best models.

\begin{figure}[h!]
\centering
\begin{tikzpicture}
\begin{axis}[
    height=0.5\textwidth,
    width=0.9\textwidth,
    ybar=2pt,
    ymajorgrids=true,
    grid style={dashed, gray!50},
    bar width=7pt,
    yminorgrids=true,
    minor y tick num=1,
    xtick pos=left,
    ytick pos=left,
    ylabel={$\Delta$ nDCG@10},
    extra y ticks={0},
    extra y tick labels={},
    extra y tick style={grid=major, grid style={solid, black}},
    yticklabel style={
        /pgf/number format/fixed,
        /pgf/number format/precision=2,
        /pgf/number format/fixed zerofill,
    },
    symbolic x coords={
        Tip-of-the-tongue,
        \stackexchange,
        \paperretrieval,
        SetOps,
        Clinical Trial,
        Legal QA,
        Theorem Retrieval,
        Code Retrieval,
        Average,
    },
    xtick=data,
    xticklabel style={
        rotate=45, 
        anchor=east,
        font=\footnotesize,
    },
    legend style={
        at={(0.5, -0.4)},
        anchor=north,
        legend columns=-1,
        font=\footnotesize,
        fill=none,
        /tikz/every even column/.append style={column sep=0.24cm},
        /tikz/every odd column/.append style={column sep=0.05cm},
        draw=white,
    },
    legend image code/.code={
        \fill[#1] (0cm,-0.1cm) rectangle (0.25cm,0.2cm);
    },
]

\addplot[mybarstyle={colorMain}{}] coordinates { ({Tip-of-the-tongue}, -0.0400) ({\stackexchange}, -0.0340) ({\paperretrieval}, 0.0150) ({SetOps}, 0.0130) ({Clinical Trial}, 0.0050) ({Legal QA}, -0.0230) ({Theorem Retrieval}, -0.0580) ({Code Retrieval}, 0.0010) ({Average}, -0.0151) };
\addplot[mybarstyle={colorQTA}{north west lines}] coordinates { ({Tip-of-the-tongue}, 0.0140) ({\stackexchange}, -0.0350) ({\paperretrieval}, -0.0090) ({SetOps}, -0.0170) ({Clinical Trial}, 0.0530) ({Legal QA}, -0.0300) ({Theorem Retrieval}, -0.0600) ({Code Retrieval}, -0.0100) ({Average}, -0.0117) };
\addplot[mybarstyle={colorQTD}{crosshatch dots}] coordinates { ({Tip-of-the-tongue}, -0.0180) ({\stackexchange}, -0.0650) ({\paperretrieval}, -0.0070) ({SetOps}, -0.0700) ({Clinical Trial}, 0.0520) ({Legal QA}, -0.0640) ({Theorem Retrieval}, -0.1410) ({Code Retrieval}, -0.0690) ({Average}, -0.0478) };

\legend{
    Promptriever,
    GTE Qwen 1.5B,
    GTE Qwen 7B,
}

\end{axis}
\end{tikzpicture}
\vspace{-25pt}
\caption{Comparison of change in nDCG@10 when a generic instruction is used instead of a task-specific instruction. The result is found by taking $(nDCG_{generic} - nDCG_{task-specific})$, where $nDCG_{generic}$ is the nDCG@10 with the generic instruction and $nDCG_{task-specific}$ is the nDCG@10 with the task-specific instruction.}
\label{fig:generic_instructions_ndcg}
\end{figure}
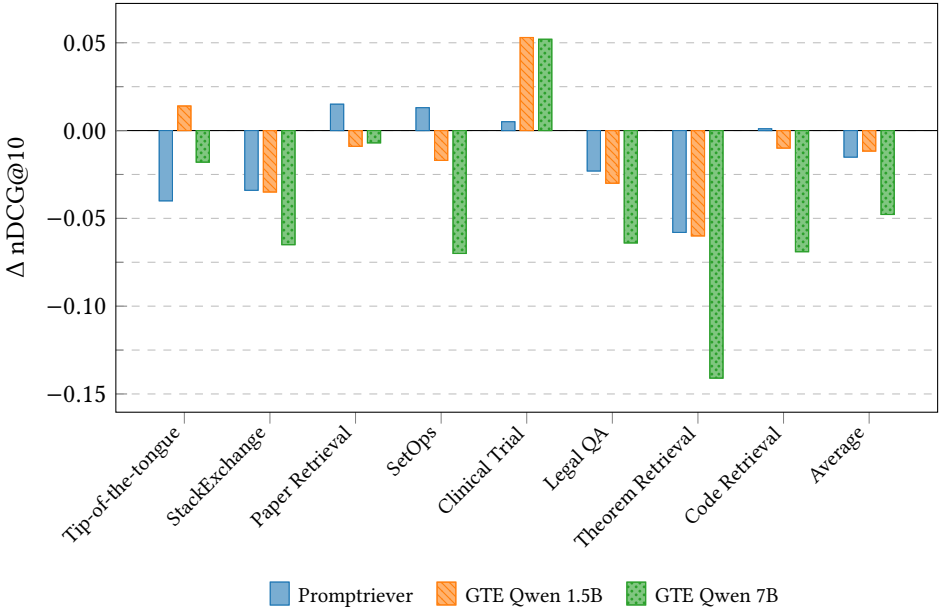

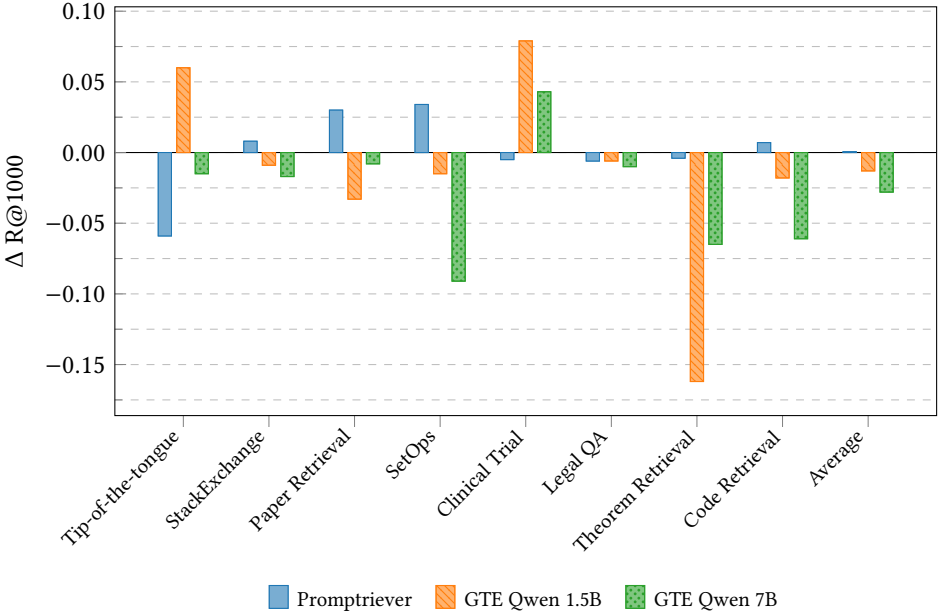
\begin{figure}[h!]
\centering
\begin{tikzpicture}
\begin{axis}[
    height=0.5\textwidth,
    width=0.9\textwidth,
    ybar=2pt,
    ymajorgrids=true,
    yminorgrids=true,
    minor y tick num=1,
    grid style={dashed, gray!50},
    bar width=7pt,
    xtick pos=left,
    ytick pos=left,
    ylabel={$\Delta$ R@1000},
    extra y ticks={0},
    extra y tick labels={},
    extra y tick style={grid=major, grid style={solid, black}},
    yticklabel style={
        /pgf/number format/fixed,
        /pgf/number format/precision=2,
        /pgf/number format/fixed zerofill,
    },
    symbolic x coords={
        Tip-of-the-tongue,
        \stackexchange,
        \paperretrieval,
        SetOps,
        Clinical Trial,
        Legal QA,
        Theorem Retrieval,
        Code Retrieval,
        Average,
    },
    xtick=data,
    xticklabel style={
        rotate=45, 
        anchor=east,
        font=\footnotesize,
    },
    legend style={
        at={(0.5, -0.4)},
        anchor=north,
        legend columns=-1,
        font=\footnotesize,
        fill=none,
        /tikz/every even column/.append style={column sep=0.24cm},
        /tikz/every odd column/.append style={column sep=0.05cm},
        draw=white,
    },
    legend image code/.code={
        \fill[#1] (0cm,-0.1cm) rectangle (0.25cm,0.2cm);
    },
]

\addplot[mybarstyle={colorMain}{}] coordinates { ({Tip-of-the-tongue}, -0.0590) ({\stackexchange}, 0.0080) ({\paperretrieval}, 0.0300) ({SetOps}, 0.0340) ({Clinical Trial}, -0.0050) ({Legal QA}, -0.0060) ({Theorem Retrieval}, -0.0040) ({Code Retrieval}, 0.0070) ({Average}, 0.0006) };
\addplot[mybarstyle={colorQTA}{north west lines}] coordinates { ({Tip-of-the-tongue}, 0.0600) ({\stackexchange}, -0.0090) ({\paperretrieval}, -0.0330) ({SetOps}, -0.0150) ({Clinical Trial}, 0.0790) ({Legal QA}, -0.0060) ({Theorem Retrieval}, -0.1620) ({Code Retrieval}, -0.0180) ({Average}, -0.0130) };
\addplot[mybarstyle={colorQTD}{crosshatch dots}] coordinates { ({Tip-of-the-tongue}, -0.0150) ({\stackexchange}, -0.0170) ({\paperretrieval}, -0.0080) ({SetOps}, -0.0910) ({Clinical Trial}, 0.0430) ({Legal QA}, -0.0100) ({Theorem Retrieval}, -0.0650) ({Code Retrieval}, -0.0610) ({Average}, -0.0280) };

\legend{
    Promptriever,
    GTE Qwen 1.5B,
    GTE Qwen 7B,
}

\end{axis}
\end{tikzpicture}
\vspace{-25pt}
\caption{Comparison of change in R@1000 when a generic instruction is used instead of a task-specific instruction. The result is found by taking $(R_{generic} - R_{task-specific})$, where $R_{generic}$ is the R@1000 with the generic instruction and $R_{task-specific}$ is the R@1000 with the task-specific instruction.}
\label{fig:generic_instructions_recall}
\end{figure}

The results can be seen in Figure \ref{fig:generic_instructions_ndcg} and Figure \ref{fig:generic_instructions_recall}, for nDCG@10 and R@1000 respectively. Both figures show how performance changes when task-specific instructions are swapped with generic instructions. Starting with Figure \ref{fig:generic_instructions_ndcg} we see that on average removing task-specific hurts nDCG@10 for both GTE Models and Promptriever. The amount varies with GTE Qwen 7B having the most profound decrease, followed by Promptriver, and GTE Qwen 1.5B. Interestingly, the difference in performance is highly dependent on the dataset, with Theorem Retrieval having the most overall performance decrease with generic instructions and Clinical Trial seeing gains with generic instructions for all models. Clinical Trial is the only task which sees all models improve with generic instructions, while Legal QA and \stackexchange join Theorem Retrieval in worse performance across the board with generic instructions. 

Promptriever is the model that improves on the largest number of tasks when generic instructions are used which might suggest that its ability to follow instructions to changes its query-time behavior may not be the main factor in its improvement over the similar Lion models. Looking at the average performance with generic instructions, Promptriever only loses around 0.015 nDCG which still puts it well ahead of the Lion models.

Moving to the recall changes in Figure \ref{fig:generic_instructions_recall}, we find that on average both Qwen models are hurt when using generic instructions, while Promptriever is slightly better with generic instructions. We see that Legal QA and Theorem Retrieval both still have worse performance across all models with generic instructions, suggesting that these tasks see benefits with instruction regardless of the metric used. In general, the task-level trends are similar to those found with nDCG. In fact, for both Qwen models, when generic instructions hurt or help is exactly the same with R@1000 as it is for nDCG@10. Like the Qwen models, the recall changes for Promptriever largely follow the same trend as nDCG, but with flips on the \stackexchange, Clinical Trial, and Code Retrieval tasks. Although for both Clinical Trial and Code Retrieval the change is pretty minimal and could be noise. The average slight gain with instructions does suggest that for Promptriever instructions are less helpful for improving recall. This makes intuitive sense, as trying to put a relevant document in the top 10 positions has far less room for error than in the top 1000. Instructions likely reduce this error, but might not be fully needed for doing well on recall.

In general, the impact of instructions varies by model and task. On average, the GTE Qwen models, especially the 7B variant, are most affected by a change to generic instructions. Promptriever is surprisingly minimally affected by the switch to generic instructions, especially for recall, where there is actually a slight gain with generic instructions. This may be in part because the instructions used for training Promptriever are longer and more specific than the ones we provided. Perhaps our instructions are more inline with the instructions used by GTE Qwen which could explain the differences between Promptriever and the Qwen models. Further, the GTE Qwen models were trained on far more varied data and thus are likely better able to adapt retrieval strategies for different domains based on the instructions.

\section{Discussion and Analysis}
\label{sec:discussion}
From the results it is clear that retrieval models tend to struggle with complex tasks and there are some interesting findings on what might contribute to model success. In this section, we do further analysis of which groups of \dataset tasks do models struggle with the most and what this says about the limitations of these models and what areas might need to be further improved. We also discuss the importance and take away from the query augmentation experiments as well as discuss the limitations of our data.

\subsection{Where do Models Struggle?}
Our results in Table \ref{tab:main_results_table} reveal several patterns about how retrieval models handle complex tasks. The consistently lower performance compared to standard benchmarks like BEIR and TREC DL Tracks demonstrates that current retrieval models struggle when faced with queries that have complex information needs. This fact is highlighted by the substantially lower average nDCG@10 of 0.346 for the best model (GTE Qwen 7B) than what these same models achieve on traditional benchmarks.

It is clear that models on average perform poorly, but what are the underlying causes that lead to these bad results? One major takeaway is that where and how models struggle is not uniform across tasks and models. For instance, on the Code Retrieval task, the top three models have better nDCG than their average nDCG while the bottom models are far worse than their average. These groups of models tend to correlate through the tasks and metrics so we will do an analysis of each group of models separately to find what trends they each follow and what they struggle with.

\subsubsection{Where do the Stronger Models Struggle?}
Starting with the set of the best models, GTE Qwen 7B, Promptriever, and GTE Qwen 1.5B we find that all of their nDCG scores are lower than their average on Tip-of-the-tongue and SetOps. Promptriever also sees lower than average performance for Theorem Retrieval, Legal QA, \stackexchange; GTE Qwen 1.5B also sees lower performance on Theorem Retrieval and Legal QA; while GTE Qwen 7B also has lower performance on \stackexchange. 

Looking at the tasks where all models performed worse than average on nDCG, SetOps and Tip-of-the-tongue, we find two very different datasets. Tip-of-the-tongue has long multi-part queries with minimal keyword overlap with relevant documents, while SetOps has the shortest queries in the collection which often contain exact or near exact terms from the documents. One thing that might explain the poor nDCG performance for both is that keyword matching is likely not enough to do well on both of them, but over reliance on keywords can lead to poor outcomes for both. In Tip-of-the-tongue, there may not always be a lot of term overlap between the query and relevant document, but there are a lot of terms which might show up in movie descriptions. Things like genres, time periods, etc. If a model focuses too much on these superficial matches without understanding the larger picture the query is painting, it would often be led astray. SetOps seems to not fit in the same category as many of the terms, such as entities or time periods, are very similar between the query and documents, but as the queries have set operations often matching just these aspects would lead to incorrect documents. For example, if the query is \textit{movies from the 90s that aren't action movies} it is likely that many retrieval models would find action movies from the 90s if they are not able to understand or represent the negation. The property of often having misleading lexical or semantic overlap could account for the lower scores on these datasets.

Another similarity is that both tasks use Wikipedia as a corpus. Though this is likely not a direct cause, it does mean that the documents are long articles that are split into multiple chunks. Another possible explanation for the models poor performance could be that both tasks have queries that refer to several aspects of the document and that chunking could lead to relevant aspects being split into separate chunks. This might result in the model finding many chunks for a relevant document, but due to the MaxP operation used only the maximum scoring chunk is retained, which might mean documents with only a partial match coming from a single chunk end up with similar scores to documents with multiple chunk matches. If this is the case, it may indicate that the retrieval models are more capable than they appear, but it still leaves a practical problem of how to chunk documents and combine the scores of chunks in a robust way.
The other datasets in which the models perform poorly do not seem to have a clear pattern, as each model seems to struggle in different ways. The exception is Promptriever that struggles on all of the datasets which one of the two Qwen models also struggles with. This might be caused by the lack of diversity in Promptriever's training data, as all three datasets require specific domain knowledge. 

Code Retrieval, \paperretrieval, and Clinical Trial have higher than average nDCG across all the best models. These datasets share several characteristics with some of the datasets where models perform badly, such as being in specialized domains and requiring some reasoning to understand. One other observation is that these datasets make up three of the four datasets with the highest number of relevant documents per query. This might mean that the increase in nDCG is not necessarily because the models are stronger in these domains, but that the chance of retrieving something relevant is inherently higher. 

Moving to R@1000, we find that all three models underperform in Code Retrieval, Clinical Trial, and SetOps. Interestingly, these also make up three of the four datasets with the most relevant documents. A potential cause which we mentioned in the per-task results might be that a large number of relevant documents likely correlates with a more diverse set of relevant documents which causes models to fail as they are only able to target a subset of the relevant document space. Otherwise, the collections have minimal in common with each other. It is interesting that the only dataset which all models struggle with for both R@1000 and nDCG is SetOps. Perhaps the large number of possible "distractor" documents, which have certain attributes but not others, is to blame, as well as the inability of the models to understand the set-based logic.

Each of the best models also struggled with one of these tasks \paperretrieval, Tip-of-the-tongue, and Theorem Retrieval tasks. As each model had a unique task that it struggled with, it is hard to draw many general conclusions. For reasons why each model may have struggled on these particular tasks you can refer to the task-result sections in Section \ref{sec:perfromance_per_task}.

The datasets where all of the best models perform well are \stackexchange and Legal QA. Both of these have a small number of relevant documents which might contribute to the good performance. \stackexchange also has a fairly small document collection, which may also be a contributing factor. 

\subsubsection{Where do the Weaker Models Struggle?}

Looking at nDCG for the worse performing models, we find that three tasks have below average performance across the board: Code Retrieval, Theorem Retrieval, and Tip-of-the-tongue. All three of these datasets have low term overlap and also have minimal semantic overlap. This suggests that the models may be overly dependent on semantic or lexical matching and are not able to perform higher-level reasoning-based or abstractive retrieval. All but Lion SB 1B also performed below average on Legal QA which also has minimal semantic and lexical overlap. Additionally, all the non-sparse models also perform below average on Set Ops suggesting that perhaps they struggle to properly represent the tail entities and aspects that appear in many of the queries.

Looking at R@1000 all models do worse than average on Code Retrieval and Clinical Trial with almost all doing worse on Theorem Retrieval with the exception of Snowflake. The worst datasets are shared between the best and worst performing models suggesting that these datasets might have some universally difficult features. Though the better performing models also all underperformed on Set Ops while only Snowflake performed below average for the second-tier models. The below average performance on Theorem Retrieval for most of the models is interesting given that the other datasets where the models performed worse have larger number of relevant documents while Theorem Retrieval has one of the lowest. With that said, Theorem Retrieval is one of the farthest out of domain for models trained only on MSMARCO and likely features minimal term overlap which is likely a major cause for the poor performance. Interestingly, all models but the learned sparse models also perform worse than average on Tip-of-the-tongue.

\subsection{Effectiveness of Query Augmentation Strategies}
The counterintuitive finding that LLM-based query rewriting often hurts performance, especially for stronger models, challenges current assumptions about the benefits of query augmentation. For the best-performing model, query rewriting consistently degraded performance across all techniques and metrics. This suggests that strong retrieval models may already effectively capture the semantic intent of complex queries, and additional augmentation introduces noise rather than useful information.
The differential impact on weaker versus stronger models reveals an important transition point where query augmentation helps. Weaker models benefit from query rewriting because the augmented queries provide additional semantic context that their limited training likely did not include. However, stronger models appear to suffer from information overload or distribution shift when queries are heavily modified from their training distribution.
The specific patterns of improvement - Query-to-Answer helping with precision (nDCG@10) while Query-to-Doc improving recall (R@1000) - suggest that different augmentation strategies serve different purposes. Answer-focused augmentation creates targeted semantic signals that help with ranking, while document-focused augmentation adds broader context that improves candidate retrieval. Understanding these trade-offs could inform more sophisticated adaptive augmentation strategies.
It is also worth noting that the rewriting model used almost certainly impacts the exact performance. As larger models store more information, they would likely be capable of providing the correct answer for prompts like Query-to-Answer more than smaller LLMs such as the one we used. This would almost certainly increase performance for tasks with a clear answer such as Tip-of-the-tongue, but it is unclear the extent to which LLM augmentation can help for many of the other tasks. As an example, for SetOps the retrieval model must be capable of executing set-based logic. It seems nearly impossible for an LLM to rewrite a query to make this possible when the underlying retriever is incapable of executing the necessary logic. Similarly, in tasks with many possible relevant documents like Code Retrieval or Clinical Trial an LLM may be able to include a near perfect representation of the document but it may only match one of dozens of relevant documents. Perhaps explicit query decomposition may help, but this would likely require some additional logic from the LLM to weigh and combine the components which goes a bit beyond traditional query rewriting. We hope \dataset can help drive more research in this direction.

\subsection{Limitations of Dataset}
No dataset is fully representative, and very few are without potential shortcomings. To help promote future work and be fully transparent, we will go over some of the limitations of our dataset. When constructing our data we made choices to limit the scope to maintain feasibility and produce a focused dataset. These limitations include:
\begin{itemize}
    \item \textbf{English Only} - Our dataset focuses only on tasks in the English language. This was for a few reasons: (1) we (the authors) understand English so we can better evaluate the quality of both the datasets selected and the results of our experiments (2) some of the most commonly used evaluation sets are in English so there is historical precedent to use English (3) related to this point most people in the academic community know English as it is the main language used in publications, which means it is likely the most universal language which allows more people to do analysis of the data (4) likely due to some of the prior reasons there are more data resources in English which made assembling this dataset easier.
    \item \textbf{Uni-Modal} - Our dataset focuses on text-only retrieval tasks. We made this choice for a few reasons: (1) adding new modalities adds complexity to distributing data which might make it less accessible (2) although new approaches which support multi-modal inputs are becoming more common, the vast majority of retrieval systems are built for text only thus making a multi-modal dataset would exclude large numbers of existing and future retrieval systems.
    \item \textbf{Single-Hop} - Many of the existing complex retrieval tasks fall into the multi-hop category, though all of the data in our datasets is single-hop. The reason for this is multi-hop questions often require an iterative approach where the steps - rewrite the query, retrieve - are repeated several times until the final answer is obtained. This process is useful for many real world search tasks, but it requires more complexity and as mentioned in the Uni-Modal bullet it would limit the types of models that could be used for our dataset. For this reason, we focused on tasks that could be solved with a single retrieval. Note iterative systems can still be applied, they just are not required. 
    \item \textbf{Subset of Possible Complex Tasks} - Our dataset only contains eight complex tasks, though this only accounts for a small subset of all the potential complex retrieval tasks. This is because: (1) it is time consuming to standardize datasets especially given the strong emphasis we gave to produce high-quality document collections for each dataset (2) finding high quality datasets which are complex is hard and there is a limited supply; although more datasets are produced every day, evaluating, standardizing, and running benchmarks on them is resource and time intensive (3) there are diminishing returns, more tasks likely mean there are overlaps in task features and thus less useful information is gained with each new task (4) as we would like our benchmark to be used by others, reducing the size of the dataset makes it more likely to be adopted and enables those with minimal computation resources to use it.
\end{itemize}
These limitations are consequences of our decisions and desires for the dataset, there are also some inherent limitations that come from the datasets we choose to include and how we modified them for this benchmark. We will now elucidate any limitations we think those using our dataset should be aware of for each task:

\textbf{Tip-of-the-tongue} - This dataset has a minimal chance that a document marked as relevant is not relevant due to how the data was collected. With that said, there may be a chance that another document could plausibly also exactly match the description provided, though this is unlikely. This is because it is assumed only the actual document the user was looking for is correct even if others are equally plausible. Although the marked relative documents are likely correct, they may not always contain enough information to know they are relevant. Additionally, for the short document version the chunking may split relevant information across chunks which may impact retrieval. Also, the chunked version does not have meaningful per-chunk labels as we only have per-document labels, thus all chunks from a document are considered relevant. In our experiments we used MaxP to overcome this limitation. For more details on why and how to use MaxP see Section \ref{sec:retrieval_model_implementation}.

\textbf{\stackexchange} - This data was derived from BRIGHT \cite{bright} so it inherits the issues from BRIGHT though these are minimal. A major limitation is the small number of documents collected for each topic area, we partially remedy this by combining the document collections from all topics. However, this does increase the risk of an unlabeled positive, but we believe this risk is small, as each topic is pretty distinct so documents from one topic are unlikely to even be topically relevant to queries from another topic. Another issue is that we created our own document collection to improve quality and unify the data format, this required aligning our collection with the original collection. For the full documents this was straightforward, but for the chunked collection it was more difficult. We tried to ensure our chunk-level labels were still valid by ensuring an original BRIGHT-chunk was fully encapsulated by one of our chunks. We only kept queries if all relevant chunks could be fully encapsulated which we believe makes our data fully valid. For more details see the section introduce the \stackexchange task \ref{sec:overview_datasets}.

\textbf{\paperretrieval} - This dataset was derived from DORIS-MAE \cite{doris_mae} and most limitations come from the method used for labeling the original dataset. Documents were labeled using GPT 3.5 by checking how many topics and subtopics from the deconstructed query were satisfied. Due to cost, every document was not checked for every query only a limited candidate pool. The candidate pool used a pooling approach which combined lexical and semantic retrieval models and citation information. The limited scope of judgments could mean that relevant papers are unlabeled and the labeled papers may also have biases towards certain types of models used to create the original candidate pool. Additionally, the authors also exclude papers used as reference when writing the queries from the candidate pool. These documents are likely highly relevant and may not be labeled as such. It is unclear from the released data whether these reference documents were removed entirely from the corpus or whether they are included in the judged documents. The labels produced by GPT 3.5 may also be incorrect at times, though the labels produced were closer to a three-rater majority than an original set of individual human labels. This indicates the task may be quite difficult even for humans, but that GPT's ratings are likely reasonable. Beyond the limitation inherited from DORIS-MAE we also did additional filtering of queries and converted the topic and sub-topic judgments to query-level labels. The full procedure is described in Section \ref{sec:overview_datasets}. Though our approach is well motivated, it weighs all aspects equally in terms of importance and may be sensitive to slight errors in GPT's labels. Also, the choice to only mark documents as relevant in the binary setting when all aspects are met may be overly harsh and again may be sensitive to small errors in GPT's labeling. Though sensitive per-query, in aggregate these variances likely have minimal impact especially if GPT's mistakes are uniformly distributed. Despite these limitations, the various approaches to source documents for labeling likely minimizes bias and covers a large quantity of relevant documents and as mentioned previously, GPT's labels correlated well with human raters suggesting that although imperfect its labels are likely good enough to produce meaningful results.

\textbf{SetOps} - This dataset is based on QUEST \cite{quest} which uses Wikipedia categories to create "atomic" aspects which were combined to produce queries. Each category has a set of documents associated with it which was used to find the initial set of relevant documents. In the subset of queries we selected, the relevant documents were also judged by human raters as relevant. Thus, the documents that are marked as relevant have a high chance of being relevant and containing the necessary relevant information. Though there is still a risk that there are some documents which were not under the initial categories and were excluded for the candidate pool. We believe the risk of this is likely small, and would likely have minimal bias for or against certain types of pages. The use of templates to produce queries based on categories results in some unusual combinations of query aspects. To mitigate this effect we only selecting queries that human raters judged as having high naturalness and fluency. Despite this, some queries may have odd combinations of aspects, but (1) it is hard to know what a true distribution of users information needs would include so these queries may be plausible (2) even if they have information needs that are unrealistic they are still valid in testing systems on what is otherwise a realistic search task.

\textbf{Clinical Trial} - This dataset is based on the TREC Clinical Trial tracks and as such has fairly comprehensive labels. As TREC is experienced in producing high quality test collections the chance of incorrect labels is low though there may still be some unlabeled relevant documents, these likely would not change the relative rankings of search systems. Our conversion of documents from the original XML to markdown may have resulted in some irregular formatting for edge-cases. The original documents also had some irregularities due to the formatting of the original Clinical Trial text, though this is likely a realistic evaluation of real world text.

\textbf{Legal QA} - The quality of query-document labels for this dataset is likely high as it is unlikely there are many sections in a state's statutes that are relevant in answering the same question as this kind of redundancy would make understanding the laws difficult. Still, there may be some sections that are relevant and unlabeled. The labeling was done by legal professionals so their judgments should be sound. To specify the states for each query we added natural language text at the end of each question which is the same for each query (with the exception of the state name changing). This added context may lack naturalness, but is similar to how additional requirements are often added to keyword queries so we believe it is a reasonable inclusion.

\textbf{Theorem Retrieval} - This dataset is derived from BRIGHT's Theorem retrieval \cite{bright} task. The queries are originally from the TheoremQA dataset \cite{theorem_qa_theorem_question_answering_dataset} which the BRIGHT authors rewrite with GPT-4 and manually check to remove superficial term overlap. These queries are then paired with documents from ProofWiki using the name of the theorem which is provided by the original TheoremQA dataset. The authors look for theorems that contain the exact substring and that are in the top 10 results for BM25. The candidate documents are then scored using GPT-4. The authors show that GPT-4 has a high correlation with human raters. The methods used by the BRIGHT authors may result in a few potential problems, first the GPT rewritten queries may have issues though each query is manually checked which reduces this risk. Second, the alignment and labeling has room to produce some problems though the use of the proof name likely guarantees the relevant theorem is found in most cases. The authors also discard queries which have no relevant labels which would exclude queries where the approaches to find theorems failed. Overall, the approach has enough human checks to produce confident in the results.

\textbf{Code Retrieval} - Based on the dataset APPs \cite{apps_measuring_coding_challenge_competence} this is the only task that does not come from an existing retrieval dataset. The original dataset was meant to test code generation but includes a number of possible solutions along with code problems. To produce a retrieval dataset we treat the code problems as queries and code solutions as relevant documents. As the solutions have all been verified to pass problem-specific tests they are known to be correct, with that said there may be a chance one problem's code solves one of the other  problems. Although many problems use the same techniques, a major focus with these types of coding problems is to obscure the exact technique that is necessary, thus the chance the inputs and outputs for two problems is identical and the underlying algorithm are identical is low. Another potential issue is that the code solutions may include problem specific terms that makes the task unrealistic. Our qualitative investigation found that most of the code solutions had minimal term overlap with the questions and usually used short uninformative variable names and did not include documentation. This is confirmed by the low nDCG@10 of BM25. We found one instance of a document including the name of the problem, but only saw that one instance. Overall, we believe the nature of the original dataset makes the conversion to a retrieval task valid due to the low likelihood of unlabeled positives and minimal lexical similarity between queries and documents.

\subsection{Citing Direction}
As our benchmark consists of many other datasets that made our work possible, we ask that when using our benchmark, in addition to citing this paper, you also cite the original datasets. For convenience, all the citations are cited here \cite{bright, apps_measuring_coding_challenge_competence, zheng2025reasoningfocused, doris_mae, quest, trec_tot_2023, roberts2021overviewct, roberts2022overviewct}.

\section{Conclusion}
\label{sec:conclusion}
In this work, we compile a diverse and complex set of retrieval tasks to benchmark how a wide selection of powerful neural retrieval models perform on complex tasks. We find that across the board even the best models struggle to produce satisfactory nDCG@10 values suggesting that additional work is still needed to have general-purpose retrieval models for complex retrieval tasks. Recall is also often low, especially for tasks with several relevant documents. This finding implies that even using powerful reranking models overall ranking performance will still be poor in many cases due to weak first-stage candidates.
Our experiments give some direction for future improvements and find some key attributes of the most successful models: (1) the best models use large backbone models that have been trained with vast quantities of data (2) they also use large amounts of varied retrieval specific data to pretrain the retrieval models though the success of Promptriever shows that this is not always necessary for a high quality model (3) there is reason to believe that using an instruction tuned backbone LLM results in better performance than using the base LLM, though more results are still needed to exactly quantify this effect (4) they are capable of modifying their search behavior based on natural language instructions. We hope that by releasing \dataset, a high-quality diverse and complex retrieval benchmark, we can provide a meaningful measure of retrieval progress and spur new innovations.

\begin{acks}
The authors thank Sriharsha Hatwar for his substantial help in producing the data for the \stackexchange task. He helped to download and process the web pages that were used for the document collection, as well as assisting with aligning the document chunks and manually verifying the alignment quality.
We thank Varun Pininty and Madhav Jhawar for their help adding and checking citations, and Harrie Oosterhuis for his early feedback. 

Furthermore, we acknowledge the use of the large language model Gemini 2.5 Pro for various purposes throughout the development and drafting of this work, including assisting with figure creation, table creation and design, code generation, and paper feedback. This work was supported in part by the Center for Intelligent Information Retrieval, in part by NSF grant number 2402873, in part by the Office of Naval Research contract number N000142412612, and in part by the NSF Graduate Research Fellowships Program (GRFP) Award number 1938059. Any opinions, findings, and conclusions or recommendations expressed in this material are those of the authors and do not necessarily reflect those of the sponsor.
\end{acks}

\appendix

\section{Prompts for LLM Query Rewriting} \label{sec:appendix_prompts}
This section provides the detailed prompts used for the LLM query rewriting techniques discussed in Section \ref{sec:QueryRewriting}.

\begin{figure*}[h!]
\begin{prompt}{Query-as-Doc Prompt}
Follow the below steps to produce a document that is relevant to the main information request based on what document type is requested -- Make sure to do each step:

1. Identify the type of document the user is trying to retrieve (and thus what should be output).

2. Analyze the request, what kinds of information is the user looking for? What kinds of documents would exist that would satisfy the request?

3. Do additional brainstorming about what a relevant document would be.

4. You must do the brainstorming and analysis steps! Once the full analysis and brainstorming steps have been completed output \#\#final output followed by the final document (and no additional comments, information, or anything else) on a new line which would be relevant based on the information request. The document should be in the style and format mentioned in the first line of the information request. For example, if the information request is looking for relevant Wikipedia pages the document should resemble the style and formatting (in markdown) of a Wikipedia page. Include the same amount of text as a real document would contain, do not include placeholders. Only output a single document even if there are multiple that might be relevant.

\vspace{10pt}
Information Request: \texttt{<instruction>}

\texttt{<query>}
\end{prompt}
\caption{Prompt used for Query-as-Doc query rewriting.}
\label{fig:prompt-query-as-answer}
\end{figure*}

\begin{figure*}[h!]
\begin{prompt}{Query-as-Answer Prompt}
Follow the below steps to produce an answer to an information request -- make sure to do each step:

1. Analyze the request, what kinds of information is the user looking for? What kinds of information would a complete answer have?

2. Do additional brainstorming about what might be relevant and what might not be.

3. You must do the brainstorming and analysis steps! Once the full analysis and brainstorming steps have been completed output \#\#final output followed by the final answer (and no additional comments, information, or anything else) on a new line which should answer the information request.

\vspace{10pt}
Information Request:  \texttt{<instruction>}

\texttt{<query>}
\end{prompt}
\caption{Prompt used for Query-as-Answer query rewriting.}
\label{fig:prompt-query-as-answer}
\end{figure*}

\begin{figure*}[h!]
\begin{prompt}{Query-as-Reasoning-Step Prompt}
Follow the below steps to produce an answer to an information request -- make sure to do each step:

1. Identify the essential problem in the information request.

2. Think step by step to reason about what should be included in relevant documents.

3. Draft an answer.

\vspace{10pt}
Information Request: \texttt{<instruction>}

\texttt{<query>}
\end{prompt}
\caption{Prompt used for Query-as-Reasoning-Step query rewriting.}
\label{fig:prompt-query-as-reasoning-step}
\end{figure*}

\section{Dataset Examples}
\label{sec:appendix_dataset_examples}
This section provides illustrative examples (query and a relevant document chunk) for each dataset in our benchmark.

\begin{table*}[htbp]
\centering
\caption{Example query and relevant passage for the Clinical Trial task.}
\label{tab:example_clinical_trial}
\begin{tabular}{L{0.9\linewidth}}
\toprule
\textbf{Query} \\
\midrule
\vspace{-8pt}
\begin{lstlisting}
Patient is a 45-year-old man with a history of anaplastic astrocytoma of the spine complicated by severe lower extremity weakness and urinary retention s/p Foley catheter, high-dose steroids, hypertension, and chronic pain. The tumor is located in the T-L spine, unresectable anaplastic astrocytoma s/p radiation. Complicated by progressive lower extremity weakness and urinary retention. Patient initially presented with RLE weakness where his right knee gave out with difficulty walking and right anterior thigh numbness. MRI showed a spinal cord conus mass which was biopsied and found to be anaplastic astrocytoma. Therapy included field radiation t10-l1 followed by 11 cycles of temozolomide 7 days on and 7 days off. This was followed by CPT-11 Weekly x4 with Avastin Q2 weeks/ 2 weeks rest and repeat cycle.
\end{lstlisting}
\\ \toprule
\textbf{Document} \\
\midrule
\vspace{-8pt}
\begin{lstlisting}
# A Phase II Trial of Paclitaxel and Topotecan With Filgrastim in Patients With Recurrent or Refractory Glioblastoma Multiforme or Anaplastic Astrocytoma
## Eligibility
gender: All
minimum age: 18 Years
maximum age: N/A
healthy volunteers: No

## Conditions
- Brain and Central Nervous System Tumors

## Interventions
Type: Biological, Name: filgrastim
Type: Drug, Name: paclitaxel
Type: Drug, Name: topotecan hydrochloride

## Criteria
DISEASE CHARACTERISTICS: Biopsy proven glioblastoma multiforme or anaplastic astrocytoma Central pathologic review at Dartmouth-Hitchcock Medical Center, including assay for tumor p53 expression No anaplastic oligodendroglioma No mixed oligodendroastrocytoma Recurrent or progressive disease following radiotherapy documented by CT or MRI within 2 weeks of entryPATIENT CHARACTERISTICS: Age: 18 and over Performance status: Karnofsky 60%
\end{lstlisting}
\\ \bottomrule
\end{tabular}
\end{table*}

\begin{table*}[htbp]
\centering
\caption{Example query and relevant passage for the Code Retrieval task.}
\label{tab:example_code_retrieval}
\begin{tabular}{L{0.9\linewidth}}
\toprule
\textbf{Query} \\
\midrule
\vspace{-8pt}
\begin{lstlisting}
On the way to school, Karen became fixated on the puzzle game on her phone! [Image] 

The game is played as follows. In each level, you have a grid with n rows and m columns. Each cell originally contains the number 0.

One move consists of choosing one row or column, and adding 1 to all of the cells in that row or column.

To win the level, after all the moves, the number in the cell at the i-th row and j-th column should be equal to g_{i}, j.

Karen is stuck on one level, and wants to know a way to beat this level using the minimum number of moves. Please, help her with this task!
\end{lstlisting}
\\ \toprule
\textbf{Document} \\
\midrule
\vspace{-8pt}
\begin{lstlisting}
import sys
#sys.stdin=open("data.txt")
input=sys.stdin.readline

n,m=list(map(int,input().split()))

realg=[list(map(int,input().split())) for _ in range(n)]
g=[[0]*m for _ in range(n)]

ans=[]

# get differences
f1=min([realg[0][i] for i in range(m)])
for i in range(m):
    for _ in range(realg[0][i]-f1):
        ans.append("col %
    for j in range(n):
        g[j][i]+=realg[0][i]-f1

f2=min([realg[i][0] for i in range(n)])
for i in range(n):
    for _ in range(realg[i][0]-f2):
        ans.append("row %
    for j in range(m):
        g[i][j]+=realg[i][0]-f2
        
[truncated for brevity]

\end{lstlisting}
\\ \bottomrule
\end{tabular}
\end{table*}

\begin{table*}[htbp]
\centering
\caption{Example query and relevant passage for the Legal QA task.}
\label{tab:example_legal_qa}
\begin{tabular}{L{0.9\linewidth}}
\toprule
\textbf{Query} \\
\midrule
\vspace{-8pt}
\begin{lstlisting}
Are eviction cases first heard in high court? In the state of Tennessee
\end{lstlisting}
\\ \toprule
\textbf{Document} \\
\midrule
\vspace{-8pt}
\begin{lstlisting}
# 2021 Tennessee Code
## Title 29 - Remedies and Special Proceedings
### Chapter 18 - Forcible Entry and Detainer
#### § 29-18-106. Alternative Actions
 Where the action is to recover real property, ejectment, or forcible or unlawful entry or detainer may be brought.   Code 1858, § 2750; Shan., § 4441; Code 1932, § 8567; T.C.A. (orig. ed.), § 23-1606.  
#### § 29-18-107. Jurisdiction of General Sessions Judge
 All cases of forcible entry and detainer, forcible detainer, and unlawful detainer, may be tried before any one (1) judge of the court of general sessions of the county in which the acts are committed, who shall decide the particular case, and all questions of law and fact arising.   Code 1858, § 3346 (deriv. Acts 1841-1842, ch. 186, § 1); Acts 1879, ch. 23; Shan., § 5095; Code 1932, § 9249; impl. am. Acts 1979, ch. 68, § 3; T.C.A. (orig. ed.), § 23-1607.  
#### § 29-18-108. Original Jurisdiction of Circuit Court
 The action for the recovery of the possession of land, given in this chapter, may also be originally instituted in the circuit court, the same forms being substantially pursued as those prescribed, the process being issued by the clerk, the plaintiff first giving bond and security to answer costs and damages as provided in § 29-18-111.   Code 1858, § 3366 (deriv. Acts 1841-1842, ch. 186, § 8); Shan., § 5115; Code 1932, § 9270; T.C.A. (orig. ed.), § 23-1608.
\end{lstlisting}
\\ \bottomrule
\end{tabular}
\end{table*}

\begin{table*}[htbp]
\centering
\caption{Example query and relevant passage for the \paperretrieval task.}
\label{tab:example_scientific_paper}
\begin{tabular}{L{0.9\linewidth}}
\toprule
\textbf{Query} \\
\midrule
\vspace{-8pt}
\begin{lstlisting}
My goal is to develop a learning model that can handle multiple tasks simultaneously. This proposed learning model will function as a sub-model selector, meaning that when presented with a new task, it will determine the most suitable sub-model to learn this task. Additionally, my learning model could also operate as a sub-model constructor. If it determines that no existing sub-model can learn the task, it will generate innovative model architectures to construct a new sub-model suitable for the task. At present, I am considering the application of a combination of reinforcement learning and model architecture search algorithms to achieve this. To effectively evaluate the performance, I require a comprehensive benchmark that includes a variety of datasets that can be used as different sub-tasks. My aim is for my learning model to achieve state-of-the-art performance on this benchmark. In general, I am open to any methodologies that enhance the performance and adaptability of my learning model across different problem domains.
\end{lstlisting}
\\ \toprule
\textbf{Document} \\
\midrule
\vspace{-8pt}
\begin{lstlisting}
# Continual Learning with Adaptive Weights (CLAW)
## Categories
- Machine Learning
## Abstract
Approaches to continual learning aim to successfully learn a set of related tasks that arrive in an online manner. Recently, several frameworks have been developed which enable deep learning to be deployed in this learning scenario. A key modelling decision is to what extent the architecture should be shared across tasks. On the one hand, separately modelling each task avoids catastrophic forgetting but it does not support transfer learning and leads to large models. On the other hand, rigidly specifying a shared component and a task-specific part enables task transfer and limits the model size, but it is vulnerable to catastrophic forgetting and restricts the form of task-transfer that can occur. Ideally, the network should adaptively identify which parts of the network to share in a data driven way. Here we introduce such an approach called Continual Learning with Adaptive Weights (CLAW), which is based on probabilistic modelling and variational inference. Experiments show that CLAW achieves state-of-the-art performance on six benchmarks in terms of overall continual learning performance, as measured by classification accuracy, and in terms of addressing catastrophic forgetting.
\end{lstlisting}
\\ \bottomrule
\end{tabular}
\end{table*}

\begin{table*}[htbp]
\centering
\caption{Example query and relevant passage for the SetOps task.}
\label{tab:example_set_ops}
\begin{tabular}{L{0.9\linewidth}}
\toprule
\textbf{Query} \\
\midrule
\vspace{-8pt}
\begin{lstlisting}
German spy comedy films, or 2000s comedy-drama mystery films.
\end{lstlisting}
\\ \toprule
\textbf{Document} \\
\midrule
\vspace{-8pt}
\begin{lstlisting}
# The Mimosa Wants to Blossom Too
***The Mimosa Wants to Blossom Too** ''(German:***Auch Mimosen wollen blühen**'') is a 1976 West German comedy spy film directed by Helmut Meewes and starring Curd Jürgens, Eric Pohlmann and Horst Frank.
The film's sets were designed by the art director Peter Rothe.
- Curd Jürgens as Josef Popov
- Eric Pohlmann as Iwan Pederenko
- Horst Frank as Oberst Oschenko
- Susi Nicoletti as Emily Hopkins
- Heinz Reincke as Obdachloser
- Barbara Nielsen as Ludmilla
- Chiquita Gordon as Miss Ly
- Erich Padalewski as Mr. Gate
- Ljuba Welitsch as Lady Shots
- Harry Hardt as Sir Shots
- Bock, Hans-Michael & Bergfelder, Tim. *The Concise CineGraph. Encyclopedia of German Cinema*. Berghahn Books, 2009.
\end{lstlisting}
\\ \bottomrule
\end{tabular}
\end{table*}

\begin{table*}[htbp]
\centering
\caption{Example query and relevant passage for the \stackexchange task.}
\label{tab:example_stackexchange}
\begin{tabular}{L{0.9\linewidth}}
\toprule
\textbf{Query} \\
\midrule
\vspace{-8pt}
\begin{lstlisting}
Does the genetic sequence of SARS-CoV-2 end with 33 A's?
Looking at the DNA (or RNA?) sequence of the Covid-19 virus here: https://www.ncbi.nlm.nih.gov/nuccore/MN908947.3
I notice it ends in 33 a's. Does the virus really look like that, or is it some artifact of the sequencing process, or some sort of padding to round out numbers?
Here's the last few lines:
29761 acagtgaaca atgctaggga gagctgccta tatggaagag ccctaatgtg taaaattaat
29821 tttagtagtg ctatccccat gtgattttaa tagcttctta ggagaatgac aaaaaaaaaa
29881 aaaaaaaaaa aaaaaaaaaa aaa
\end{lstlisting}
\\ \toprule
\textbf{Document} \\
\midrule
\vspace{-8pt}
\begin{lstlisting}
Polyadenylation is the addition of a poly(A) tail to an RNA transcript, typically a messenger RNA (mRNA). The poly(A) tail consists of multiple adenosine monophosphates; in other words, it is a stretch of RNA that has only adenine bases. In eukaryotes, polyadenylation is part of the process that produces mature mRNA for translation. In many bacteria, the poly(A) tail promotes degradation of the mRNA. It, therefore, forms part of the larger process of gene expression.
The process of polyadenylation begins as the transcription of a gene terminates. The 3′-most segment of the newly made pre-mRNA is first cleaved off by a set of proteins; these proteins then synthesize the poly(A) tail at the RNA's 3′ end. In some genes these proteins add a poly(A) tail at one of several possible sites. Therefore, polyadenylation can produce more than one transcript from a single gene (alternative polyadenylation), similar to alternative splicing.
The poly(A) tail is important for the nuclear export, translation and stability of mRNA. The tail is shortened over time, and, when it is short enough, the mRNA is enzymatically degraded. However, in a few cell types, mRNAs with short poly(A) tails are stored for later activation by re-polyadenylation in the cytosol. In contrast, when polyadenylation occurs in bacteria, it promotes RNA degradation. This is also sometimes the case for eukaryotic non-coding RNAs.
mRNA molecules in both prokaryotes and eukaryotes have polyadenylated 3′-ends, with the prokaryotic poly(A) tails generally shorter and fewer mRNA molecules polyadenylated.
\end{lstlisting}
\\ \bottomrule
\end{tabular}
\end{table*}

\begin{table*}[htbp]
\centering
\caption{Example query and relevant passage for the Theorem Retrieval task.}
\label{tab:example_theorem_retrieval}
\begin{tabular}{L{0.9\linewidth}}
\toprule
\textbf{Query} \\
\midrule
\vspace{-8pt}
\begin{lstlisting}
Imagine you have a digital scale that can measure the weight of an infinite number of infinitely small digital dots. Each dot can either be on or off, and their weights are determined by a specific pattern: the first dot weighs 1/4 gram, the second dot weighs 1/16 grams, the third dot weighs 1/64 grams, and so on, with each subsequent dot weighing exactly one-fourth the weight of the previous dot. If you can create any combination of these dots being on (weighing their specified amount) or off (weighing nothing), and you can also add a whole number of grams to the scale (from a set of weights that are whole grams), what would be the total possible weight measured by the scale? (specifically, Lebesgue measure of this sum)
\end{lstlisting}
\\ \toprule
\textbf{Document} \\
\midrule
\vspace{-8pt}
\begin{lstlisting}
\section{Lebesgue Measure is Diffuse}
Tags: Measure Theory, Lebesgue Measure, Diffuse Measures

\begin{theorem}
Let $\lambda^n$ be Lebesgue measure on $\R^n$.
Then $\lambda^n$ is a diffuse measure.
\end{theorem}

\begin{proof}
A singleton $\set {\mathbf x} \subseteq \R^n$ is seen to be closed by combining:
:Euclidean Space is Complete Metric Space
:Metric Space is Hausdorff
:Corollary to Compact Subspace of Hausdorff Space is Closed
Hence by Closed Set Measurable in Borel Sigma-Algebra:
:$\set {\mathbf x} \in \map \BB {\R^n}$
where $\map \BB {\R^n}$ is the Borel $\sigma$-algebra on $\R^n$.
Write $\mathbf x + \epsilon = \tuple {x_1 + \epsilon, \ldots, x_n + \epsilon}$ for $\epsilon > 0$.
Then:
:$\ds \set {\mathbf x} = \bigcap_{m \mathop \in \N} \horectr {\mathbf x} {\mathbf x + \frac 1 m}$
where $\\horectr {\mathbf x} {\mathbf x + \dfrac 1 m}$ is a half-open $n$-rectangle.
{{handwaving|justify equality}}
By definition of Lebesgue measure, we have (for all $m \in \N$):
:$\ds \map {\lambda^n} {\horectr {\mathbf x} {\mathbf x + \frac 1 m} } = \prod_{i \mathop = 1}^n \frac 1 m = m^{-n}$
From Characterization of Measures, it follows that:
:$\ds \map {\lambda^n} {\set {\mathbf x} } = \lim_{m \mathop \to \infty} m^{-n}$
which equals $0$ from Sequence of Powers of Reciprocals is Null Sequence.
Therefore, for each $\mathbf x \in \R^n$:
:$\map {\lambda^n} {\set {\mathbf x} } = 0$
that is, $\lambda^n$ is a diffuse measure.
{{qed}}
[truncated for brevity]

\end{lstlisting}
\\ \bottomrule
\end{tabular}
\end{table*}

\begin{table*}[htbp]
\centering
\caption{Example query and relevant passage for the Tip-of-the-tongue task.}
\label{tab:example_tot}
\begin{tabular}{L{0.9\linewidth}}
\toprule
\textbf{Query} \\
\midrule
\vspace{-8pt}
\begin{lstlisting}
Cop's son needs blood transfusion from an inmate .
 A cop's son is very sick and needs a certain type of blood transfusion. He finds out an inmate has the same type and once agreed the inmate takes advantage of the situation in the hospital and escapes. Then a long chasing odyssey till the end when the kid finally gets cured
\end{lstlisting}
\\ \toprule
\textbf{Document} \\
\midrule
\vspace{-8pt}
\begin{lstlisting}
# Desperate Measures (film)
## Plot
Conner and Hawkins make their way to McCabe and convince him to let them inside so that Hawkins can attend to Matt. As McCabe watches Conner on the security cameras, he realizes that his nemesis is a truly devoted father, and develops a grudging respect for him. Conner intervenes when McCabe is about to ambush Cassidy and his SWAT team with a set of tanks of cyclopropane. Cassidy is furious that Conner continues to aid an escaped convict, while McCabe is angry that Conner foiled his plan. He kidnaps Matt and descends to the sub-levels of the building.
Matt tries to wound McCabe to give his father a better chance; impressed, McCabe spares Matt and leaves him at the hospital for Conner to find. McCabe then escapes into San Francisco, where he steals a car. Conner chases McCabe to a bridge, still needing him captured alive. Cassidy and his men arrive in a helicopter and a sniper opens fire. Conner again shields McCabe and is wounded in the arm. McCabe attempts to flee, but Conner is determined not to let him go. Conner wounds McCabe, sending him off the bridge and into the bay. Conner then dives in and saves him.
Back in the hospital, a wounded McCabe agrees to the transplant, which saves Matt\'s life. Even though his career is clearly over, Conner is overjoyed that his son will live. McCabe is informed by a guard that the surgery went well. As the bed reclines upwards and McCabe looks at the guard menacingly, the guard suddenly realizes that his gun is gone. McCabe holds it over the guard and asks, \"What kind of car do you have?\".
\end{lstlisting}
\\ \bottomrule
\end{tabular}
\end{table*}

\bibliographystyle{ACM-Reference-Format}
\bibliography{custom}


\begin{thebibliography}{83}


\ifx \showCODEN    \undefined \def \showCODEN     #1{\unskip}     \fi
\ifx \showISBNx    \undefined \def \showISBNx     #1{\unskip}     \fi
\ifx \showISBNxiii \undefined \def \showISBNxiii  #1{\unskip}     \fi
\ifx \showISSN     \undefined \def \showISSN      #1{\unskip}     \fi
\ifx \showLCCN     \undefined \def \showLCCN      #1{\unskip}     \fi
\ifx \shownote     \undefined \def \shownote      #1{#1}          \fi
\ifx \showarticletitle \undefined \def \showarticletitle #1{#1}   \fi
\ifx \showURL      \undefined \def \showURL       {\relax}        \fi
\providecommand\bibfield[2]{#2}
\providecommand\bibinfo[2]{#2}
\providecommand\natexlab[1]{#1}
\providecommand\showeprint[2][]{arXiv:#2}

\bibitem[ope(2025)]%
        {open_ai_o3_and_o4_mini}
 \bibinfo{year}{2025}\natexlab{}.
\newblock
\urldef\tempurl%
\url{https://openai.com/index/introducing-o3-and-o4-mini/}
\showURL{%
\tempurl}


\bibitem[Alaofi et~al\mbox{.}(2022)]%
        {where_do_queries_come_from}
\bibfield{author}{\bibinfo{person}{Marwah Alaofi}, \bibinfo{person}{Luke Gallagher}, \bibinfo{person}{Dana Mckay}, \bibinfo{person}{Lauren~L. Saling}, \bibinfo{person}{Mark Sanderson}, \bibinfo{person}{Falk Scholer}, \bibinfo{person}{Damiano Spina}, {and} \bibinfo{person}{Ryen~W. White}.} \bibinfo{year}{2022}\natexlab{}.
\newblock \showarticletitle{Where {Do} {Queries} {Come} {From}?}. In \bibinfo{booktitle}{\emph{Proceedings of the 45th {International} {ACM} {SIGIR} {Conference} on {Research} and {Development} in {Information} {Retrieval}}}. \bibinfo{publisher}{ACM}, \bibinfo{address}{Madrid Spain}, \bibinfo{pages}{2850--2862}.
\newblock
\showISBNx{978-1-4503-8732-3}
\href{https://doi.org/10.1145/3477495.3531711}{doi:\nolinkurl{10.1145/3477495.3531711}}


\bibitem[Allan et~al\mbox{.}(2017)]%
        {trec_common_core_2017_overview}
\bibfield{author}{\bibinfo{person}{James Allan}, \bibinfo{person}{Donna Harman}, \bibinfo{person}{Evangelos Kanoulas}, \bibinfo{person}{Dan Li}, \bibinfo{person}{Christophe~Van Gysel}, {and} \bibinfo{person}{Ellen~M. Voorhees}.} \bibinfo{year}{2017}\natexlab{}.
\newblock \showarticletitle{{TREC} 2017 Common Core Track Overview}. In \bibinfo{booktitle}{\emph{Proceedings of The Twenty-Sixth Text REtrieval Conference, {TREC} 2017, Gaithersburg, Maryland, USA, November 15-17, 2017}} \emph{(\bibinfo{series}{{NIST} Special Publication}, Vol.~\bibinfo{volume}{500-324})}, \bibfield{editor}{\bibinfo{person}{Ellen~M. Voorhees} {and} \bibinfo{person}{Angela Ellis}} (Eds.). \bibinfo{publisher}{National Institute of Standards and Technology {(NIST)}}.
\newblock
\urldef\tempurl%
\url{https://trec.nist.gov/pubs/trec26/papers/Overview-CC.pdf}
\showURL{%
\tempurl}


\bibitem[Arguello et~al\mbox{.}(2025)]%
        {trec_tot_2024}
\bibfield{author}{\bibinfo{person}{Jaime Arguello}, \bibinfo{person}{Samarth Bhargav}, \bibinfo{person}{Fernando Diaz}, \bibinfo{person}{Evangelos Kanoulas}, \bibinfo{person}{To~Eun Kim}, \bibinfo{person}{Yifan He}, {and} \bibinfo{person}{Bhaskar Mitra}.} \bibinfo{year}{2025}\natexlab{}.
\newblock \showarticletitle{Overview of the TREC 2024 Tip-of-the-Tongue Track}. In \bibinfo{booktitle}{\emph{Proceedings of the Thirty-Third Text REtrieval Conference}}.
\newblock


\bibitem[Arguello et~al\mbox{.}(2023)]%
        {trec_tot_2023}
\bibfield{author}{\bibinfo{person}{Jaime Arguello}, \bibinfo{person}{Samarth Bhargav}, \bibinfo{person}{Fernando Diaz}, \bibinfo{person}{Evangelos Kanoulas}, {and} \bibinfo{person}{Bhaskar Mitra}.} \bibinfo{year}{2023}\natexlab{}.
\newblock \showarticletitle{Overview of the {TREC} 2023 Tip-of-the-Tongue Track}. In \bibinfo{booktitle}{\emph{The Thirty-Second Text REtrieval Conference Proceedings {(TREC} 2023), Gaithersburg, MD, USA, November 14-17, 2023}} \emph{(\bibinfo{series}{{NIST} Special Publication}, Vol.~\bibinfo{volume}{500-xxx})}, \bibfield{editor}{\bibinfo{person}{Ian Soboroff} {and} \bibinfo{person}{Angela Ellis}} (Eds.). \bibinfo{publisher}{National Institute of Standards and Technology {(NIST)}}.
\newblock
\urldef\tempurl%
\url{https://trec.nist.gov/pubs/trec32/papers/Overview\_tot.pdf}
\showURL{%
\tempurl}


\bibitem[Arguello et~al\mbox{.}(2021)]%
        {tip_of_the_tongue_original}
\bibfield{author}{\bibinfo{person}{Jaime Arguello}, \bibinfo{person}{Adam Ferguson}, \bibinfo{person}{Emery Fine}, \bibinfo{person}{Bhaskar Mitra}, \bibinfo{person}{Hamed Zamani}, {and} \bibinfo{person}{Fernando Diaz}.} \bibinfo{year}{2021}\natexlab{}.
\newblock \showarticletitle{Tip of the Tongue Known-Item Retrieval: {A} Case Study in Movie Identification}. In \bibinfo{booktitle}{\emph{{CHIIR} '21: {ACM} {SIGIR} Conference on Human Information Interaction and Retrieval, Canberra, ACT, Australia, March 14-19, 2021}}, \bibfield{editor}{\bibinfo{person}{Falk Scholer}, \bibinfo{person}{Paul Thomas}, \bibinfo{person}{David Elsweiler}, \bibinfo{person}{Hideo Joho}, \bibinfo{person}{Noriko Kando}, {and} \bibinfo{person}{Catherine Smith}} (Eds.). \bibinfo{publisher}{{ACM}}, \bibinfo{pages}{5--14}.
\newblock
\href{https://doi.org/10.1145/3406522.3446021}{doi:\nolinkurl{10.1145/3406522.3446021}}


\bibitem[Asai et~al\mbox{.}(2023)]%
        {task_aware_retrieval_with_instructions}
\bibfield{author}{\bibinfo{person}{Akari Asai}, \bibinfo{person}{Timo Schick}, \bibinfo{person}{Patrick Lewis}, \bibinfo{person}{Xilun Chen}, \bibinfo{person}{Gautier Izacard}, \bibinfo{person}{Sebastian Riedel}, \bibinfo{person}{Hannaneh Hajishirzi}, {and} \bibinfo{person}{Wen{-}tau Yih}.} \bibinfo{year}{2023}\natexlab{}.
\newblock \showarticletitle{Task-aware Retrieval with Instructions}. In \bibinfo{booktitle}{\emph{Findings of the Association for Computational Linguistics: {ACL} 2023, Toronto, Canada, July 9-14, 2023}}, \bibfield{editor}{\bibinfo{person}{Anna Rogers}, \bibinfo{person}{Jordan~L. Boyd{-}Graber}, {and} \bibinfo{person}{Naoaki Okazaki}} (Eds.). \bibinfo{publisher}{Association for Computational Linguistics}, \bibinfo{pages}{3650--3675}.
\newblock
\href{https://doi.org/10.18653/V1/2023.FINDINGS-ACL.225}{doi:\nolinkurl{10.18653/V1/2023.FINDINGS-ACL.225}}


\bibitem[Barbaresi(2021)]%
        {trafilatura_web_scraping_library}
\bibfield{author}{\bibinfo{person}{Adrien Barbaresi}.} \bibinfo{year}{2021}\natexlab{}.
\newblock \showarticletitle{{Trafilatura: A Web Scraping Library and Command-Line Tool for Text Discovery and Extraction}}. In \bibinfo{booktitle}{\emph{Proceedings of the Joint Conference of the 59th Annual Meeting of the Association for Computational Linguistics and the 11th International Joint Conference on Natural Language Processing: System Demonstrations}}. \bibinfo{publisher}{Association for Computational Linguistics}, \bibinfo{pages}{122--131}.
\newblock
\urldef\tempurl%
\url{https://aclanthology.org/2021.acl-demo.15}
\showURL{%
\tempurl}


\bibitem[Bashir and Rauber(2011)]%
        {relationship_between_query_characterisitics_and_retrieval_bias}
\bibfield{author}{\bibinfo{person}{Shariq Bashir} {and} \bibinfo{person}{Andreas Rauber}.} \bibinfo{year}{2011}\natexlab{}.
\newblock \showarticletitle{On the relationship between query characteristics and {IR} functions retrieval bias}.
\newblock \bibinfo{journal}{\emph{Journal of the American Society for Information Science and Technology}} \bibinfo{volume}{62}, \bibinfo{number}{8} (\bibinfo{year}{2011}), \bibinfo{pages}{1515--1532}.
\newblock
\showISSN{1532-2890}
\href{https://doi.org/10.1002/asi.21549}{doi:\nolinkurl{10.1002/asi.21549}}
\newblock
\shownote{\_eprint: https://onlinelibrary.wiley.com/doi/pdf/10.1002/asi.21549}.


\bibitem[Chen et~al\mbox{.}(2023)]%
        {theorem_qa_theorem_question_answering_dataset}
\bibfield{author}{\bibinfo{person}{Wenhu Chen}, \bibinfo{person}{Ming Yin}, \bibinfo{person}{Max Ku}, \bibinfo{person}{Pan Lu}, \bibinfo{person}{Yixin Wan}, \bibinfo{person}{Xueguang Ma}, \bibinfo{person}{Jianyu Xu}, \bibinfo{person}{Xinyi Wang}, {and} \bibinfo{person}{Tony Xia}.} \bibinfo{year}{2023}\natexlab{}.
\newblock \showarticletitle{{T}heorem{QA}: A Theorem-driven Question Answering Dataset}. In \bibinfo{booktitle}{\emph{Proceedings of the 2023 Conference on Empirical Methods in Natural Language Processing}}, \bibfield{editor}{\bibinfo{person}{Houda Bouamor}, \bibinfo{person}{Juan Pino}, {and} \bibinfo{person}{Kalika Bali}} (Eds.). \bibinfo{publisher}{Association for Computational Linguistics}, \bibinfo{address}{Singapore}, \bibinfo{pages}{7889--7901}.
\newblock
\href{https://doi.org/10.18653/v1/2023.emnlp-main.489}{doi:\nolinkurl{10.18653/v1/2023.emnlp-main.489}}


\bibitem[Chen et~al\mbox{.}(2020)]%
        {hybridqa_multihop_tabular_text_dataset}
\bibfield{author}{\bibinfo{person}{Wenhu Chen}, \bibinfo{person}{Hanwen Zha}, \bibinfo{person}{Zhiyu Chen}, \bibinfo{person}{Wenhan Xiong}, \bibinfo{person}{Hong Wang}, {and} \bibinfo{person}{William~Yang Wang}.} \bibinfo{year}{2020}\natexlab{}.
\newblock \showarticletitle{{H}ybrid{QA}: A Dataset of Multi-Hop Question Answering over Tabular and Textual Data}. In \bibinfo{booktitle}{\emph{Findings of the Association for Computational Linguistics: EMNLP 2020}}, \bibfield{editor}{\bibinfo{person}{Trevor Cohn}, \bibinfo{person}{Yulan He}, {and} \bibinfo{person}{Yang Liu}} (Eds.). \bibinfo{publisher}{Association for Computational Linguistics}, \bibinfo{address}{Online}, \bibinfo{pages}{1026--1036}.
\newblock
\href{https://doi.org/10.18653/v1/2020.findings-emnlp.91}{doi:\nolinkurl{10.18653/v1/2020.findings-emnlp.91}}


\bibitem[Collins{-}Thompson et~al\mbox{.}(2013)]%
        {trec_web_track_2013_overview}
\bibfield{author}{\bibinfo{person}{Kevyn Collins{-}Thompson}, \bibinfo{person}{Paul~N. Bennett}, \bibinfo{person}{Fernando Diaz}, \bibinfo{person}{Charlie Clarke}, {and} \bibinfo{person}{Ellen~M. Voorhees}.} \bibinfo{year}{2013}\natexlab{}.
\newblock \showarticletitle{{TREC} 2013 Web Track Overview}. In \bibinfo{booktitle}{\emph{Proceedings of The Twenty-Second Text REtrieval Conference, {TREC} 2013, Gaithersburg, Maryland, USA, November 19-22, 2013}} \emph{(\bibinfo{series}{{NIST} Special Publication}, Vol.~\bibinfo{volume}{500-302})}, \bibfield{editor}{\bibinfo{person}{Ellen~M. Voorhees}} (Ed.). \bibinfo{publisher}{National Institute of Standards and Technology {(NIST)}}.
\newblock
\urldef\tempurl%
\url{http://trec.nist.gov/pubs/trec22/papers/WEB.OVERVIEW.pdf}
\showURL{%
\tempurl}


\bibitem[Collins{-}Thompson et~al\mbox{.}(2014)]%
        {trec_web_track_2014_overview}
\bibfield{author}{\bibinfo{person}{Kevyn Collins{-}Thompson}, \bibinfo{person}{Craig Macdonald}, \bibinfo{person}{Paul~N. Bennett}, \bibinfo{person}{Fernando Diaz}, {and} \bibinfo{person}{Ellen~M. Voorhees}.} \bibinfo{year}{2014}\natexlab{}.
\newblock \showarticletitle{{TREC} 2014 Web Track Overview}. In \bibinfo{booktitle}{\emph{Proceedings of The Twenty-Third Text REtrieval Conference, {TREC} 2014, Gaithersburg, Maryland, USA, November 19-21, 2014}} \emph{(\bibinfo{series}{{NIST} Special Publication}, Vol.~\bibinfo{volume}{500-308})}, \bibfield{editor}{\bibinfo{person}{Ellen~M. Voorhees} {and} \bibinfo{person}{Angela Ellis}} (Eds.). \bibinfo{publisher}{National Institute of Standards and Technology {(NIST)}}.
\newblock
\urldef\tempurl%
\url{http://trec.nist.gov/pubs/trec23/papers/overview-web.pdf}
\showURL{%
\tempurl}


\bibitem[Craswell et~al\mbox{.}(2020a)]%
        {craswell2021overviewdl}
\bibfield{author}{\bibinfo{person}{Nick Craswell}, \bibinfo{person}{Bhaskar Mitra}, \bibinfo{person}{Emine Yilmaz}, {and} \bibinfo{person}{Daniel Campos}.} \bibinfo{year}{2020}\natexlab{a}.
\newblock \showarticletitle{Overview of the {TREC} 2020 Deep Learning Track}. In \bibinfo{booktitle}{\emph{Proceedings of the Twenty-Ninth Text REtrieval Conference, {TREC} 2020, Virtual Event [Gaithersburg, Maryland, USA], November 16-20, 2020}} \emph{(\bibinfo{series}{{NIST} Special Publication}, Vol.~\bibinfo{volume}{1266})}, \bibfield{editor}{\bibinfo{person}{Ellen~M. Voorhees} {and} \bibinfo{person}{Angela Ellis}} (Eds.). \bibinfo{publisher}{National Institute of Standards and Technology {(NIST)}}.
\newblock
\urldef\tempurl%
\url{https://trec.nist.gov/pubs/trec29/papers/OVERVIEW.DL.pdf}
\showURL{%
\tempurl}


\bibitem[Craswell et~al\mbox{.}(2020b)]%
        {craswell2020overview}
\bibfield{author}{\bibinfo{person}{Nick Craswell}, \bibinfo{person}{Bhaskar Mitra}, \bibinfo{person}{Emine Yilmaz}, \bibinfo{person}{Daniel Campos}, {and} \bibinfo{person}{Ellen~M. Voorhees}.} \bibinfo{year}{2020}\natexlab{b}.
\newblock \showarticletitle{Overview of the {TREC} 2019 deep learning track}.
\newblock \bibinfo{journal}{\emph{CoRR}}  \bibinfo{volume}{abs/2003.07820} (\bibinfo{year}{2020}).
\newblock
\showeprint[arXiv]{2003.07820}
\urldef\tempurl%
\url{https://arxiv.org/abs/2003.07820}
\showURL{%
\tempurl}


\bibitem[Dai and Callan(2019)]%
        {max_p_original}
\bibfield{author}{\bibinfo{person}{Zhuyun Dai} {and} \bibinfo{person}{Jamie Callan}.} \bibinfo{year}{2019}\natexlab{}.
\newblock \showarticletitle{Deeper Text Understanding for {IR} with Contextual Neural Language Modeling}. In \bibinfo{booktitle}{\emph{Proceedings of the 42nd International {ACM} {SIGIR} Conference on Research and Development in Information Retrieval, {SIGIR} 2019, Paris, France, July 21-25, 2019}}, \bibfield{editor}{\bibinfo{person}{Benjamin Piwowarski}, \bibinfo{person}{Max Chevalier}, \bibinfo{person}{{\'{E}}ric Gaussier}, \bibinfo{person}{Yoelle Maarek}, \bibinfo{person}{Jian{-}Yun Nie}, {and} \bibinfo{person}{Falk Scholer}} (Eds.). \bibinfo{publisher}{{ACM}}, \bibinfo{pages}{985--988}.
\newblock
\href{https://doi.org/10.1145/3331184.3331303}{doi:\nolinkurl{10.1145/3331184.3331303}}


\bibitem[Dai et~al\mbox{.}(2023)]%
        {dai2023promptagator}
\bibfield{author}{\bibinfo{person}{Zhuyun Dai}, \bibinfo{person}{Vincent~Y. Zhao}, \bibinfo{person}{Ji Ma}, \bibinfo{person}{Yi Luan}, \bibinfo{person}{Jianmo Ni}, \bibinfo{person}{Jing Lu}, \bibinfo{person}{Anton Bakalov}, \bibinfo{person}{Kelvin Guu}, \bibinfo{person}{Keith~B. Hall}, {and} \bibinfo{person}{Ming{-}Wei Chang}.} \bibinfo{year}{2023}\natexlab{}.
\newblock \showarticletitle{Promptagator: Few-shot Dense Retrieval From 8 Examples}. In \bibinfo{booktitle}{\emph{The Eleventh International Conference on Learning Representations, {ICLR} 2023, Kigali, Rwanda, May 1-5, 2023}}. \bibinfo{publisher}{OpenReview.net}.
\newblock
\urldef\tempurl%
\url{https://openreview.net/forum?id=gmL46YMpu2J}
\showURL{%
\tempurl}


\bibitem[DeepSeek{-}AI et~al\mbox{.}(2025)]%
        {deepseek_r1}
\bibfield{author}{\bibinfo{person}{DeepSeek{-}AI}, \bibinfo{person}{Daya Guo}, \bibinfo{person}{Dejian Yang}, \bibinfo{person}{Haowei Zhang}, \bibinfo{person}{Junxiao Song}, \bibinfo{person}{Ruoyu Zhang}, \bibinfo{person}{Runxin Xu}, \bibinfo{person}{Qihao Zhu}, \bibinfo{person}{Shirong Ma}, \bibinfo{person}{Peiyi Wang}, \bibinfo{person}{Xiao Bi}, \bibinfo{person}{Xiaokang Zhang}, \bibinfo{person}{Xingkai Yu}, \bibinfo{person}{Yu Wu}, \bibinfo{person}{Z.~F. Wu}, \bibinfo{person}{Zhibin Gou}, \bibinfo{person}{Zhihong Shao}, \bibinfo{person}{Zhuoshu Li}, \bibinfo{person}{Ziyi Gao}, \bibinfo{person}{Aixin Liu}, \bibinfo{person}{Bing Xue}, \bibinfo{person}{Bingxuan Wang}, \bibinfo{person}{Bochao Wu}, \bibinfo{person}{Bei Feng}, \bibinfo{person}{Chengda Lu}, \bibinfo{person}{Chenggang Zhao}, \bibinfo{person}{Chengqi Deng}, \bibinfo{person}{Chenyu Zhang}, \bibinfo{person}{Chong Ruan}, \bibinfo{person}{Damai Dai}, \bibinfo{person}{Deli Chen}, \bibinfo{person}{Dongjie Ji}, \bibinfo{person}{Erhang Li},
  \bibinfo{person}{Fangyun Lin}, \bibinfo{person}{Fucong Dai}, \bibinfo{person}{Fuli Luo}, \bibinfo{person}{Guangbo Hao}, \bibinfo{person}{Guanting Chen}, \bibinfo{person}{Guowei Li}, \bibinfo{person}{H. Zhang}, \bibinfo{person}{Han Bao}, \bibinfo{person}{Hanwei Xu}, \bibinfo{person}{Haocheng Wang}, \bibinfo{person}{Honghui Ding}, \bibinfo{person}{Huajian Xin}, \bibinfo{person}{Huazuo Gao}, \bibinfo{person}{Hui Qu}, \bibinfo{person}{Hui Li}, \bibinfo{person}{Jianzhong Guo}, \bibinfo{person}{Jiashi Li}, \bibinfo{person}{Jiawei Wang}, \bibinfo{person}{Jingchang Chen}, \bibinfo{person}{Jingyang Yuan}, \bibinfo{person}{Junjie Qiu}, \bibinfo{person}{Junlong Li}, \bibinfo{person}{J.~L. Cai}, \bibinfo{person}{Jiaqi Ni}, \bibinfo{person}{Jian Liang}, \bibinfo{person}{Jin Chen}, \bibinfo{person}{Kai Dong}, \bibinfo{person}{Kai Hu}, \bibinfo{person}{Kaige Gao}, \bibinfo{person}{Kang Guan}, \bibinfo{person}{Kexin Huang}, \bibinfo{person}{Kuai Yu}, \bibinfo{person}{Lean Wang}, \bibinfo{person}{Lecong Zhang},
  \bibinfo{person}{Liang Zhao}, \bibinfo{person}{Litong Wang}, \bibinfo{person}{Liyue Zhang}, \bibinfo{person}{Lei Xu}, \bibinfo{person}{Leyi Xia}, \bibinfo{person}{Mingchuan Zhang}, \bibinfo{person}{Minghua Zhang}, \bibinfo{person}{Minghui Tang}, \bibinfo{person}{Meng Li}, \bibinfo{person}{Miaojun Wang}, \bibinfo{person}{Mingming Li}, \bibinfo{person}{Ning Tian}, \bibinfo{person}{Panpan Huang}, \bibinfo{person}{Peng Zhang}, \bibinfo{person}{Qiancheng Wang}, \bibinfo{person}{Qinyu Chen}, \bibinfo{person}{Qiushi Du}, \bibinfo{person}{Ruiqi Ge}, \bibinfo{person}{Ruisong Zhang}, \bibinfo{person}{Ruizhe Pan}, \bibinfo{person}{Runji Wang}, \bibinfo{person}{R.~J. Chen}, \bibinfo{person}{R.~L. Jin}, \bibinfo{person}{Ruyi Chen}, \bibinfo{person}{Shanghao Lu}, \bibinfo{person}{Shangyan Zhou}, \bibinfo{person}{Shanhuang Chen}, \bibinfo{person}{Shengfeng Ye}, \bibinfo{person}{Shiyu Wang}, \bibinfo{person}{Shuiping Yu}, \bibinfo{person}{Shunfeng Zhou}, \bibinfo{person}{Shuting Pan}, {and} \bibinfo{person}{S.~S. Li}.}
  \bibinfo{year}{2025}\natexlab{}.
\newblock \showarticletitle{DeepSeek-R1: Incentivizing Reasoning Capability in LLMs via Reinforcement Learning}.
\newblock \bibinfo{journal}{\emph{CoRR}}  \bibinfo{volume}{abs/2501.12948} (\bibinfo{year}{2025}).
\newblock
\showeprint[arXiv]{2501.12948}
\href{https://doi.org/10.48550/ARXIV.2501.12948}{doi:\nolinkurl{10.48550/ARXIV.2501.12948}}


\bibitem[Devlin et~al\mbox{.}(2019)]%
        {bert}
\bibfield{author}{\bibinfo{person}{Jacob Devlin}, \bibinfo{person}{Ming{-}Wei Chang}, \bibinfo{person}{Kenton Lee}, {and} \bibinfo{person}{Kristina Toutanova}.} \bibinfo{year}{2019}\natexlab{}.
\newblock \showarticletitle{{BERT:} Pre-training of Deep Bidirectional Transformers for Language Understanding}. In \bibinfo{booktitle}{\emph{Proceedings of the 2019 Conference of the North American Chapter of the Association for Computational Linguistics: Human Language Technologies, {NAACL-HLT} 2019, Minneapolis, MN, USA, June 2-7, 2019, Volume 1 (Long and Short Papers)}}, \bibfield{editor}{\bibinfo{person}{Jill Burstein}, \bibinfo{person}{Christy Doran}, {and} \bibinfo{person}{Thamar Solorio}} (Eds.). \bibinfo{publisher}{Association for Computational Linguistics}, \bibinfo{pages}{4171--4186}.
\newblock
\href{https://doi.org/10.18653/V1/N19-1423}{doi:\nolinkurl{10.18653/V1/N19-1423}}


\bibitem[Dubey et~al\mbox{.}(2024)]%
        {llama3herd2024}
\bibfield{author}{\bibinfo{person}{Abhimanyu Dubey} {et~al\mbox{.}}} \bibinfo{year}{2024}\natexlab{}.
\newblock \showarticletitle{The Llama 3 Herd of Models}.
\newblock \bibinfo{journal}{\emph{CoRR}}  \bibinfo{volume}{abs/2407.21783} (\bibinfo{year}{2024}).
\newblock
\showeprint[arXiv]{2407.21783}
\href{https://doi.org/10.48550/ARXIV.2407.21783}{doi:\nolinkurl{10.48550/ARXIV.2407.21783}}


\bibitem[Enevoldsen et~al\mbox{.}(2025)]%
        {multilingual_mteb}
\bibfield{author}{\bibinfo{person}{Kenneth~C. Enevoldsen}, \bibinfo{person}{Isaac Chung}, \bibinfo{person}{Imene Kerboua}, \bibinfo{person}{M{\'{a}}rton Kardos}, \bibinfo{person}{Ashwin Mathur}, \bibinfo{person}{David Stap}, \bibinfo{person}{Jay Gala}, \bibinfo{person}{Wissam Siblini}, \bibinfo{person}{Dominik Krzeminski}, \bibinfo{person}{Genta~Indra Winata}, \bibinfo{person}{Saba Sturua}, \bibinfo{person}{Saiteja Utpala}, \bibinfo{person}{Mathieu Ciancone}, \bibinfo{person}{Marion Schaeffer}, \bibinfo{person}{Diganta Misra}, \bibinfo{person}{Shreeya Dhakal}, \bibinfo{person}{Jonathan Rystr{\o}m}, \bibinfo{person}{Roman Solomatin}, \bibinfo{person}{{\"{O}}mer~Veysel {\c{C}}agatan}, \bibinfo{person}{Akash Kundu}, {and} \bibinfo{person}{et al.}} \bibinfo{year}{2025}\natexlab{}.
\newblock \showarticletitle{{MMTEB:} Massive Multilingual Text Embedding Benchmark}. In \bibinfo{booktitle}{\emph{The Thirteenth International Conference on Learning Representations, {ICLR} 2025, Singapore, April 24-28, 2025}}. \bibinfo{publisher}{OpenReview.net}.
\newblock
\urldef\tempurl%
\url{https://openreview.net/forum?id=zl3pfz4VCV}
\showURL{%
\tempurl}


\bibitem[Formal et~al\mbox{.}(2021)]%
        {splade_v1}
\bibfield{author}{\bibinfo{person}{Thibault Formal}, \bibinfo{person}{Benjamin Piwowarski}, {and} \bibinfo{person}{St{\'{e}}phane Clinchant}.} \bibinfo{year}{2021}\natexlab{}.
\newblock \showarticletitle{{SPLADE:} Sparse Lexical and Expansion Model for First Stage Ranking}. In \bibinfo{booktitle}{\emph{{SIGIR} '21: The 44th International {ACM} {SIGIR} Conference on Research and Development in Information Retrieval, Virtual Event, Canada, July 11-15, 2021}}, \bibfield{editor}{\bibinfo{person}{Fernando Diaz}, \bibinfo{person}{Chirag Shah}, \bibinfo{person}{Torsten Suel}, \bibinfo{person}{Pablo Castells}, \bibinfo{person}{Rosie Jones}, {and} \bibinfo{person}{Tetsuya Sakai}} (Eds.). \bibinfo{publisher}{{ACM}}, \bibinfo{pages}{2288--2292}.
\newblock
\href{https://doi.org/10.1145/3404835.3463098}{doi:\nolinkurl{10.1145/3404835.3463098}}


\bibitem[Han et~al\mbox{.}(2025)]%
        {atab_evaluating_embeddings_advanced_nlp_tasks}
\bibfield{author}{\bibinfo{person}{Simeng Han}, \bibinfo{person}{Frank~Palma Gomez}, \bibinfo{person}{Tu Vu}, \bibinfo{person}{Zefei Li}, \bibinfo{person}{Daniel Cer}, \bibinfo{person}{Hansi Zeng}, \bibinfo{person}{Chris Tar}, \bibinfo{person}{Arman Cohan}, {and} \bibinfo{person}{Gustavo~Hernandez Abrego}.} \bibinfo{year}{2025}\natexlab{}.
\newblock \bibinfo{title}{{ATEB}: {Evaluating} and {Improving} {Advanced} {NLP} {Tasks} for {Text} {Embedding} {Models}}.
\newblock
\href{https://doi.org/10.48550/arXiv.2502.16766}{doi:\nolinkurl{10.48550/arXiv.2502.16766}}
\newblock
\shownote{arXiv:2502.16766 [cs]}.


\bibitem[Hendrycks et~al\mbox{.}(2021)]%
        {apps_measuring_coding_challenge_competence}
\bibfield{author}{\bibinfo{person}{Dan Hendrycks}, \bibinfo{person}{Steven Basart}, \bibinfo{person}{Saurav Kadavath}, \bibinfo{person}{Mantas Mazeika}, \bibinfo{person}{Akul Arora}, \bibinfo{person}{Ethan Guo}, \bibinfo{person}{Collin Burns}, \bibinfo{person}{Samir Puranik}, \bibinfo{person}{Horace He}, \bibinfo{person}{Dawn Song}, {and} \bibinfo{person}{Jacob Steinhardt}.} \bibinfo{year}{2021}\natexlab{}.
\newblock \showarticletitle{Measuring Coding Challenge Competence With {APPS}}. In \bibinfo{booktitle}{\emph{Proceedings of the Neural Information Processing Systems Track on Datasets and Benchmarks 1, NeurIPS Datasets and Benchmarks 2021, December 2021, virtual}}, \bibfield{editor}{\bibinfo{person}{Joaquin Vanschoren} {and} \bibinfo{person}{Sai{-}Kit Yeung}} (Eds.).
\newblock
\urldef\tempurl%
\url{https://datasets-benchmarks-proceedings.neurips.cc/paper/2021/hash/c24cd76e1ce41366a4bbe8a49b02a028-Abstract-round2.html}
\showURL{%
\tempurl}


\bibitem[Izacard et~al\mbox{.}(2023)]%
        {izacard2022atlas}
\bibfield{author}{\bibinfo{person}{Gautier Izacard}, \bibinfo{person}{Patrick Lewis}, \bibinfo{person}{Maria Lomeli}, \bibinfo{person}{Lucas Hosseini}, \bibinfo{person}{Fabio Petroni}, \bibinfo{person}{Timo Schick}, \bibinfo{person}{Jane Dwivedi{-}Yu}, \bibinfo{person}{Armand Joulin}, \bibinfo{person}{Sebastian Riedel}, {and} \bibinfo{person}{Edouard Grave}.} \bibinfo{year}{2023}\natexlab{}.
\newblock \showarticletitle{Atlas: Few-shot Learning with Retrieval Augmented Language Models}.
\newblock \bibinfo{journal}{\emph{J. Mach. Learn. Res.}}  \bibinfo{volume}{24} (\bibinfo{year}{2023}), \bibinfo{pages}{251:1--251:43}.
\newblock
\urldef\tempurl%
\url{https://jmlr.org/papers/v24/23-0037.html}
\showURL{%
\tempurl}


\bibitem[Jagerman et~al\mbox{.}(2023)]%
        {query_expansion_by_prompting}
\bibfield{author}{\bibinfo{person}{Rolf Jagerman}, \bibinfo{person}{Honglei Zhuang}, \bibinfo{person}{Zhen Qin}, \bibinfo{person}{Xuanhui Wang}, {and} \bibinfo{person}{Michael Bendersky}.} \bibinfo{year}{2023}\natexlab{}.
\newblock \showarticletitle{Query Expansion by Prompting Large Language Models}.
\newblock \bibinfo{journal}{\emph{CoRR}}  \bibinfo{volume}{abs/2305.03653} (\bibinfo{year}{2023}).
\newblock
\showeprint[arXiv]{2305.03653}
\href{https://doi.org/10.48550/ARXIV.2305.03653}{doi:\nolinkurl{10.48550/ARXIV.2305.03653}}


\bibitem[J{\"{a}}rvelin and Kek{\"{a}}l{\"{a}}inen(2002)]%
        {ndcg}
\bibfield{author}{\bibinfo{person}{Kalervo J{\"{a}}rvelin} {and} \bibinfo{person}{Jaana Kek{\"{a}}l{\"{a}}inen}.} \bibinfo{year}{2002}\natexlab{}.
\newblock \showarticletitle{Cumulated gain-based evaluation of {IR} techniques}.
\newblock \bibinfo{journal}{\emph{{ACM} Trans. Inf. Syst.}} \bibinfo{volume}{20}, \bibinfo{number}{4} (\bibinfo{year}{2002}), \bibinfo{pages}{422--446}.
\newblock
\href{https://doi.org/10.1145/582415.582418}{doi:\nolinkurl{10.1145/582415.582418}}


\bibitem[Jin et~al\mbox{.}(2025)]%
        {search_r1}
\bibfield{author}{\bibinfo{person}{Bowen Jin}, \bibinfo{person}{Hansi Zeng}, \bibinfo{person}{Zhenrui Yue}, \bibinfo{person}{Dong Wang}, \bibinfo{person}{Hamed Zamani}, {and} \bibinfo{person}{Jiawei Han}.} \bibinfo{year}{2025}\natexlab{}.
\newblock \showarticletitle{Search-R1: Training LLMs to Reason and Leverage Search Engines with Reinforcement Learning}.
\newblock \bibinfo{journal}{\emph{CoRR}}  \bibinfo{volume}{abs/2503.09516} (\bibinfo{year}{2025}).
\newblock
\showeprint[arXiv]{2503.09516}
\href{https://doi.org/10.48550/ARXIV.2503.09516}{doi:\nolinkurl{10.48550/ARXIV.2503.09516}}


\bibitem[Ju and Dong(2025)]%
        {mirb_mathmatical_benchmark_ir}
\bibfield{author}{\bibinfo{person}{Haocheng Ju} {and} \bibinfo{person}{Bin Dong}.} \bibinfo{year}{2025}\natexlab{}.
\newblock \showarticletitle{{MIRB:} Mathematical Information Retrieval Benchmark}.
\newblock \bibinfo{journal}{\emph{CoRR}}  \bibinfo{volume}{abs/2505.15585} (\bibinfo{year}{2025}).
\newblock
\showeprint[arXiv]{2505.15585}
\href{https://doi.org/10.48550/ARXIV.2505.15585}{doi:\nolinkurl{10.48550/ARXIV.2505.15585}}


\bibitem[Kamath et~al\mbox{.}(2025)]%
        {gemma_3_technical_report}
\bibfield{author}{\bibinfo{person}{Aishwarya Kamath}, \bibinfo{person}{Johan Ferret}, \bibinfo{person}{Shreya Pathak}, \bibinfo{person}{Nino Vieillard}, \bibinfo{person}{Ramona Merhej}, \bibinfo{person}{Sarah Perrin}, \bibinfo{person}{Tatiana Matejovicova}, \bibinfo{person}{Alexandre Ram{\'{e}}}, \bibinfo{person}{Morgane Rivi{\`{e}}re}, \bibinfo{person}{Louis Rouillard}, \bibinfo{person}{Thomas Mesnard}, \bibinfo{person}{Geoffrey Cideron}, \bibinfo{person}{Jean{-}Bastien Grill}, \bibinfo{person}{Sabela Ramos}, \bibinfo{person}{Edouard Yvinec}, \bibinfo{person}{Michelle Casbon}, \bibinfo{person}{Etienne Pot}, \bibinfo{person}{Ivo Penchev}, \bibinfo{person}{Ga{\"{e}}l Liu}, \bibinfo{person}{Francesco Visin}, \bibinfo{person}{Kathleen Kenealy}, \bibinfo{person}{Lucas Beyer}, \bibinfo{person}{Xiaohai Zhai}, \bibinfo{person}{Anton Tsitsulin}, \bibinfo{person}{R{\'{o}}bert Busa{-}Fekete}, \bibinfo{person}{Alex Feng}, \bibinfo{person}{Noveen Sachdeva}, \bibinfo{person}{Benjamin Coleman}, \bibinfo{person}{Yi Gao},
  \bibinfo{person}{Basil Mustafa}, \bibinfo{person}{Iain Barr}, \bibinfo{person}{Emilio Parisotto}, \bibinfo{person}{David Tian}, \bibinfo{person}{Matan Eyal}, \bibinfo{person}{Colin Cherry}, \bibinfo{person}{Jan{-}Thorsten Peter}, \bibinfo{person}{Danila Sinopalnikov}, \bibinfo{person}{Surya Bhupatiraju}, \bibinfo{person}{Rishabh Agarwal}, \bibinfo{person}{Mehran Kazemi}, \bibinfo{person}{Dan Malkin}, \bibinfo{person}{Ravin Kumar}, \bibinfo{person}{David Vilar}, \bibinfo{person}{Idan Brusilovsky}, \bibinfo{person}{Jiaming Luo}, \bibinfo{person}{Andreas Steiner}, \bibinfo{person}{Abe Friesen}, \bibinfo{person}{Abhanshu Sharma}, \bibinfo{person}{Abheesht Sharma}, \bibinfo{person}{Adi~Mayrav Gilady}, \bibinfo{person}{Adrian Goedeckemeyer}, \bibinfo{person}{Alaa Saade}, \bibinfo{person}{Alexander Kolesnikov}, \bibinfo{person}{Alexei Bendebury}, \bibinfo{person}{Alvin Abdagic}, \bibinfo{person}{Amit Vadi}, \bibinfo{person}{Andr{\'{a}}s Gy{\"{o}}rgy}, \bibinfo{person}{Andr{\'{e}}~Susano Pinto},
  \bibinfo{person}{Anil Das}, \bibinfo{person}{Ankur Bapna}, \bibinfo{person}{Antoine Miech}, \bibinfo{person}{Antoine Yang}, \bibinfo{person}{Antonia Paterson}, \bibinfo{person}{Ashish Shenoy}, \bibinfo{person}{Ayan Chakrabarti}, \bibinfo{person}{Bilal Piot}, \bibinfo{person}{Bo Wu}, \bibinfo{person}{Bobak Shahriari}, \bibinfo{person}{Bryce Petrini}, \bibinfo{person}{Charlie Chen}, \bibinfo{person}{Charline~Le Lan}, \bibinfo{person}{Christopher~A. Choquette{-}Choo}, \bibinfo{person}{CJ Carey}, \bibinfo{person}{Cormac Brick}, \bibinfo{person}{Daniel Deutsch}, \bibinfo{person}{Danielle Eisenbud}, \bibinfo{person}{Dee Cattle}, \bibinfo{person}{Derek Cheng}, \bibinfo{person}{Dimitris Paparas}, \bibinfo{person}{Divyashree~Shivakumar Sreepathihalli}, \bibinfo{person}{Doug Reid}, \bibinfo{person}{Dustin Tran}, \bibinfo{person}{Dustin Zelle}, \bibinfo{person}{Eric Noland}, \bibinfo{person}{Erwin Huizenga}, \bibinfo{person}{Eugene Kharitonov}, \bibinfo{person}{Frederick Liu}, \bibinfo{person}{Gagik Amirkhanyan},
  \bibinfo{person}{Glenn Cameron}, \bibinfo{person}{Hadi Hashemi}, \bibinfo{person}{Hanna Klimczak{-}Plucinska}, \bibinfo{person}{Harman Singh}, \bibinfo{person}{Harsh Mehta}, \bibinfo{person}{Harshal~Tushar Lehri}, \bibinfo{person}{Hussein Hazimeh}, \bibinfo{person}{Ian Ballantyne}, \bibinfo{person}{Idan Szpektor}, \bibinfo{person}{Ivan Nardini}, \bibinfo{person}{Jean Pouget{-}Abadie}, \bibinfo{person}{Jetha Chan}, \bibinfo{person}{Joe Stanton}, \bibinfo{person}{John Wieting}, \bibinfo{person}{Jonathan Lai}, \bibinfo{person}{Jordi Orbay}, \bibinfo{person}{Joseph Fernandez}, \bibinfo{person}{Josh Newlan}, \bibinfo{person}{Ju{-}yeong Ji}, \bibinfo{person}{Jyotinder Singh}, \bibinfo{person}{Kat Black}, \bibinfo{person}{Kathy Yu}, \bibinfo{person}{Kevin Hui}, \bibinfo{person}{Kiran Vodrahalli}, \bibinfo{person}{Klaus Greff}, \bibinfo{person}{Linhai Qiu}, \bibinfo{person}{Marcella Valentine}, \bibinfo{person}{Marina Coelho}, \bibinfo{person}{Marvin Ritter}, \bibinfo{person}{Matt Hoffman}, \bibinfo{person}{Matthew
  Watson}, \bibinfo{person}{Mayank Chaturvedi}, \bibinfo{person}{Michael Moynihan}, \bibinfo{person}{Min Ma}, \bibinfo{person}{Nabila Babar}, \bibinfo{person}{Natasha Noy}, \bibinfo{person}{Nathan Byrd}, \bibinfo{person}{Nick Roy}, \bibinfo{person}{Nikola Momchev}, \bibinfo{person}{Nilay Chauhan}, \bibinfo{person}{Oskar Bunyan}, \bibinfo{person}{Pankil Botarda}, \bibinfo{person}{Paul Caron}, \bibinfo{person}{Paul~Kishan Rubenstein}, \bibinfo{person}{Phil Culliton}, \bibinfo{person}{Philipp Schmid}, \bibinfo{person}{Pier~Giuseppe Sessa}, \bibinfo{person}{Pingmei Xu}, \bibinfo{person}{Piotr Stanczyk}, \bibinfo{person}{Pouya Tafti}, \bibinfo{person}{Rakesh Shivanna}, \bibinfo{person}{Renjie Wu}, \bibinfo{person}{Renke Pan}, \bibinfo{person}{Reza Rokni}, \bibinfo{person}{Rob Willoughby}, \bibinfo{person}{Rohith Vallu}, \bibinfo{person}{Ryan Mullins}, \bibinfo{person}{Sammy Jerome}, \bibinfo{person}{Sara Smoot}, \bibinfo{person}{Sertan Girgin}, \bibinfo{person}{Shariq Iqbal}, \bibinfo{person}{Shashir Reddy},
  \bibinfo{person}{Shruti Sheth}, \bibinfo{person}{Siim P{\~{o}}der}, \bibinfo{person}{Sijal Bhatnagar}, \bibinfo{person}{Sindhu~Raghuram Panyam}, \bibinfo{person}{Sivan Eiger}, \bibinfo{person}{Susan Zhang}, \bibinfo{person}{Tianqi Liu}, \bibinfo{person}{Trevor Yacovone}, \bibinfo{person}{Tyler Liechty}, \bibinfo{person}{Uday Kalra}, \bibinfo{person}{Utku Evci}, \bibinfo{person}{Vedant Misra}, \bibinfo{person}{Vincent Roseberry}, \bibinfo{person}{Vlad Feinberg}, \bibinfo{person}{Vlad Kolesnikov}, \bibinfo{person}{Woohyun Han}, \bibinfo{person}{Woosuk Kwon}, \bibinfo{person}{Xi Chen}, \bibinfo{person}{Yinlam Chow}, \bibinfo{person}{Yuvein Zhu}, \bibinfo{person}{Zichuan Wei}, \bibinfo{person}{Zoltan Egyed}, \bibinfo{person}{Victor Cotruta}, \bibinfo{person}{Minh Giang}, \bibinfo{person}{Phoebe Kirk}, \bibinfo{person}{Anand Rao}, \bibinfo{person}{Jessica Lo}, \bibinfo{person}{Erica Moreira}, \bibinfo{person}{Luiz~Gustavo Martins}, \bibinfo{person}{Omar Sanseviero}, \bibinfo{person}{Lucas Gonzalez},
  \bibinfo{person}{Zach Gleicher}, \bibinfo{person}{Tris Warkentin}, \bibinfo{person}{Vahab Mirrokni}, \bibinfo{person}{Evan Senter}, \bibinfo{person}{Eli Collins}, \bibinfo{person}{Joelle Barral}, \bibinfo{person}{Zoubin Ghahramani}, \bibinfo{person}{Raia Hadsell}, \bibinfo{person}{Yossi Matias}, \bibinfo{person}{D. Sculley}, \bibinfo{person}{Slav Petrov}, \bibinfo{person}{Noah Fiedel}, \bibinfo{person}{Noam Shazeer}, \bibinfo{person}{Oriol Vinyals}, \bibinfo{person}{Jeff Dean}, \bibinfo{person}{Demis Hassabis}, \bibinfo{person}{Koray Kavukcuoglu}, \bibinfo{person}{Cl{\'{e}}ment Farabet}, \bibinfo{person}{Elena Buchatskaya}, \bibinfo{person}{Jean{-}Baptiste Alayrac}, \bibinfo{person}{Rohan Anil}, \bibinfo{person}{Dmitry~(Dima) Lepikhin}, \bibinfo{person}{Sebastian Borgeaud}, \bibinfo{person}{Olivier Bachem}, \bibinfo{person}{Armand Joulin}, \bibinfo{person}{Alek Andreev}, \bibinfo{person}{Cassidy Hardin}, \bibinfo{person}{Robert Dadashi}, {and} \bibinfo{person}{L{\'{e}}onard Hussenot}.}
  \bibinfo{year}{2025}\natexlab{}.
\newblock \showarticletitle{Gemma 3 Technical Report}.
\newblock \bibinfo{journal}{\emph{CoRR}}  \bibinfo{volume}{abs/2503.19786} (\bibinfo{year}{2025}).
\newblock
\showeprint[arXiv]{2503.19786}
\href{https://doi.org/10.48550/ARXIV.2503.19786}{doi:\nolinkurl{10.48550/ARXIV.2503.19786}}


\bibitem[Killingback et~al\mbox{.}(2025)]%
        {hypencoder}
\bibfield{author}{\bibinfo{person}{Julian Killingback}, \bibinfo{person}{Hansi Zeng}, {and} \bibinfo{person}{Hamed Zamani}.} \bibinfo{year}{2025}\natexlab{}.
\newblock \showarticletitle{Hypencoder: Hypernetworks for Information Retrieval}. In \bibinfo{booktitle}{\emph{Proceedings of the 48th International ACM SIGIR Conference on Research and Development in Information Retrieval}} (Padua, Italy) \emph{(\bibinfo{series}{SIGIR '25})}. \bibinfo{publisher}{Association for Computing Machinery}, \bibinfo{address}{New York, NY, USA}, \bibinfo{pages}{2372–2383}.
\newblock
\showISBNx{9798400715921}
\href{https://doi.org/10.1145/3726302.3729983}{doi:\nolinkurl{10.1145/3726302.3729983}}


\bibitem[Kwiatkowski et~al\mbox{.}(2019)]%
        {natural_questions}
\bibfield{author}{\bibinfo{person}{Tom Kwiatkowski}, \bibinfo{person}{Jennimaria Palomaki}, \bibinfo{person}{Olivia Redfield}, \bibinfo{person}{Michael Collins}, \bibinfo{person}{Ankur Parikh}, \bibinfo{person}{Chris Alberti}, \bibinfo{person}{Danielle Epstein}, \bibinfo{person}{Illia Polosukhin}, \bibinfo{person}{Jacob Devlin}, \bibinfo{person}{Kenton Lee}, \bibinfo{person}{Kristina Toutanova}, \bibinfo{person}{Llion Jones}, \bibinfo{person}{Matthew Kelcey}, \bibinfo{person}{Ming-Wei Chang}, \bibinfo{person}{Andrew~M. Dai}, \bibinfo{person}{Jakob Uszkoreit}, \bibinfo{person}{Quoc Le}, {and} \bibinfo{person}{Slav Petrov}.} \bibinfo{year}{2019}\natexlab{}.
\newblock \showarticletitle{Natural Questions: A Benchmark for Question Answering Research}.
\newblock \bibinfo{journal}{\emph{Transactions of the Association for Computational Linguistics}}  \bibinfo{volume}{7} (\bibinfo{year}{2019}), \bibinfo{pages}{452--466}.
\newblock
\href{https://doi.org/10.1162/tacl_a_00276}{doi:\nolinkurl{10.1162/tacl_a_00276}}


\bibitem[Li et~al\mbox{.}(2025b)]%
        {r2med_reasoning_driven_benchmark_medical_ir}
\bibfield{author}{\bibinfo{person}{Lei Li}, \bibinfo{person}{Xiao Zhou}, {and} \bibinfo{person}{Zheng Liu}.} \bibinfo{year}{2025}\natexlab{b}.
\newblock \showarticletitle{{R2MED:} {A} Benchmark for Reasoning-Driven Medical Retrieval}.
\newblock \bibinfo{journal}{\emph{CoRR}}  \bibinfo{volume}{abs/2505.14558} (\bibinfo{year}{2025}).
\newblock
\showeprint[arXiv]{2505.14558}
\href{https://doi.org/10.48550/ARXIV.2505.14558}{doi:\nolinkurl{10.48550/ARXIV.2505.14558}}


\bibitem[Li et~al\mbox{.}(2025a)]%
        {coir_code_benchmark_ir}
\bibfield{author}{\bibinfo{person}{Xiangyang Li}, \bibinfo{person}{Kuicai Dong}, \bibinfo{person}{Yi~Quan Lee}, \bibinfo{person}{Wei Xia}, \bibinfo{person}{Hao Zhang}, \bibinfo{person}{Xinyi Dai}, \bibinfo{person}{Yasheng Wang}, {and} \bibinfo{person}{Ruiming Tang}.} \bibinfo{year}{2025}\natexlab{a}.
\newblock \showarticletitle{CoIR: {A} Comprehensive Benchmark for Code Information Retrieval Models}. In \bibinfo{booktitle}{\emph{Proceedings of the 63rd Annual Meeting of the Association for Computational Linguistics (Volume 1: Long Papers), {ACL} 2025, Vienna, Austria, July 27 - August 1, 2025}}, \bibfield{editor}{\bibinfo{person}{Wanxiang Che}, \bibinfo{person}{Joyce Nabende}, \bibinfo{person}{Ekaterina Shutova}, {and} \bibinfo{person}{Mohammad~Taher Pilehvar}} (Eds.). \bibinfo{publisher}{Association for Computational Linguistics}, \bibinfo{pages}{22074--22091}.
\newblock
\urldef\tempurl%
\url{https://aclanthology.org/2025.acl-long.1072/}
\showURL{%
\tempurl}


\bibitem[Li et~al\mbox{.}(2023)]%
        {li2023towardsgte}
\bibfield{author}{\bibinfo{person}{Zehan Li}, \bibinfo{person}{Xin Zhang}, \bibinfo{person}{Yanzhao Zhang}, \bibinfo{person}{Dingkun Long}, \bibinfo{person}{Pengjun Xie}, {and} \bibinfo{person}{Meishan Zhang}.} \bibinfo{year}{2023}\natexlab{}.
\newblock \showarticletitle{Towards General Text Embeddings with Multi-stage Contrastive Learning}.
\newblock \bibinfo{journal}{\emph{CoRR}}  \bibinfo{volume}{abs/2308.03281} (\bibinfo{year}{2023}).
\newblock
\showeprint[arXiv]{2308.03281}
\href{https://doi.org/10.48550/ARXIV.2308.03281}{doi:\nolinkurl{10.48550/ARXIV.2308.03281}}


\bibitem[Lin et~al\mbox{.}(2021)]%
        {pyserini}
\bibfield{author}{\bibinfo{person}{Jimmy Lin}, \bibinfo{person}{Xueguang Ma}, \bibinfo{person}{Sheng-Chieh Lin}, \bibinfo{person}{Jheng-Hong Yang}, \bibinfo{person}{Ronak Pradeep}, {and} \bibinfo{person}{Rodrigo Nogueira}.} \bibinfo{year}{2021}\natexlab{}.
\newblock \showarticletitle{{Pyserini}: A {Python} Toolkit for Reproducible Information Retrieval Research with Sparse and Dense Representations}. In \bibinfo{booktitle}{\emph{Proceedings of the 44th Annual International ACM SIGIR Conference on Research and Development in Information Retrieval (SIGIR 2021)}}. \bibinfo{pages}{2356--2362}.
\newblock


\bibitem[Lin et~al\mbox{.}(2023)]%
        {whats_that_book_decomposing_complex_queries_for_tot_retrieval}
\bibfield{author}{\bibinfo{person}{Kevin Lin}, \bibinfo{person}{Kyle Lo}, \bibinfo{person}{Joseph Gonzalez}, {and} \bibinfo{person}{Dan Klein}.} \bibinfo{year}{2023}\natexlab{}.
\newblock \showarticletitle{Decomposing Complex Queries for Tip-of-the-tongue Retrieval}. In \bibinfo{booktitle}{\emph{Findings of the Association for Computational Linguistics: EMNLP 2023}}, \bibfield{editor}{\bibinfo{person}{Houda Bouamor}, \bibinfo{person}{Juan Pino}, {and} \bibinfo{person}{Kalika Bali}} (Eds.). \bibinfo{publisher}{Association for Computational Linguistics}, \bibinfo{address}{Singapore}, \bibinfo{pages}{5521--5533}.
\newblock
\href{https://doi.org/10.18653/v1/2023.findings-emnlp.367}{doi:\nolinkurl{10.18653/v1/2023.findings-emnlp.367}}


\bibitem[Malaviya et~al\mbox{.}(2023)]%
        {quest}
\bibfield{author}{\bibinfo{person}{Chaitanya Malaviya}, \bibinfo{person}{Peter Shaw}, \bibinfo{person}{Ming{-}Wei Chang}, \bibinfo{person}{Kenton Lee}, {and} \bibinfo{person}{Kristina Toutanova}.} \bibinfo{year}{2023}\natexlab{}.
\newblock \showarticletitle{{QUEST:} {A} Retrieval Dataset of Entity-Seeking Queries with Implicit Set Operations}. In \bibinfo{booktitle}{\emph{Proceedings of the 61st Annual Meeting of the Association for Computational Linguistics (Volume 1: Long Papers), {ACL} 2023, Toronto, Canada, July 9-14, 2023}}, \bibfield{editor}{\bibinfo{person}{Anna Rogers}, \bibinfo{person}{Jordan~L. Boyd{-}Graber}, {and} \bibinfo{person}{Naoaki Okazaki}} (Eds.). \bibinfo{publisher}{Association for Computational Linguistics}, \bibinfo{pages}{14032--14047}.
\newblock
\href{https://doi.org/10.18653/V1/2023.ACL-LONG.784}{doi:\nolinkurl{10.18653/V1/2023.ACL-LONG.784}}


\bibitem[Malon and Bai(2020)]%
        {generating_followup_questions_for_multihop_qa}
\bibfield{author}{\bibinfo{person}{Christopher Malon} {and} \bibinfo{person}{Bing Bai}.} \bibinfo{year}{2020}\natexlab{}.
\newblock \showarticletitle{Generating Followup Questions for Interpretable Multi-hop Question Answering}.
\newblock \bibinfo{journal}{\emph{CoRR}}  \bibinfo{volume}{abs/2002.12344} (\bibinfo{year}{2020}).
\newblock
\showeprint[arXiv]{2002.12344}
\urldef\tempurl%
\url{https://arxiv.org/abs/2002.12344}
\showURL{%
\tempurl}


\bibitem[Mansourian and Ford(2007)]%
        {web_searchers_attributions_of_success_and_failure}
\bibfield{author}{\bibinfo{person}{Yazdan Mansourian} {and} \bibinfo{person}{Nigel Ford}.} \bibinfo{year}{2007}\natexlab{}.
\newblock \showarticletitle{Web searchers' attributions of success and failure: an empirical study}.
\newblock \bibinfo{journal}{\emph{Journal of Documentation}} \bibinfo{volume}{63}, \bibinfo{number}{5} (\bibinfo{date}{Sept.} \bibinfo{year}{2007}), \bibinfo{pages}{659--679}.
\newblock
\showISSN{0022-0418}
\href{https://doi.org/10.1108/00220410710827745}{doi:\nolinkurl{10.1108/00220410710827745}}


\bibitem[Merola and Singh(2025)]%
        {advanced_chunking_strategies}
\bibfield{author}{\bibinfo{person}{Carlo Merola} {and} \bibinfo{person}{Jaspinder Singh}.} \bibinfo{year}{2025}\natexlab{}.
\newblock \showarticletitle{Reconstructing Context: Evaluating Advanced Chunking Strategies for Retrieval-Augmented Generation}.
\newblock \bibinfo{journal}{\emph{CoRR}}  \bibinfo{volume}{abs/2504.19754} (\bibinfo{year}{2025}).
\newblock
\showeprint[arXiv]{2504.19754}
\href{https://doi.org/10.48550/ARXIV.2504.19754}{doi:\nolinkurl{10.48550/ARXIV.2504.19754}}


\bibitem[Muennighoff et~al\mbox{.}(2023)]%
        {mteb}
\bibfield{author}{\bibinfo{person}{Niklas Muennighoff}, \bibinfo{person}{Nouamane Tazi}, \bibinfo{person}{Lo{\"{\i}}c Magne}, {and} \bibinfo{person}{Nils Reimers}.} \bibinfo{year}{2023}\natexlab{}.
\newblock \showarticletitle{{MTEB:} Massive Text Embedding Benchmark}. In \bibinfo{booktitle}{\emph{Proceedings of the 17th Conference of the European Chapter of the Association for Computational Linguistics, {EACL} 2023, Dubrovnik, Croatia, May 2-6, 2023}}, \bibfield{editor}{\bibinfo{person}{Andreas Vlachos} {and} \bibinfo{person}{Isabelle Augenstein}} (Eds.). \bibinfo{publisher}{Association for Computational Linguistics}, \bibinfo{pages}{2006--2029}.
\newblock
\href{https://doi.org/10.18653/V1/2023.EACL-MAIN.148}{doi:\nolinkurl{10.18653/V1/2023.EACL-MAIN.148}}


\bibitem[Nguyen et~al\mbox{.}(2016)]%
        {msmarco}
\bibfield{author}{\bibinfo{person}{Tri Nguyen}, \bibinfo{person}{Mir Rosenberg}, \bibinfo{person}{Xia Song}, \bibinfo{person}{Jianfeng Gao}, \bibinfo{person}{Saurabh Tiwary}, \bibinfo{person}{Rangan Majumder}, {and} \bibinfo{person}{Li Deng}.} \bibinfo{year}{2016}\natexlab{}.
\newblock \showarticletitle{{MS} {MARCO:} {A} Human Generated MAchine Reading COmprehension Dataset}. In \bibinfo{booktitle}{\emph{Proceedings of the Workshop on Cognitive Computation: Integrating neural and symbolic approaches 2016 co-located with the 30th Annual Conference on Neural Information Processing Systems {(NIPS} 2016), Barcelona, Spain, December 9, 2016}} \emph{(\bibinfo{series}{{CEUR} Workshop Proceedings}, Vol.~\bibinfo{volume}{1773})}, \bibfield{editor}{\bibinfo{person}{Tarek~Richard Besold}, \bibinfo{person}{Antoine Bordes}, \bibinfo{person}{Artur~S. d'Avila Garcez}, {and} \bibinfo{person}{Greg Wayne}} (Eds.). \bibinfo{publisher}{CEUR-WS.org}.
\newblock
\urldef\tempurl%
\url{https://ceur-ws.org/Vol-1773/CoCoNIPS\_2016\_paper9.pdf}
\showURL{%
\tempurl}


\bibitem[Paszke et~al\mbox{.}(2019)]%
        {pytorch}
\bibfield{author}{\bibinfo{person}{Adam Paszke}, \bibinfo{person}{Sam Gross}, \bibinfo{person}{Francisco Massa}, \bibinfo{person}{Adam Lerer}, \bibinfo{person}{James Bradbury}, \bibinfo{person}{Gregory Chanan}, \bibinfo{person}{Trevor Killeen}, \bibinfo{person}{Zeming Lin}, \bibinfo{person}{Natalia Gimelshein}, \bibinfo{person}{Luca Antiga}, \bibinfo{person}{Alban Desmaison}, \bibinfo{person}{Andreas K{\"{o}}pf}, \bibinfo{person}{Edward~Z. Yang}, \bibinfo{person}{Zachary DeVito}, \bibinfo{person}{Martin Raison}, \bibinfo{person}{Alykhan Tejani}, \bibinfo{person}{Sasank Chilamkurthy}, \bibinfo{person}{Benoit Steiner}, \bibinfo{person}{Lu Fang}, \bibinfo{person}{Junjie Bai}, {and} \bibinfo{person}{Soumith Chintala}.} \bibinfo{year}{2019}\natexlab{}.
\newblock \showarticletitle{PyTorch: An Imperative Style, High-Performance Deep Learning Library}. In \bibinfo{booktitle}{\emph{Advances in Neural Information Processing Systems 32: Annual Conference on Neural Information Processing Systems 2019, NeurIPS 2019, December 8-14, 2019, Vancouver, BC, Canada}}, \bibfield{editor}{\bibinfo{person}{Hanna~M. Wallach}, \bibinfo{person}{Hugo Larochelle}, \bibinfo{person}{Alina Beygelzimer}, \bibinfo{person}{Florence d'Alch{\'{e}}{-}Buc}, \bibinfo{person}{Emily~B. Fox}, {and} \bibinfo{person}{Roman Garnett}} (Eds.). \bibinfo{pages}{8024--8035}.
\newblock
\urldef\tempurl%
\url{https://proceedings.neurips.cc/paper/2019/hash/bdbca288fee7f92f2bfa9f7012727740-Abstract.html}
\showURL{%
\tempurl}


\bibitem[Pham et~al\mbox{.}(2025)]%
        {vietnamese_mteb}
\bibfield{author}{\bibinfo{person}{Loc Pham}, \bibinfo{person}{Tung Luu}, \bibinfo{person}{Thu Vo}, \bibinfo{person}{Minh Nguyen}, {and} \bibinfo{person}{Viet Hoang}.} \bibinfo{year}{2025}\natexlab{}.
\newblock \bibinfo{title}{VN-MTEB: Vietnamese Massive Text Embedding Benchmark}.
\newblock
\showeprint[arxiv]{2507.21500}~[cs.CL]
\urldef\tempurl%
\url{https://arxiv.org/abs/2507.21500}
\showURL{%
\tempurl}


\bibitem[Poswiata et~al\mbox{.}(2024)]%
        {polish_mteb}
\bibfield{author}{\bibinfo{person}{Rafal Poswiata}, \bibinfo{person}{Slawomir Dadas}, {and} \bibinfo{person}{Michal Perelkiewicz}.} \bibinfo{year}{2024}\natexlab{}.
\newblock \showarticletitle{{PL-MTEB:} Polish Massive Text Embedding Benchmark}.
\newblock \bibinfo{journal}{\emph{CoRR}}  \bibinfo{volume}{abs/2405.10138} (\bibinfo{year}{2024}).
\newblock
\showeprint[arXiv]{2405.10138}
\href{https://doi.org/10.48550/ARXIV.2405.10138}{doi:\nolinkurl{10.48550/ARXIV.2405.10138}}


\bibitem[Reimers and Gurevych(2019)]%
        {sentence_transformers}
\bibfield{author}{\bibinfo{person}{Nils Reimers} {and} \bibinfo{person}{Iryna Gurevych}.} \bibinfo{year}{2019}\natexlab{}.
\newblock \showarticletitle{Sentence-BERT: Sentence Embeddings using Siamese BERT-Networks}. In \bibinfo{booktitle}{\emph{Proceedings of the 2019 Conference on Empirical Methods in Natural Language Processing}}. \bibinfo{publisher}{Association for Computational Linguistics}.
\newblock
\urldef\tempurl%
\url{https://arxiv.org/abs/1908.10084}
\showURL{%
\tempurl}


\bibitem[Research(2024)]%
        {snowflake_arctic_embed_l_v2}
\bibfield{author}{\bibinfo{person}{Snowflake~AI Research}.} \bibinfo{year}{2024}\natexlab{}.
\newblock \bibinfo{title}{Arctic embed: The best open source embedding models}.
\newblock \bibinfo{howpublished}{Snowflake Blog}.
\newblock
\urldef\tempurl%
\url{https://www.snowflake.com/blog/arctic-embed-open-source-embedding-models/}
\showURL{%
\tempurl}
\newblock
\shownote{Accessed: June 11, 2025. See also arXiv:2412.04506 for technical details.}.


\bibitem[Roberts et~al\mbox{.}(2021)]%
        {roberts2021overviewct}
\bibfield{author}{\bibinfo{person}{Kirk Roberts}, \bibinfo{person}{Tasmeer Alam}, \bibinfo{person}{Steven Bedrick}, \bibinfo{person}{Dina Demner{-}Fushman}, \bibinfo{person}{Kyle Lo}, \bibinfo{person}{Ian Soboroff}, \bibinfo{person}{Ellen~M. Voorhees}, \bibinfo{person}{Lucy~Lu Wang}, {and} \bibinfo{person}{William~R. Hersh}.} \bibinfo{year}{2021}\natexlab{}.
\newblock \showarticletitle{Searching for scientific evidence in a pandemic: An overview of {TREC-COVID}}.
\newblock \bibinfo{journal}{\emph{J. Biomed. Informatics}}  \bibinfo{volume}{121} (\bibinfo{year}{2021}), \bibinfo{pages}{103865}.
\newblock
\href{https://doi.org/10.1016/J.JBI.2021.103865}{doi:\nolinkurl{10.1016/J.JBI.2021.103865}}


\bibitem[Roberts et~al\mbox{.}(2022)]%
        {roberts2022overviewct}
\bibfield{author}{\bibinfo{person}{Kirk Roberts}, \bibinfo{person}{Dina Demner{-}Fushman}, \bibinfo{person}{Ellen~M. Voorhees}, \bibinfo{person}{Steven Bedrick}, {and} \bibinfo{person}{William~R. Hersh}.} \bibinfo{year}{2022}\natexlab{}.
\newblock \showarticletitle{Overview of the {TREC} 2022 Clinical Trials Track}. In \bibinfo{booktitle}{\emph{Proceedings of the Thirty-First Text REtrieval Conference, {TREC} 2022, online, November 15-19, 2022}} \emph{(\bibinfo{series}{{NIST} Special Publication}, Vol.~\bibinfo{volume}{500-338})}, \bibfield{editor}{\bibinfo{person}{Ian Soboroff} {and} \bibinfo{person}{Angela Ellis}} (Eds.). \bibinfo{publisher}{National Institute of Standards and Technology {(NIST)}}.
\newblock
\urldef\tempurl%
\url{https://trec.nist.gov/pubs/trec31/papers/Overview\_trials.pdf}
\showURL{%
\tempurl}


\bibitem[Robertson and Soboroff(2001)]%
        {trec_filtering_track_2001}
\bibfield{author}{\bibinfo{person}{Stephen~E. Robertson} {and} \bibinfo{person}{Ian Soboroff}.} \bibinfo{year}{2001}\natexlab{}.
\newblock \showarticletitle{The {TREC} 2001 Filtering Track Report}. In \bibinfo{booktitle}{\emph{Proceedings of The Tenth Text REtrieval Conference, {TREC} 2001, Gaithersburg, Maryland, USA, November 13-16, 2001}} \emph{(\bibinfo{series}{{NIST} Special Publication}, Vol.~\bibinfo{volume}{500-250})}, \bibfield{editor}{\bibinfo{person}{Ellen~M. Voorhees} {and} \bibinfo{person}{Donna~K. Harman}} (Eds.). \bibinfo{publisher}{National Institute of Standards and Technology {(NIST)}}.
\newblock
\urldef\tempurl%
\url{http://trec.nist.gov/pubs/trec10/papers/filtering\_track.pdf}
\showURL{%
\tempurl}


\bibitem[Robertson and Walker(1994)]%
        {bm25}
\bibfield{author}{\bibinfo{person}{S.~E. Robertson} {and} \bibinfo{person}{S. Walker}.} \bibinfo{year}{1994}\natexlab{}.
\newblock \showarticletitle{Some simple effective approximations to the 2-Poisson model for probabilistic weighted retrieval}. In \bibinfo{booktitle}{\emph{Proceedings of the 17th Annual International ACM SIGIR Conference on Research and Development in Information Retrieval}} (Dublin, Ireland) \emph{(\bibinfo{series}{SIGIR '94})}. \bibinfo{publisher}{Springer-Verlag}, \bibinfo{address}{Berlin, Heidelberg}, \bibinfo{pages}{232–241}.
\newblock
\showISBNx{038719889X}


\bibitem[Shao et~al\mbox{.}(2025)]%
        {reason_ir}
\bibfield{author}{\bibinfo{person}{Rulin Shao}, \bibinfo{person}{Rui Qiao}, \bibinfo{person}{Varsha Kishore}, \bibinfo{person}{Niklas Muennighoff}, \bibinfo{person}{Xi~Victoria Lin}, \bibinfo{person}{Daniela Rus}, \bibinfo{person}{Bryan Kian~Hsiang Low}, \bibinfo{person}{Sewon Min}, \bibinfo{person}{Wen{-}tau Yih}, \bibinfo{person}{Pang~Wei Koh}, {and} \bibinfo{person}{Luke Zettlemoyer}.} \bibinfo{year}{2025}\natexlab{}.
\newblock \showarticletitle{ReasonIR: Training Retrievers for Reasoning Tasks}.
\newblock \bibinfo{journal}{\emph{CoRR}}  \bibinfo{volume}{abs/2504.20595} (\bibinfo{year}{2025}).
\newblock
\showeprint[arXiv]{2504.20595}
\href{https://doi.org/10.48550/ARXIV.2504.20595}{doi:\nolinkurl{10.48550/ARXIV.2504.20595}}


\bibitem[Song et~al\mbox{.}(2025)]%
        {r1_searcher}
\bibfield{author}{\bibinfo{person}{Huatong Song}, \bibinfo{person}{Jinhao Jiang}, \bibinfo{person}{Yingqian Min}, \bibinfo{person}{Jie Chen}, \bibinfo{person}{Zhipeng Chen}, \bibinfo{person}{Wayne~Xin Zhao}, \bibinfo{person}{Lei Fang}, {and} \bibinfo{person}{Ji{-}Rong Wen}.} \bibinfo{year}{2025}\natexlab{}.
\newblock \showarticletitle{R1-Searcher: Incentivizing the Search Capability in LLMs via Reinforcement Learning}.
\newblock \bibinfo{journal}{\emph{CoRR}}  \bibinfo{volume}{abs/2503.05592} (\bibinfo{year}{2025}).
\newblock
\showeprint[arXiv]{2503.05592}
\href{https://doi.org/10.48550/ARXIV.2503.05592}{doi:\nolinkurl{10.48550/ARXIV.2503.05592}}


\bibitem[Su et~al\mbox{.}(2023)]%
        {one_embedder_instruction_finetuned_embeddings}
\bibfield{author}{\bibinfo{person}{Hongjin Su}, \bibinfo{person}{Weijia Shi}, \bibinfo{person}{Jungo Kasai}, \bibinfo{person}{Yizhong Wang}, \bibinfo{person}{Yushi Hu}, \bibinfo{person}{Mari Ostendorf}, \bibinfo{person}{Wen-tau Yih}, \bibinfo{person}{Noah~A. Smith}, \bibinfo{person}{Luke Zettlemoyer}, {and} \bibinfo{person}{Tao Yu}.} \bibinfo{year}{2023}\natexlab{}.
\newblock \showarticletitle{One Embedder, Any Task: Instruction-Finetuned Text Embeddings}. In \bibinfo{booktitle}{\emph{Findings of the Association for Computational Linguistics: ACL 2023}}, \bibfield{editor}{\bibinfo{person}{Anna Rogers}, \bibinfo{person}{Jordan Boyd-Graber}, {and} \bibinfo{person}{Naoaki Okazaki}} (Eds.). \bibinfo{publisher}{Association for Computational Linguistics}, \bibinfo{address}{Toronto, Canada}, \bibinfo{pages}{1102--1121}.
\newblock
\href{https://doi.org/10.18653/v1/2023.findings-acl.71}{doi:\nolinkurl{10.18653/v1/2023.findings-acl.71}}


\bibitem[Su et~al\mbox{.}(2025)]%
        {bright}
\bibfield{author}{\bibinfo{person}{Hongjin Su}, \bibinfo{person}{Howard Yen}, \bibinfo{person}{Mengzhou Xia}, \bibinfo{person}{Weijia Shi}, \bibinfo{person}{Niklas Muennighoff}, \bibinfo{person}{Han{-}yu Wang}, \bibinfo{person}{Haisu Liu}, \bibinfo{person}{Quan Shi}, \bibinfo{person}{Zachary~S. Siegel}, \bibinfo{person}{Michael Tang}, \bibinfo{person}{Ruoxi Sun}, \bibinfo{person}{Jinsung Yoon}, \bibinfo{person}{Sercan~{\"{O}}. Arik}, \bibinfo{person}{Danqi Chen}, {and} \bibinfo{person}{Tao Yu}.} \bibinfo{year}{2025}\natexlab{}.
\newblock \showarticletitle{{BRIGHT:} {A} Realistic and Challenging Benchmark for Reasoning-Intensive Retrieval}. In \bibinfo{booktitle}{\emph{The Thirteenth International Conference on Learning Representations, {ICLR} 2025, Singapore, April 24-28, 2025}}. \bibinfo{publisher}{OpenReview.net}.
\newblock
\urldef\tempurl%
\url{https://openreview.net/forum?id=ykuc5q381b}
\showURL{%
\tempurl}


\bibitem[Sun et~al\mbox{.}(2024)]%
        {mair}
\bibfield{author}{\bibinfo{person}{Weiwei Sun}, \bibinfo{person}{Zhengliang Shi}, \bibinfo{person}{Wu Long}, \bibinfo{person}{Lingyong Yan}, \bibinfo{person}{Xinyu Ma}, \bibinfo{person}{Yiding Liu}, \bibinfo{person}{Min Cao}, \bibinfo{person}{Dawei Yin}, {and} \bibinfo{person}{Zhaochun Ren}.} \bibinfo{year}{2024}\natexlab{}.
\newblock \showarticletitle{{MAIR:} {A} Massive Benchmark for Evaluating Instructed Retrieval}. In \bibinfo{booktitle}{\emph{Proceedings of the 2024 Conference on Empirical Methods in Natural Language Processing, {EMNLP} 2024, Miami, FL, USA, November 12-16, 2024}}, \bibfield{editor}{\bibinfo{person}{Yaser Al{-}Onaizan}, \bibinfo{person}{Mohit Bansal}, {and} \bibinfo{person}{Yun{-}Nung Chen}} (Eds.). \bibinfo{publisher}{Association for Computational Linguistics}, \bibinfo{pages}{14044--14067}.
\newblock
\href{https://doi.org/10.18653/V1/2024.EMNLP-MAIN.778}{doi:\nolinkurl{10.18653/V1/2024.EMNLP-MAIN.778}}


\bibitem[Talmor and Berant(2018)]%
        {complex_web_questions}
\bibfield{author}{\bibinfo{person}{Alon Talmor} {and} \bibinfo{person}{Jonathan Berant}.} \bibinfo{year}{2018}\natexlab{}.
\newblock \showarticletitle{The Web as a Knowledge-Base for Answering Complex Questions}. In \bibinfo{booktitle}{\emph{Proceedings of the 2018 Conference of the North {A}merican Chapter of the Association for Computational Linguistics: Human Language Technologies, Volume 1 (Long Papers)}}, \bibfield{editor}{\bibinfo{person}{Marilyn Walker}, \bibinfo{person}{Heng Ji}, {and} \bibinfo{person}{Amanda Stent}} (Eds.). \bibinfo{publisher}{Association for Computational Linguistics}, \bibinfo{address}{New Orleans, Louisiana}, \bibinfo{pages}{641--651}.
\newblock
\href{https://doi.org/10.18653/v1/N18-1059}{doi:\nolinkurl{10.18653/v1/N18-1059}}


\bibitem[Thai et~al\mbox{.}(2022)]%
        {relic_retrieving_evidence_for_literary_claims}
\bibfield{author}{\bibinfo{person}{Katherine Thai}, \bibinfo{person}{Yapei Chang}, \bibinfo{person}{Kalpesh Krishna}, {and} \bibinfo{person}{Mohit Iyyer}.} \bibinfo{year}{2022}\natexlab{}.
\newblock \showarticletitle{{REL}i{C}: Retrieving Evidence for Literary Claims}. In \bibinfo{booktitle}{\emph{Proceedings of the 60th Annual Meeting of the Association for Computational Linguistics (Volume 1: Long Papers)}}, \bibfield{editor}{\bibinfo{person}{Smaranda Muresan}, \bibinfo{person}{Preslav Nakov}, {and} \bibinfo{person}{Aline Villavicencio}} (Eds.). \bibinfo{publisher}{Association for Computational Linguistics}, \bibinfo{address}{Dublin, Ireland}, \bibinfo{pages}{7500--7518}.
\newblock
\href{https://doi.org/10.18653/v1/2022.acl-long.517}{doi:\nolinkurl{10.18653/v1/2022.acl-long.517}}


\bibitem[Thakur et~al\mbox{.}(2021)]%
        {beir}
\bibfield{author}{\bibinfo{person}{Nandan Thakur}, \bibinfo{person}{Nils Reimers}, \bibinfo{person}{Andreas R{\"u}ckl{\'e}}, \bibinfo{person}{Abhishek Srivastava}, {and} \bibinfo{person}{Iryna Gurevych}.} \bibinfo{year}{2021}\natexlab{}.
\newblock \showarticletitle{{BEIR}: A Heterogeneous Benchmark for Zero-shot Evaluation of Information Retrieval Models}. In \bibinfo{booktitle}{\emph{Thirty-fifth Conference on Neural Information Processing Systems Datasets and Benchmarks Track (Round 2)}}.
\newblock
\urldef\tempurl%
\url{https://openreview.net/forum?id=wCu6T5xFjeJ}
\showURL{%
\tempurl}


\bibitem[Trippas et~al\mbox{.}(2024)]%
        {what_do_users_really_ask_large_language_models}
\bibfield{author}{\bibinfo{person}{Johanne~R. Trippas}, \bibinfo{person}{Sara Fahad Dawood~Al Lawati}, \bibinfo{person}{Joel Mackenzie}, {and} \bibinfo{person}{Luke Gallagher}.} \bibinfo{year}{2024}\natexlab{}.
\newblock \showarticletitle{What do Users Really Ask Large Language Models? An Initial Log Analysis of Google Bard Interactions in the Wild}. In \bibinfo{booktitle}{\emph{Proceedings of the 47th International {ACM} {SIGIR} Conference on Research and Development in Information Retrieval, {SIGIR} 2024, Washington DC, USA, July 14-18, 2024}}, \bibfield{editor}{\bibinfo{person}{Grace~Hui Yang}, \bibinfo{person}{Hongning Wang}, \bibinfo{person}{Sam Han}, \bibinfo{person}{Claudia Hauff}, \bibinfo{person}{Guido Zuccon}, {and} \bibinfo{person}{Yi~Zhang}} (Eds.). \bibinfo{publisher}{{ACM}}, \bibinfo{pages}{2703--2707}.
\newblock
\href{https://doi.org/10.1145/3626772.3657914}{doi:\nolinkurl{10.1145/3626772.3657914}}


\bibitem[van~den Elsen et~al\mbox{.}(2025)]%
        {reproducing_nevir}
\bibfield{author}{\bibinfo{person}{Coen van~den Elsen}, \bibinfo{person}{Francien Barkhof}, \bibinfo{person}{Thijmen Nijdam}, \bibinfo{person}{Simon Lupart}, {and} \bibinfo{person}{Mohammad Aliannejadi}.} \bibinfo{year}{2025}\natexlab{}.
\newblock \showarticletitle{Reproducing NevIR: Negation in Neural Information Retrieval}. In \bibinfo{booktitle}{\emph{Proceedings of the 48th International {ACM} {SIGIR} Conference on Research and Development in Information Retrieval, {SIGIR} 2025, Padua, Italy, July 13-18, 2025}}, \bibfield{editor}{\bibinfo{person}{Nicola Ferro}, \bibinfo{person}{Maria Maistro}, \bibinfo{person}{Gabriella Pasi}, \bibinfo{person}{Omar Alonso}, \bibinfo{person}{Andrew Trotman}, {and} \bibinfo{person}{Suzan Verberne}} (Eds.). \bibinfo{publisher}{{ACM}}, \bibinfo{pages}{3346--3356}.
\newblock
\href{https://doi.org/10.1145/3726302.3730294}{doi:\nolinkurl{10.1145/3726302.3730294}}


\bibitem[Vemuru et~al\mbox{.}(2021)]%
        {handling_complex_queries_using_query_trees}
\bibfield{author}{\bibinfo{person}{Srihari Vemuru}, \bibinfo{person}{Eric John}, {and} \bibinfo{person}{Shrisha Rao}.} \bibinfo{year}{2021}\natexlab{}.
\newblock \bibinfo{title}{Handling Complex Queries Using Query Trees}.
\newblock
\href{https://doi.org/10.36227/techrxiv.14845212.v1}{doi:\nolinkurl{10.36227/techrxiv.14845212.v1}}


\bibitem[Voorhees(2005)]%
        {trec_robust_2004_overview}
\bibfield{author}{\bibinfo{person}{Ellen Voorhees}.} \bibinfo{year}{2005}\natexlab{}.
\newblock \bibinfo{title}{Overview of the TREC 2004 Robust Retrieval Track}.
\newblock
\href{https://doi.org/10.6028/NIST.SP.500-261}{doi:\nolinkurl{10.6028/NIST.SP.500-261}}


\bibitem[Wang et~al\mbox{.}(2023a)]%
        {doris_mae}
\bibfield{author}{\bibinfo{person}{Jianyou~(Andre) Wang}, \bibinfo{person}{Kaicheng Wang}, \bibinfo{person}{Xiaoyue Wang}, \bibinfo{person}{Prudhviraj Naidu}, \bibinfo{person}{Leon Bergen}, {and} \bibinfo{person}{Ramamohan Paturi}.} \bibinfo{year}{2023}\natexlab{a}.
\newblock \showarticletitle{Scientific Document Retrieval using Multi-level Aspect-based Queries}. In \bibinfo{booktitle}{\emph{Advances in Neural Information Processing Systems}}, \bibfield{editor}{\bibinfo{person}{A.~Oh}, \bibinfo{person}{T.~Naumann}, \bibinfo{person}{A.~Globerson}, \bibinfo{person}{K.~Saenko}, \bibinfo{person}{M.~Hardt}, {and} \bibinfo{person}{S.~Levine}} (Eds.), Vol.~\bibinfo{volume}{36}. \bibinfo{publisher}{Curran Associates, Inc.}, \bibinfo{pages}{38404--38419}.
\newblock
\urldef\tempurl%
\url{https://proceedings.neurips.cc/paper_files/paper/2023/file/78f9c04bdcb06f1ada3902912d8b64ba-Paper-Datasets_and_Benchmarks.pdf}
\showURL{%
\tempurl}


\bibitem[Wang et~al\mbox{.}(2023c)]%
        {query_to_doc}
\bibfield{author}{\bibinfo{person}{Liang Wang}, \bibinfo{person}{Nan Yang}, {and} \bibinfo{person}{Furu Wei}.} \bibinfo{year}{2023}\natexlab{c}.
\newblock \showarticletitle{Query2doc: Query Expansion with Large Language Models}. In \bibinfo{booktitle}{\emph{Proceedings of the 2023 Conference on Empirical Methods in Natural Language Processing, {EMNLP} 2023, Singapore, December 6-10, 2023}}, \bibfield{editor}{\bibinfo{person}{Houda Bouamor}, \bibinfo{person}{Juan Pino}, {and} \bibinfo{person}{Kalika Bali}} (Eds.). \bibinfo{publisher}{Association for Computational Linguistics}, \bibinfo{pages}{9414--9423}.
\newblock
\href{https://doi.org/10.18653/V1/2023.EMNLP-MAIN.585}{doi:\nolinkurl{10.18653/V1/2023.EMNLP-MAIN.585}}


\bibitem[Wang et~al\mbox{.}(2024)]%
        {birco}
\bibfield{author}{\bibinfo{person}{Xiaoyue Wang}, \bibinfo{person}{Jianyou Wang}, \bibinfo{person}{Weili Cao}, \bibinfo{person}{Kaicheng Wang}, \bibinfo{person}{Ramamohan Paturi}, {and} \bibinfo{person}{Leon Bergen}.} \bibinfo{year}{2024}\natexlab{}.
\newblock \showarticletitle{{BIRCO:} {A} Benchmark of Information Retrieval Tasks with Complex Objectives}.
\newblock \bibinfo{journal}{\emph{CoRR}}  \bibinfo{volume}{abs/2402.14151} (\bibinfo{year}{2024}).
\newblock
\showeprint[arXiv]{2402.14151}
\href{https://doi.org/10.48550/ARXIV.2402.14151}{doi:\nolinkurl{10.48550/ARXIV.2402.14151}}


\bibitem[Wang et~al\mbox{.}(2023b)]%
        {wang2022selfconsistency}
\bibfield{author}{\bibinfo{person}{Xuezhi Wang}, \bibinfo{person}{Jason Wei}, \bibinfo{person}{Dale Schuurmans}, \bibinfo{person}{Quoc~V. Le}, \bibinfo{person}{Ed~H. Chi}, \bibinfo{person}{Sharan Narang}, \bibinfo{person}{Aakanksha Chowdhery}, {and} \bibinfo{person}{Denny Zhou}.} \bibinfo{year}{2023}\natexlab{b}.
\newblock \showarticletitle{Self-Consistency Improves Chain of Thought Reasoning in Language Models}.
\newblock  (\bibinfo{year}{2023}).
\newblock
\urldef\tempurl%
\url{https://openreview.net/forum?id=1PL1NIMMrw}
\showURL{%
\tempurl}


\bibitem[Wazzan et~al\mbox{.}(2024)]%
        {comparing_traditional_and_llm_based_search}
\bibfield{author}{\bibinfo{person}{Albatool Wazzan}, \bibinfo{person}{Stephen MacNeil}, {and} \bibinfo{person}{Richard Souvenir}.} \bibinfo{year}{2024}\natexlab{}.
\newblock \showarticletitle{Comparing Traditional and LLM-based Search for Image Geolocation}. In \bibinfo{booktitle}{\emph{Proceedings of the 2024 {ACM} {SIGIR} Conference on Human Information Interaction and Retrieval, {CHIIR} 2024, Sheffield, United Kingdom, March 10-14, 2024}}, \bibfield{editor}{\bibinfo{person}{Paul~D. Clough}, \bibinfo{person}{Morgan Harvey}, {and} \bibinfo{person}{Frank Hopfgartner}} (Eds.). \bibinfo{publisher}{{ACM}}, \bibinfo{pages}{291--302}.
\newblock
\href{https://doi.org/10.1145/3627508.3638305}{doi:\nolinkurl{10.1145/3627508.3638305}}


\bibitem[Wei et~al\mbox{.}(2022)]%
        {wei2022chainofthought}
\bibfield{author}{\bibinfo{person}{Jason Wei}, \bibinfo{person}{Xuezhi Wang}, \bibinfo{person}{Dale Schuurmans}, \bibinfo{person}{Maarten Bosma}, \bibinfo{person}{Brian Ichter}, \bibinfo{person}{Fei Xia}, \bibinfo{person}{Ed~H. Chi}, \bibinfo{person}{Quoc~V. Le}, {and} \bibinfo{person}{Denny Zhou}.} \bibinfo{year}{2022}\natexlab{}.
\newblock \showarticletitle{Chain-of-Thought Prompting Elicits Reasoning in Large Language Models}. In \bibinfo{booktitle}{\emph{Advances in Neural Information Processing Systems 35: Annual Conference on Neural Information Processing Systems 2022, NeurIPS 2022, New Orleans, LA, USA, November 28 - December 9, 2022}}, \bibfield{editor}{\bibinfo{person}{Sanmi Koyejo}, \bibinfo{person}{S.~Mohamed}, \bibinfo{person}{A.~Agarwal}, \bibinfo{person}{Danielle Belgrave}, \bibinfo{person}{K.~Cho}, {and} \bibinfo{person}{A.~Oh}} (Eds.).
\newblock
\urldef\tempurl%
\url{http://papers.nips.cc/paper\_files/paper/2022/hash/9d5609613524ecf4f15af0f7b31abca4-Abstract-Conference.html}
\showURL{%
\tempurl}


\bibitem[Welbl et~al\mbox{.}(2018)]%
        {qangaroo}
\bibfield{author}{\bibinfo{person}{Johannes Welbl}, \bibinfo{person}{Pontus Stenetorp}, {and} \bibinfo{person}{Sebastian Riedel}.} \bibinfo{year}{2018}\natexlab{}.
\newblock \showarticletitle{Constructing Datasets for Multi-hop Reading Comprehension Across Documents}.
\newblock \bibinfo{journal}{\emph{Transactions of the Association for Computational Linguistics}}  \bibinfo{volume}{6} (\bibinfo{year}{2018}), \bibinfo{pages}{287--302}.
\newblock
\href{https://doi.org/10.1162/tacl_a_00021}{doi:\nolinkurl{10.1162/tacl_a_00021}}


\bibitem[Weller et~al\mbox{.}(2025a)]%
        {followir}
\bibfield{author}{\bibinfo{person}{Orion Weller}, \bibinfo{person}{Benjamin Chang}, \bibinfo{person}{Sean MacAvaney}, \bibinfo{person}{Kyle Lo}, \bibinfo{person}{Arman Cohan}, \bibinfo{person}{Benjamin~Van Durme}, \bibinfo{person}{Dawn~J. Lawrie}, {and} \bibinfo{person}{Luca Soldaini}.} \bibinfo{year}{2025}\natexlab{a}.
\newblock \showarticletitle{FollowIR: Evaluating and Teaching Information Retrieval Models to Follow Instructions}. In \bibinfo{booktitle}{\emph{Proceedings of the 2025 Conference of the Nations of the Americas Chapter of the Association for Computational Linguistics: Human Language Technologies, {NAACL} 2025 - Volume 1: Long Papers, Albuquerque, New Mexico, USA, April 29 - May 4, 2025}}, \bibfield{editor}{\bibinfo{person}{Luis Chiruzzo}, \bibinfo{person}{Alan Ritter}, {and} \bibinfo{person}{Lu~Wang}} (Eds.). \bibinfo{publisher}{Association for Computational Linguistics}, \bibinfo{pages}{11926--11942}.
\newblock
\href{https://doi.org/10.18653/V1/2025.NAACL-LONG.597}{doi:\nolinkurl{10.18653/V1/2025.NAACL-LONG.597}}


\bibitem[Weller et~al\mbox{.}(2024a)]%
        {promptriever}
\bibfield{author}{\bibinfo{person}{Orion Weller}, \bibinfo{person}{Benjamin~Van Durme}, \bibinfo{person}{Dawn~J. Lawrie}, \bibinfo{person}{Ashwin Paranjape}, \bibinfo{person}{Yuhao Zhang}, {and} \bibinfo{person}{Jack Hessel}.} \bibinfo{year}{2024}\natexlab{a}.
\newblock \showarticletitle{Promptriever: Instruction-Trained Retrievers Can Be Prompted Like Language Models}.
\newblock \bibinfo{journal}{\emph{CoRR}}  \bibinfo{volume}{abs/2409.11136} (\bibinfo{year}{2024}).
\newblock
\showeprint[arXiv]{2409.11136}
\href{https://doi.org/10.48550/ARXIV.2409.11136}{doi:\nolinkurl{10.48550/ARXIV.2409.11136}}


\bibitem[Weller et~al\mbox{.}(2024b)]%
        {negation_in_neural_information_retrieval}
\bibfield{author}{\bibinfo{person}{Orion Weller}, \bibinfo{person}{Dawn~J. Lawrie}, {and} \bibinfo{person}{Benjamin~Van Durme}.} \bibinfo{year}{2024}\natexlab{b}.
\newblock \showarticletitle{NevIR: Negation in Neural Information Retrieval}. In \bibinfo{booktitle}{\emph{Proceedings of the 18th Conference of the European Chapter of the Association for Computational Linguistics, {EACL} 2024 - Volume 1: Long Papers, St. Julian's, Malta, March 17-22, 2024}}, \bibfield{editor}{\bibinfo{person}{Yvette Graham} {and} \bibinfo{person}{Matthew Purver}} (Eds.). \bibinfo{publisher}{Association for Computational Linguistics}, \bibinfo{pages}{2274--2287}.
\newblock
\urldef\tempurl%
\url{https://aclanthology.org/2024.eacl-long.139}
\showURL{%
\tempurl}


\bibitem[Weller et~al\mbox{.}(2025b)]%
        {rank1}
\bibfield{author}{\bibinfo{person}{Orion Weller}, \bibinfo{person}{Kathryn Ricci}, \bibinfo{person}{Eugene Yang}, \bibinfo{person}{Andrew Yates}, \bibinfo{person}{Dawn~J. Lawrie}, {and} \bibinfo{person}{Benjamin~Van Durme}.} \bibinfo{year}{2025}\natexlab{b}.
\newblock \showarticletitle{Rank1: Test-Time Compute for Reranking in Information Retrieval}.
\newblock \bibinfo{journal}{\emph{CoRR}}  \bibinfo{volume}{abs/2502.18418} (\bibinfo{year}{2025}).
\newblock
\showeprint[arXiv]{2502.18418}
\href{https://doi.org/10.48550/ARXIV.2502.18418}{doi:\nolinkurl{10.48550/ARXIV.2502.18418}}


\bibitem[Wolf et~al\mbox{.}(2019)]%
        {huggingface_transformers}
\bibfield{author}{\bibinfo{person}{Thomas Wolf}, \bibinfo{person}{Lysandre Debut}, \bibinfo{person}{Victor Sanh}, \bibinfo{person}{Julien Chaumond}, \bibinfo{person}{Clement Delangue}, \bibinfo{person}{Anthony Moi}, \bibinfo{person}{Pierric Cistac}, \bibinfo{person}{Tim Rault}, \bibinfo{person}{R{\'{e}}mi Louf}, \bibinfo{person}{Morgan Funtowicz}, {and} \bibinfo{person}{Jamie Brew}.} \bibinfo{year}{2019}\natexlab{}.
\newblock \showarticletitle{HuggingFace's Transformers: State-of-the-art Natural Language Processing}.
\newblock \bibinfo{journal}{\emph{CoRR}}  \bibinfo{volume}{abs/1910.03771} (\bibinfo{year}{2019}).
\newblock
\showeprint[arXiv]{1910.03771}
\urldef\tempurl%
\url{http://arxiv.org/abs/1910.03771}
\showURL{%
\tempurl}


\bibitem[Yang et~al\mbox{.}(2018)]%
        {hotpotqa}
\bibfield{author}{\bibinfo{person}{Zhilin Yang}, \bibinfo{person}{Peng Qi}, \bibinfo{person}{Saizheng Zhang}, \bibinfo{person}{Yoshua Bengio}, \bibinfo{person}{William Cohen}, \bibinfo{person}{Ruslan Salakhutdinov}, {and} \bibinfo{person}{Christopher~D. Manning}.} \bibinfo{year}{2018}\natexlab{}.
\newblock \showarticletitle{{H}otpot{QA}: A Dataset for Diverse, Explainable Multi-hop Question Answering}. In \bibinfo{booktitle}{\emph{Proceedings of the 2018 Conference on Empirical Methods in Natural Language Processing}}, \bibfield{editor}{\bibinfo{person}{Ellen Riloff}, \bibinfo{person}{David Chiang}, \bibinfo{person}{Julia Hockenmaier}, {and} \bibinfo{person}{Jun{'}ichi Tsujii}} (Eds.). \bibinfo{publisher}{Association for Computational Linguistics}, \bibinfo{address}{Brussels, Belgium}, \bibinfo{pages}{2369--2380}.
\newblock
\href{https://doi.org/10.18653/v1/D18-1259}{doi:\nolinkurl{10.18653/v1/D18-1259}}


\bibitem[Yu et~al\mbox{.}(2024)]%
        {yu2024arcticembed}
\bibfield{author}{\bibinfo{person}{Puxuan Yu}, \bibinfo{person}{Luke Merrick}, \bibinfo{person}{Gaurav Nuti}, {and} \bibinfo{person}{Daniel Campos}.} \bibinfo{year}{2024}\natexlab{}.
\newblock \showarticletitle{Arctic-Embed 2.0: Multilingual Retrieval Without Compromise}.
\newblock \bibinfo{journal}{\emph{CoRR}}  \bibinfo{volume}{abs/2412.04506} (\bibinfo{year}{2024}).
\newblock
\showeprint[arXiv]{2412.04506}
\href{https://doi.org/10.48550/ARXIV.2412.04506}{doi:\nolinkurl{10.48550/ARXIV.2412.04506}}


\bibitem[Zeng et~al\mbox{.}(2025)]%
        {zeng2025scalingsparsedenseretrieval}
\bibfield{author}{\bibinfo{person}{Hansi Zeng}, \bibinfo{person}{Julian Killingback}, {and} \bibinfo{person}{Hamed Zamani}.} \bibinfo{year}{2025}\natexlab{}.
\newblock \bibinfo{title}{Scaling Sparse and Dense Retrieval in Decoder-Only LLMs}.
\newblock
\showeprint[arxiv]{2502.15526}~[cs.IR]
\urldef\tempurl%
\url{https://arxiv.org/abs/2502.15526}
\showURL{%
\tempurl}


\bibitem[Zhang(2008)]%
        {students_mental_model_of_information_retrieval_systems}
\bibfield{author}{\bibinfo{person}{Yan Zhang}.} \bibinfo{year}{2008}\natexlab{}.
\newblock \showarticletitle{Undergraduate students' mental models of the Web as an information retrieval system}.
\newblock \bibinfo{journal}{\emph{J. Assoc. Inf. Sci. Technol.}} \bibinfo{volume}{59}, \bibinfo{number}{13} (\bibinfo{year}{2008}), \bibinfo{pages}{2087--2098}.
\newblock
\href{https://doi.org/10.1002/ASI.20915}{doi:\nolinkurl{10.1002/ASI.20915}}


\bibitem[Zheng et~al\mbox{.}(2025)]%
        {zheng2025reasoningfocused}
\bibfield{author}{\bibinfo{person}{Lucia Zheng}, \bibinfo{person}{Neel Guha}, \bibinfo{person}{Javokhir Arifov}, \bibinfo{person}{Sarah Zhang}, \bibinfo{person}{Michal Skreta}, \bibinfo{person}{Christopher~D. Manning}, \bibinfo{person}{Peter Henderson}, {and} \bibinfo{person}{Daniel~E. Ho}.} \bibinfo{year}{2025}\natexlab{}.
\newblock \showarticletitle{A Reasoning-Focused Legal Retrieval Benchmark}. In \bibinfo{booktitle}{\emph{Proceedings of the 2025 Symposium on Computer Science and Law, {CSLAW} 2025, Munich, Germany, March 25-27, 2025}}. \bibinfo{publisher}{{ACM}}, \bibinfo{pages}{169--193}.
\newblock
\href{https://doi.org/10.1145/3709025.3712219}{doi:\nolinkurl{10.1145/3709025.3712219}}


\bibitem[Zhong et~al\mbox{.}(2023)]%
        {romqa}
\bibfield{author}{\bibinfo{person}{Victor Zhong}, \bibinfo{person}{Weijia Shi}, \bibinfo{person}{Wen{-}tau Yih}, {and} \bibinfo{person}{Luke Zettlemoyer}.} \bibinfo{year}{2023}\natexlab{}.
\newblock \showarticletitle{RoMQA: {A} Benchmark for Robust, Multi-evidence, Multi-answer Question Answering}. In \bibinfo{booktitle}{\emph{Findings of the Association for Computational Linguistics: {EMNLP} 2023, Singapore, December 6-10, 2023}}, \bibfield{editor}{\bibinfo{person}{Houda Bouamor}, \bibinfo{person}{Juan Pino}, {and} \bibinfo{person}{Kalika Bali}} (Eds.). \bibinfo{publisher}{Association for Computational Linguistics}, \bibinfo{pages}{7055--7067}.
\newblock
\href{https://doi.org/10.18653/V1/2023.FINDINGS-EMNLP.470}{doi:\nolinkurl{10.18653/V1/2023.FINDINGS-EMNLP.470}}


\bibitem[Zhuang et~al\mbox{.}(2025)]%
        {rank_r1}
\bibfield{author}{\bibinfo{person}{Shengyao Zhuang}, \bibinfo{person}{Xueguang Ma}, \bibinfo{person}{Bevan Koopman}, \bibinfo{person}{Jimmy Lin}, {and} \bibinfo{person}{Guido Zuccon}.} \bibinfo{year}{2025}\natexlab{}.
\newblock \showarticletitle{Rank-R1: Enhancing Reasoning in LLM-based Document Rerankers via Reinforcement Learning}.
\newblock \bibinfo{journal}{\emph{CoRR}}  \bibinfo{volume}{abs/2503.06034} (\bibinfo{year}{2025}).
\newblock
\showeprint[arXiv]{2503.06034}
\href{https://doi.org/10.48550/ARXIV.2503.06034}{doi:\nolinkurl{10.48550/ARXIV.2503.06034}}


\end{thebibliography}

\end{document}